\documentclass[12pt]{article}
\pdfoutput=1

\usepackage{enumerate}
\usepackage[bottom]{footmisc}
\usepackage[T1]{fontenc}
\usepackage{tikz-feynman}
\usepackage{simpler-wick}
\tikzfeynmanset{compat=1.1.0}
\graphicspath{{img/}}
\usepackage{cite}

\usepackage[utf8]{inputenc}
\usepackage[english]{babel}
\usepackage{amsmath, amsfonts, amssymb}
\usepackage{nccmath}
\usepackage{mathrsfs}
\usepackage[mathscr]{euscript}
\usepackage{color}
\usepackage{amsthm}
\usepackage{psfrag}
\usepackage{graphicx}
\usepackage{url}
\usepackage[hyperfootnotes=false]{hyperref}
\usepackage{multirow}
\usepackage{makecell}
\usepackage{bbm}

\renewcommand{\theequation}{\arabic{section}.\arabic{equation}}
\renewcommand{\thesection}{\arabic{section}}
\renewcommand{\thesubsection}{\arabic{section}.\arabic{subsection}}
\catcode`@=11
\@addtoreset{equation}{section}
\catcode`@=12

\newcommand{\be}{\begin{equation}}
\newcommand{\ee}{\end{equation}}
\newcommand{\bl}{\begin{align}}
\newcommand{\el}{\end{align}}
\newcommand{\ba}{\begin{aligned}}
\newcommand{\ea}{\end{aligned}}
\newcommand{\beqa}{\begin{eqnarray}}
\newcommand{\eeqa}{\end{eqnarray}}
\newcommand{\LL}{{\cal L}}

\newcommand\D{\Delta}

\newcommand\s{\mathbf{s}}
\newcommand{\ve}{\varepsilon}
\renewcommand{\a}{\alpha}
\renewcommand{\b}{\beta}

\renewcommand{\P}{{\cal P}}

\def\e{{\rm e}}
\def\d{\partial}

\renewcommand{\k}{\mathbf{k}}
\newcommand{\x}{\mathbf{x}}
\newcommand{\sign}{\mathop{\rm sign}\nolimits}

\newcommand{\bseq}{\begin{subequations}}
\newcommand{\eseq}{\end{subequations}}

\newcommand{\bra}[1]{\langle #1 |}
\newcommand{\ket}[1]{| #1 \rangle}
\newcommand{\braket}[2]{\langle #1 |#2 \rangle}

\allowdisplaybreaks[4]

\begin{document}
\begin{titlepage}
\clearpage

\title{{\bf Scattering amplitudes in high-energy limit }\\
{\bf of projectable Ho\v rava gravity} }
\author{Jury I. Radkovski$^{a,b}$\footnote{radkovsj@mcmaster.ca}~ and Sergey M. Sibiryakov$^{a,b}$\footnote{ssibiryakov@perimeterinstitute.ca}\\[2mm]
{\small\it $^a$Department of Physics and Astronomy, McMaster
  University,}\\ 
{\small \it 1280 Main Street West, Hamilton, ON L8S 4M1, Canada}\\
{\small\it $^b$Perimeter Institute for Theoretical Physics, Waterloo,
  Ontario, N2L 2Y5, Canada}
}
\date{}
\maketitle

\begin{abstract}
We study the high-energy limit of projectable Ho\v rava gravity using on-shell
graviton scattering amplitudes. We compute the tree-level amplitudes
using symbolic computer algebra 
and analyze their properties in the case of collisions with zero total
momentum. 
The amplitudes grow with collision energy in the way
consistent with tree-level unitarity. 
We discuss their angular dependence and derive
the expression for the differential cross section that happens to
depend only on the essential combinations of the couplings.
One of
our key results is that the amplitudes for arbitrary kinematics are
finite when the coupling $\lambda$ in the kinetic Lagrangian is taken
to infinity --- the value corresponding to candidate asymptotically
free
ultraviolet
fixed points of the theory. We formulate a
modified action
which reproduces the same amplitudes and
is directly applicable at $\lambda=\infty$, thereby establishing that the limit
$\lambda\to\infty$ of projectable Ho\v rava gravity is
regular. As an auxiliary result, we derive the generalized Ward
identities for the amplitudes 
in non-relativistic gauge theories. 
\end{abstract} 

\thispagestyle{empty}
\end{titlepage}

\newpage

\tableofcontents

\section{Introduction}
\label{intro}
Ho\v rava gravity (HG), proposed in \cite{Horava2009}, is a metric
quantum theory of gravity realized as a power-counting renormalizable
quantum field theory (see
\cite{Blas2010,Mukohyama2010,Sotiriou2010,Wang2017,
Barvinsky2023} for reviews). The
power-counting renormalizability is achieved by separating spacetime
into space \textit{and} time: 
The theory at tree level and at high energies is
taken to be invariant
under anisotropic (Lifshitz) scaling 
\begin{equation}
    {\bf x} \rightarrow b^{-1} \, {\bf x}, \quad t \rightarrow b^{-z} \,
    t \label{eq:anis_scaling} \, , 
\end{equation}
where $b$ is a scaling parameter and $z$ is the Lifshitz exponent. In
HG the latter is taken to be equal to the number of spatial dimensions, $z=
d$.
Such a symmetry implies that we can have a Lagrangian quadratic in
first time derivatives, yet containing terms with more than two
spatial derivatives of fields. The propagators then have more powers
of momenta than of energy in the denominators, which  makes them decay
fast in the ultraviolet (UV) and improves convergence of the loop
integrals in perturbation theory. 
Since the equations of motion contain
only two time derivatives,
we do not get any problematic extra degrees of freedom 
(ghosts), in contrast to the generally covariant higher curvature
gravity 
\cite{Stelle:1976gc,Stelle:1977ry,Salvio2014,Einhorn2014}.\footnote{See 
\cite{Salvio:2015gsi,Strumia:2017dvt} and references
  therein for suggested interpretations of quantum theories with
  higher time derivatives.} 

The price to pay is the  violation of Lorentz invariance at high
energies that propagates down to low energies in the form of a
preferred spacelike foliation whose dynamics is described by a new
scalar field called {\it scalar  graviton} or {\it khronon} \cite{Blas2010}. The
violation of Lorentz invariance in the visible sector can be
sufficiently small in the {\it non-projectable} version of the theory 
\cite{Blas2009} to reproduce the observed phenomenology, albeit with some
degree of tuning \cite{EmirGumrukcuoglu:2017cfa}. From the theoretical
perspective, the non-projectable theory is complicated since it
involves large (but still finite) number of marginal couplings that
describe its behavior in the UV. Its renormalizability beyond power
counting has not yet been established, though there has been an important progress in
this direction recently 
\cite{Bellorin:2022qeu,Bellorin:2022efu}. Further analysis of its UV
properties, such as the renormalization group (RG) flow, is presently
beyond reach.   

In this paper we consider a simpler version of the theory: the {\it
  projectable} model, which has been proven to be perturbatively
renormalizable \cite{Barvinsky:2015kil,Barvinsky:2017zlx} and whose
one-loop RG flow has been computed in
\cite{Barvinsky2019,Barvinsky2021}. The flow possesses a number of UV
fixed points with vanishing gravitational constant which indicates
asymptotically free behavior. Some of these points, however, are
characterized by a divergent dimensionless coefficient 
in the kinetic
term of the action conventionally denoted by $\lambda$. Since
positive powers of
$\lambda$ appear in the interaction vertices, one may worry if this
divergence jeopardizes the 
asymptotic freedom.  

The purpose of the present paper is to address this concern by
scrutinizing the projectable HG in the limit\footnote{The
  directionality of the limit, i.e. whether $\lambda$ goes to
  $+\infty$ or $-\infty$ is unimportant, at least within the
  perturbation theory.}
\be
\label{limit}
\lambda\to\infty~,~~~~~~~\text{other couplings fixed.}
\ee
Early work \cite{Gumrukcuoglu2011_1} studied cosmological
perturbations in HG and showed that their power spectrum and cubic
interactions remain well-behaved in the limit
(\ref{limit}), suggesting that it corresponds to a regular
theory. 
More recently a similar limit in a supersymmetric version of HG has
been connected to the Perelman--Ricci flows \cite{Frenkel2020}.

We take a different approach and
use the scattering amplitudes as gauge-invariant probes of the theory.  
We compute the full set of tree-level amplitudes for $2\to2$ scattering of
transverse and scalar gravitons in the projectable HG taking into
account all marginal couplings. This calculation is of high algebraic
complexity which we overcome by making use of computer algebra
\cite{xAct,xPerm,xPert,xTras}.
The resulting expressions for the amplitudes at general kinematics are
too cumbersome to be analyzed explicitly,\footnote{They are
  available in the {\it Mathematica} \cite{Mathematica} format at 
\cite{github}.}  
so we focus in the paper on the simplest case of
scattering with vanishing net momentum, to which we refer as `head-on
collisions'.\footnote{In contrast to
  relativistic theories, this is a genuine
  restriction. Due to the absence of Lorentz invariance in HG one
  cannot set the net momentum to zero by boosting to the
  center-of-mass frame.}
We discuss the energy and angular dependence of the amplitudes and
observe that they are finite in the limit
(\ref{limit}). We 
verify the latter property for an arbitrary kinematics using our code
and encouraged by these
results develop an analytic proof of cancellation
between potentially divergent contributions. 
Further, we show that a reformulation of the theory with introduction
of an additional auxiliary field allows one to take the limit
(\ref{limit}) at the level of the action, implying that this limit is
regular beyond the
tree-level and 
$2\to 2$ processes.

The complexity of the amplitudes calls for subjecting them to various
consistency checks. An important class of such checks are requirements
of gauge invariance. In relativistic theories and for relativistic
gauges they imply two types of conditions. First, the on-shell
amplitudes for physical states must be independent of the gauge-fixing
parameters. Second, they must satisfy the Ward identities stating that
an amplitude for scattering of a gauge boson vanishes 
whenever its polarization vector (for Yang--Mills theories) or
tensor (for gravity) is replaced by a vector / tensor proportional to
the boson's four-momentum. While the first condition translates
without change to non-relativistic theories, the second is less
obvious since the four-momentum is no longer a useful object. 
To generalize the Ward identities to the case of
HG, we go back to first
principles and construct its Hilbert space using the 
Becchi--Rouet--Stora--Tyutin (BRST) quantization. The sought after
conditions then arise from the requirement of the BRST invariance of the
${\cal S}$-matrix. This approach is not restricted to HG and applies
to any non-relativistic gauge theory, as we illustrate on an example
of a Yang--Mills model with $z=2$ Lifshitz scaling in $(4+1)$
dimensions. 

It is worth noting that the phenomenological viability of projectable
HG is problematic since it does not possess a stable perturbative 
Minkowski vacuum
where gravitons would propagate with the speed of light
\cite{Koyama:2009hc,Blas2010} (see also \cite{Barvinsky2023} for
recent discussion).  
Refs.~\cite{Mukohyama2010,Izumi2011,Gumrukcuoglu2011} suggested that
it may still reproduce general relativity with an additional sector
behaving as dark matter if the khronon field is strongly coupled. We
do not attempt to add anything to this aspect of the model and focus
on its properties at high energies where it is stable and weakly
coupled. 

The paper is organized as follows. In Sec.~\ref{sec:hl gravity} we
review the projectable HG and perform its BRST quantization. 
In Sec.~\ref{sec:brst}
we derive the generalized Ward identities for
scattering
amplitudes in non-relativistic gauge theories, illustrating the
general framework on the Yang--Mills theory with Lifshitz scaling 
before applying it to HG. 
In Sec.~\ref{sec:amplitude} 
we outline the calculation of amplitudes in projectable HG and present
our results for scattering with zero total
momentum. In Sec.~\ref{sec:linf} we consider the limit (\ref{limit})
and show that the amplitudes remain finite. We also present 
an alternative formulation of the theory which allows us to
take the limit (\ref{limit}) directly at the level of the action. 
We conclude in Sec.~\ref{sec:discussion}. Lengthy formulas
are relegated to Appendices.


\section{Projectable Ho\v rava Gravity}
\label{sec:hl gravity}

\subsection{Formulating the theory}

A theory symmetric under scaling \eqref{eq:anis_scaling} cannot
be invariant under the full group of spacetime
diffeomorphisms. However, it can still be invariant under its
foliation-preserving subgroup (FDiffs):
\begin{equation}
    {\bf x} \mapsto \tilde{\bf x}(\mathbf{x},t) \, , 
\quad t \mapsto
    \tilde{t}(t) \, , 
\end{equation}
with $\tilde{t}(t)$ - monotonic function. Ho\v rava gravity
\cite{Horava2009} is a metric theory with this symmetry,
conventionally formulated using the Arnowitt--Deser--Misner (ADM)
decomposition of the spacetime line element,
\begin{equation}
    ds^2 = -N^2 dt^2 + \gamma_{ij} (dx^i + N^i dt ) (dx^j + N^j dt ) \,
    , \quad i,j = 1, 2, 3 \, ,
\end{equation}
where we have specified to three spatial dimensions. The {\it lapse},
{\it shift} and the spatial metric transform under FDiffs as 
\begin{equation}
    N \mapsto N \frac{d t}{d \tilde{t}} \, , \quad N^{i}
    \mapsto \left(N^{j} \frac{\partial \tilde{x}^{i}}{\partial
        x^{j}} - \frac{\partial \tilde{x}^i}{\partial t} \right)\frac{d
      t}{d \tilde{t}} \, , \quad \gamma_{i j} \mapsto \gamma_{k l}
    \frac{\partial x^{k}}{\partial \tilde{x}^{i}} \frac{\partial
      x^{l}}{\partial \tilde{x}^{j}} \, . \label{eq:transform_law} 
\end{equation}
These transformations
are compatible with the {\it projectability}
condition which states that the lapse $N$ is only a function of time, 
$N = N(t)$. In this case it can be set to $1$ by an appropriate choice
of the time coordinate. Equivalently, at least in perturbation theory,
we can consider a model without time reparameterizations and with unit
lapse from the start. This is the formulation we adopt in this
paper. An alternative option --- taking the lapse to be a function of both
time and space --- leads to the non-projectable~HG.

Using the remaining variables $\gamma_{ij}$ and $N^i$ we construct the
most general action with two time derivatives which is invariant under
FDiffs (\ref{eq:transform_law}) and the Lifshitz scaling
(\ref{eq:anis_scaling}) with $z=3$. To do this, we need to assign the
{\it scaling dimensions} to the metric and the shift, which will
determine how their quantum fluctuations scale in the UV. In more
detail, we say that a field $\Phi$ has scaling dimension $\dim\Phi=r$
if under the symmetry (\ref{eq:anis_scaling}) it transforms as
\be
\label{genscal}
\Phi({\bf x},t)\mapsto \Phi'(b^{-1}{\bf x},b^{-z} t)=b^r\,
\Phi({\bf x},t)\;.
\ee  
The metric $\gamma_{ij}$ enters into the action non-linearly, while
its time
derivative enters through the extrinsic
curvature of the constant-time slices which transforms covariantly
under the FDiffs,\footnote{The indices are raised
and lowered using the spatial metric $\gamma_{ij}$.}
\be
\label{eq:Kij}
    K_{ij} = \frac{1}{2}\big(
\dot{\gamma}_{ij} -\nabla_{i}N_{j}-\nabla_{j}N_{i}\big)\;.
\ee
To
preserve the homogeneous scaling of different terms in the action, we assign
the dimensions $0$ to $\gamma_{ij}$ and $2$ to $N_i$,
\be
\label{dimsgammaN}
\dim\gamma_{ij}=0~,~~~~~\dim N_i=2\;.
\ee
This leads us to the action, 
\begin{equation}
\label{eq:HG_Action}
    S = \frac{1}{2 G} \int d^3 x dt 
\sqrt{\gamma} \left( K_{ij}K^{ij} - \lambda K^2 - \mathcal{V} \right) \, ,
\end{equation}
where $G$ is the gravitational coupling controlling the overall
strength of the interactions and $K\equiv \gamma^{ij}K_{ij}$ is trace of
the extrinsic curvature. Note the free parameter $\lambda$ which 
appears in the kinetic term of HG compared to general relativity,
where it is fixed to be $\lambda=1$ by the full spacetime
diff-invariance. 

The ``potential'' term ${\cal V}$ in Eq.~(\ref{eq:HG_Action}) depends
on three-dimensional curvature invariants constructed using the spatial
metric $\gamma_{ij}$. To be compatible with the Lifshitz scaling, it
must consist of operators with scaling dimension $6$. The most general
such potential reads,
\be
\label{pothigh}
\mathcal{V}^{\dim=6}=\nu_1 R^3 + \nu_2 R R_{ij} R^{ij} + 
\nu_3 R_{ij}R^{jk}R_{k}^{i} + \nu_4 \nabla_i R \nabla^i R 
+ \nu_5 \nabla_i R_{jk} \nabla^i R^{jk}\;, 
\ee 
where $R_{ij}$ and $R$ are the three-dimensional Ricci tensor and the
scalar curvature, respectively; $\nu_a$, $a=1,\ldots,5$, are coupling
constants. Note that there are no terms with the Riemann tensor, since
in three dimensions it is not independent and can be expressed through
$R_{ij}$. 

One can also add to the potential the terms of lower scaling dimension
which represent relevant deformations of the Lifshitz scaling,
\be 
\label{potlow}
\mathcal{V}^{\dim<6} = 
2 \Lambda - \eta R + \mu_1 R^2 + \mu_2 R_{ij}R^{ij}\;. 
\ee
In fact, these terms are required for renormalizability since the
Lifshitz scaling is broken by quantum corrections, as manifested by
the RG running of the couplings \cite{Barvinsky2021}. In this paper we
disregard the 
low-dimension terms because we are interested in the high-energy
properties of the theory controlled by the marginal operators collected
in (\ref{pothigh}).

\subsection{BRST quantization}
\label{sec:BRSTHG}

The flat static metric $\gamma_{ij}=\delta_{ij}$ with vanishing shift
$N^i=0$ is a solution of the classical equations following from the
action (\ref{eq:HG_Action}) with the potential (\ref{pothigh}). We
want to quantize the theory around this background, so we introduce
the metric perturbation
\be
\label{metrpert}
h_{ij}\equiv \gamma_{ij}-\delta_{ij}\;.
\ee
Next we need to fix the gauge. This is done consistently within the
BRST formalism \cite{Becchi:1975nq,Tyutin:1975qk}; we follow here
\cite{Weinberg:1996kr,Barvinsky:2015kil}. We introduce the fermionic
Faddeev--Popov ghosts $c^i$ and anti-ghosts $\bar c_i$, bosonic
Nakanishi--Lautrup field $b_i$ and the Slavnov operator $\s$ which
implements the BRST transformations of the original and new fields,
\begin{subequations}
\label{BRSTall}
\begin{gather}
\label{BRSThN}
    \mathbf{s} h_{ij} = \partial_{i} c_{j} + \partial_{j} c_{i} 
+ \partial_{i}c^{k}h_{j k} + \partial_{j}c^{k}h_{i k} +
c^{k} \partial_{k} h_{ij}\;,~~~~ 
    \mathbf{s} N^{i} = \dot{c}^{i} - N^{j} \partial_{j}c^{i} +
    c^{j}\partial_{j}N^i\;,\\ 
\label{BRSTghosts}
 \mathbf{s}c^{i} = c^{j} \partial_{j} c^{i}\;,~~~~ 
\mathbf{s} \bar{c}_{i} = b_{i}\;,~~~~
  \mathbf{s} b_i = 0\;.
\end{gather}
\end{subequations}
Note that from now on the indices are raised and lowered with flat
background metric $\delta_{ij}$. The first two expressions here are,
of course, nothing but the infinitesimal gauge transformations of the
metric and shift with the gauge parameters replaced by the
ghosts. 
With these definitions it is straightforward to to show that the
Slavnov operator is nilpotent, i.e. the action of $\s^2$ on any field
vanishes.\footnote{In the proof one should recall that, since $\s$
  is a fermionic operator, it obeys a graded Leibniz rule: $\s
  (AB)=(\s A)B+(-1)^{|A|}A(\s B)$, where $|A|=0$ ($|A|=1$) for a bosonic
  (fermionic) field $A$.} 

The quantum tree-level action is constructed as the sum of the
original action (\ref{eq:HG_Action}) and the BRST variation of a
gauge-fixing fermion $\Psi$,
\begin{equation}
\label{quantact}
    S_q=S + \frac{1}{2G}\int d^3xdt\,\s\Psi \, .
\end{equation}
Gauge invariance of the original action and the nilpotency of the
Slavnov operator imply that $S_q$ is BRST invariant, $\s S_q=0$. The
gauge-fixing fermion is conventionally taken in the form
\begin{equation}
\label{GDfermion}
    \Psi =  2 \bar{c}_{i}\, F^{i} -  \bar{c}_{i}\, O^{i j }\, b_j \, ,
\end{equation}
where $F^i$ are the gauge-fixing functions and $O^{ij}$ is a
non-degenerate operator.

Following \cite{Barvinsky:2015kil} we adopt a family of gauges
compatible with the Lifshitz scaling and possessing two free
parameters $\sigma$, $\xi$ :
\begin{equation}
    \label{eq:GFF}
F^i = \dot{N}^{i} + \frac{1}{2} O^{i j}\big( \partial_k h_{j}^k 
- \lambda \partial_{j} h \big) \;,\qquad  
O^{ij} = -\frac{1}{\sigma} \big( \delta^{i j}\Delta^2 +
      \xi \partial^{i}\Delta \partial^j\big)\;, 
\end{equation}
where $h\equiv h_k^k$ is the trace of the metric perturbation and
$\Delta\equiv \d_k\d^k$ is the spatial Laplacian.\footnote{The
  operator $O^{ij}$ corresponds to $-\sigma^{-1}(\mathcal{O}^{-1})^{ij}$ in
the notations of \cite{Barvinsky:2015kil}. The sign difference is due to
the fact that \cite{Barvinsky:2015kil} works with the Euclidean
version of the theory obtained upon the Wick rotation, 
whereas here we work in the physical time.} 
Upon substituting these expressions into (\ref{quantact}), it is
convenient to integrate out the non-dynamical Nakanishi--Lautrup
field, and the action takes the form,\footnote{This procedure produces
a factor $(\det O^{ij})^{-1/2}$ in the path integral measure of the
theory, which is irrelevant at tree level.} 
\begin{equation}
\label{newSq}
    S_q =S+ \int d^3xdt\bigg(\frac{1}{2G}F^i O_{ij}^{-1} F^j
-\frac{1}{G} \bar c_i\,\s F^i\bigg)\;.
\end{equation}
The first term in the brackets is the gauge-fixing Lagrangian, whereas
the second term gives the Lagrangian for ghosts. Note that the
operator 
\be
\label{Oinv}
O_{ij}^{-1}=-\frac{\sigma}{\Delta^2}+\frac{\sigma\xi}{(1+\xi)}
\frac{\d_i\d_j}{\Delta^3}
\ee 
is non-local in space which, however, does not
lead to any complications since it enters the action only at the
quadratic order. 

Integrating out the Nakanishi--Lautrup field modifies the BRST
transformation of the anti-ghosts which now reads,
\begin{equation}
\label{newBRSTbarc}
    \mathbf{s} \bar{c}_{i} = O_{ij}^{-1} F^{j} \, .
\end{equation}
In other words, it is proportional to the gauge-fixing functions. This
fact will be exploited in the next section when discussing the BRST
invariance of the scattering amplitudes. Note that the nilpotency of
the transformation (\ref{newBRSTbarc}) requires $\s F^i=0$ which is
satisfied only on-shell. Indeed, this is precisely the equation of
motion for ghosts, as one can see by varying the action (\ref{newSq})
with respect to $\bar c_i$.

We are now ready to quantize the theory and define its Fock space. To
this end, we focus on the quadratic part of the Lagrangian. From the
action (\ref{newSq}) we have,
\be
\label{L2q}
\begin{split}
{\cal L}_q^{(2)}=\frac{1}{2G}\bigg\{&\frac{\dot h_{ij}^2}{4}
-\frac{\lambda \dot h^2}{4}
+\frac{\nu_5}{4}h_{ij}\D^3 h_{ij} 
+\bigg(\frac{\nu_5}{2}-\frac{1}{4\sigma}\bigg)\d_j h_{ji}\D^2\d_k
h_{ki}\\
&+\bigg(\nu_4+\frac{\nu_5}{2}+\frac{\xi}{4\sigma}\bigg)
\d_i\d_jh_{ij}\D\d_k\d_l h_{kl}
-\bigg(2\nu_4+\frac{\nu_5}{2}+\frac{\lambda(1+\xi)}{2\sigma}\bigg)
\D^2 h\,\d_i\d_j h_{ij}\\
&+\bigg(\nu_4+\frac{\nu_5}{4}+\frac{\lambda^2(1+\xi)}{4\sigma}\bigg)
h\D^3h\\
&-\dot N_i\frac{\sigma}{\D^2}\dot N_i
-\d_i\dot N_i\frac{\sigma\xi}{(1+\xi)\D^3}\d_j\dot N_j
-\frac{1}{2}N_i\D N_i+\bigg(\frac{1}{2}-\lambda\bigg)(\d_i N_i)^2
\bigg\}\\
+\frac{1}{G}\bigg\{&\dot{\bar c}_i\dot c_i
+\frac{1}{2\sigma} \bar c_i\D^3 c_i
+\frac{\xi+(1+\xi)(1-2\lambda)}{2\sigma} \bar c_i\D^2\d_i\d_jc_j\bigg\}.
\end{split}
\ee
where we have made various integrations by part and placed all
indices downwards for simplicity. We see the advantage of the gauge
(\ref{eq:GFF}): it decouples $h_{ij}$ and $N_i$ in the quadratic
action which significantly simplifies the quantization. We next
perform the helicity decomposition of the fields entering (\ref{L2q}),
diagonalize the Lagrangian and solve the respective equations of
motion. This leads us to a set of positive-frequency modes which we
label with the spatial momentum $\k$ and helicity $\alpha$ :
\bseq
\label{modeset}
\begin{align}
&h_{\k\a}~,&&\a=\pm 2\,,~\pm 1\,,~0\,,~ 0'\,,\\
&N_{\k\a}\,,~c_{\k\a}\,,~\bar c_{\k\a}~,&&\a=\pm 1\,,~0\;.
\end{align}
\eseq
The details, including the expressions for the polarization vectors /
tensors of the modes, are given in Appendix~\ref{app:decomp}.

The modes with helicities $\pm 2$ are present only in the metric and
correspond to transverse traceless (tensor) gravitons. Their on-shell
dispersion
relation is manifestly gauge invariant,
\begin{equation}
\label{omtt}
    \omega^2_{tt} = \nu_5 k^6 \,. 
\end{equation}
The stability of the
mode requires $\nu_5>0$. The modes with
helicities $\pm 1$ and $0$ are pure gauge and have dispersion
relations
\be
\label{omga}
\omega_1^2=\frac{k^6}{2\sigma}~,~~~~~
\omega_0^2=\frac{(1-\lambda)(1+\xi)}{\sigma}k^6\;,
\ee
which clearly depend on the gauge parameters. We choose the latter in
such a way that both $\omega_1^2$ and $\omega_0^2$ are positive.
Finally, an additional
scalar mode $0'$ is present in the metric. This is physical and
corresponds to a scalar graviton of HG. Its dispersion relation is
gauge invariant and reads, 
\begin{equation}
\label{oms}
\omega^2_s = \nu_s k^6~,~~~~~~~~
\nu_s =\frac{1- \lambda}{1 - 3 \lambda} (8 \nu_4 + 3 \nu_5) \, .
\end{equation}
The mode is stable provided $\nu_s>0$ which
together with the positivity of the kinetic
term (see Appendix~\ref{app:decomp}) 
implies $\lambda <1/3$ {\em or} $\lambda>1$ and $8\nu_4+3\nu_5>0$. 

Upon quantization, the coefficients of the positive-frequency modes
(\ref{modeset}) become the
annihilation operators and together with their respective creation
operators $h_{\k\a}^+$, $N_{\k\a}^+$, $\bar c_{\k\a}^+$, $c_{\k\a}^+$
generate the Fock space. The states with only the
transverse traceless and scalar $0'$ gravitons have positive norm,
whereas the gauge sector with helicities $\pm 1$ and $0$ contains both
positive and negative-norm states. As usual, the negative norm states
are eliminated by restricting to the cohomology of the BRST operator
$Q$ --- the Noether charge associated with the BRST
invariance. Importantly, we are dealing here with the action of the
BRST transformations on the asymptotic states made of free particles,
implying that the transformations are restricted to linear
order. Accordingly, the operator $Q$ is restricted to the quadratic
part, which we will highlight with the superscript `$(2)$'. Applying the
Noether theorem to the quadratic Lagrangian (\ref{L2q}) we obtain,
\be
\label{Q2}
\begin{split}
Q^{(2)}=\frac{1}{2G} \int d^3x\bigg[&\dot c_i\,(\d_j h_{ji}-\lambda \d_i
h)
-c_i \,(\d_j \dot h_{ji}-\lambda \d_i \dot h)\\
&-2\dot c_i \bigg(\frac{\sigma}{\D^2}\dot N_i
-\frac{\sigma\xi}{(1+\xi)\D^3}\d_i\d_j\dot N_j\bigg)
+c_i\D N_i+(1-2\lambda) c_i\,\d_i\d_jN_j\bigg].
\end{split}
\ee
Note that if we want the BRST charge to be Hermitian, we 
must choose the ghost field $c_i$ to be Hermitian as well. Then the
Hermiticity of the Lagrangian requires the
anti-ghost $\bar c_i$ to be anti-Hermitian.
Using the commutation relations from Appendix~\ref{app:decomp}, one
verifies that
\be
\label{Qcomm}
i[Q^{(2)},\Phi]_{\mp}=(\s \Phi)_{\rm lin}\;,
\ee
for any field $\Phi$ of the theory. Here the square brackets with
subscript $\mp$ mean commutator (anti-commutator) for bosonic
(fermionic) fields, and $(\s \Phi)_{\rm lin}$ is the linear part of
the BRST transformations (\ref{BRSTall}),
(\ref{newBRSTbarc}). Clearly, $Q^{(2)}$ is nilpotent since $\s$ is
nilpotent on-shell. 

Physical states $\ket{\psi}$ have zero ghost
number\footnote{Defined as the number of ghosts minus the number of
  anti-ghosts. It corresponds to the symmetry of the action
  (\ref{newSq}) under opposite scaling of the ghost and anti-ghost
fields and is preserved by the evolution.} 
and are $Q^{(2)}$-closed. Besides, two states are equivalent
if their difference is $Q^{(2)}$-exact. Thus, we have
\be
\label{physcond}
Q^{(2)}\ket{\psi}=0~,~~~~~~~~
\ket{\psi_1}\sim \ket{\psi_2}~~~ \leftrightarrow~~~
\ket{\psi_1}=\ket{\psi_2}+Q^{(2)}\ket{\chi}\;.
\ee
Then using the standard arguments \cite{Kugo:1977yx,Kugo:1979gm} 
one can show that each equivalence class contains a state made only of
the physical tensor
and scalar gravitons. The norm of all states in the equivalence
class coincides with the norm of this state and is positive definite.

\section{Generalized Ward Identities}
\label{sec:brst}

In this section we derive the constraints imposed on the scattering
amplitudes by the BRST invariance of the ${\cal S}$-matrix. We then
illustrate them on an example of non-relativistic Yang--Mills theory
and finally apply to the projectable HG.

\subsection{General considerations}
\label{ssec:brstgen}

We consider a gauge theory that may or may not be relativistic, the
latter case being of primary interest to us. We assume that there
exists an ${\cal S}$-matrix which establishes a map between the
asymptotic {\it in} and {\it out} states,
\be
\braket{q',out}{q,in}=\bra{q',in}{\cal S}\ket{q,in}\;,
\ee
where $q$, $q'$ stand for the collection of quantum numbers such as
particle types, 
momenta and polarizations in the initial and final
states. The space of asymptotic states is assumed to be isomorphic to
the Fock space of non-interacting theory. In what follows we will omit
the labels {\it in} when writing the ${\cal S}$-matrix elements. 

It should be noted that in making these assumptions
we disregard the infrared divergences plaguing the definition of
the ${\cal S}$-matrix in theories with massless particles. In Lifshitz
theories with $z>1$ these problems can be further aggravated due to a
softer scaling of particle energy with the momentum. Moreover, the
dispersion relation $\omega\propto k^z$ with $z>1$ kinematically
allows a single particle to split into two particles, rendering all
particles unstable and further complicating the definition of
asymptotic states. Thus, our derivation below in this subsection
should be considered as rather formal and strictly applicable only at
tree level where the above problems do not arise. Still, we believe
that, with a proper infrared regularization, the end
result should also hold beyond the tree level. We leave its rigorous
derivation for future.

The BRST transformations constitute a symmetry of the gauge-fixed
action implying that the ${\cal S}$-matrix commutes with the BRST
charge. Since the ${\cal S}$-matrix acts on the
asymptotic free-particle states, we have to restrict the charge to its
quadratic part $Q^{(2)}$ which gives,
 \begin{equation}
    [Q^{(2)}, {\cal S}] = 0 
\label{eq:sym_Smatr} \; .
\end{equation}
The restriction to $Q^{(2)}$ here is non-trivial. 
Within the interaction
picture one can think of ${\cal S}$ as the operator describing
non-linear evolution from
$t=-\infty$ to $t=+\infty$.
The full BRST
charge $Q$ commuting with the non-linear Hamiltonian
contains terms of
higher order in the fields, so one may wonder if the higher-order
terms in $Q$ must be also kept in the commutator
(\ref{eq:sym_Smatr}). 
This is not
the case, as can be shown \cite{Becchi:1996yh} 
using the Lehmann--Symanzik--Zimmermann
(LSZ) representation for the ${\cal S}$-matrix. For completeness, we
reproduce the argument in 
Appendix~\ref{app:Smatr}.

The property (\ref{eq:sym_Smatr}) implies that the ${\cal S}$-matrix
element between a physical state $\ket{\psi '}$ and any
$Q^{(2)}$-exact state vanishes,
\begin{equation}
\label{ampgm1}
    \bra{\psi'} {\cal S} Q^{(2)}\ket{\chi}=0\;.
\end{equation}
In particular, for $\ket{\chi}$ we can take a state obtained by adding
an anti-ghost to another physical state,
\be
\label{chigauge}
\ket{\chi}=\bar{c}^+_{\k\a}\ket{\psi}\;.
\ee
In general, the BRST transformation of the anti-ghost is proportional
to the gauge-fixing function and can be written as the linear
combination of gauge modes with the same momentum and helicity,
\be
\label{BRSTaggen}
i[Q^{(2)},\bar c^+_{\k\a}]_+=i\sum_a {\cal C}_a\, \Phi_{\k\a}^{a+}\;,
\ee
where $\Phi^a$ are various gauge fields in the theory and ${\cal
  C}_a$ are c-number coefficients that can depend on the momentum and
helicity. Substituting this into Eq.~(\ref{ampgm1}) and using
$Q^{(2)}\ket{\psi}=0$ we obtain,
\bseq
\label{ampgm2}
\be
\label{ampgm2a}
\sum_a {\cal C}_a\,\bra{\psi'}{\cal S}\,\Phi_{\k\a}^{a+}\ket{\psi}=0\;. 
\ee 
Similar arguments apply to the final state and give
\be
\label{ampgm2b}
\sum_a {\cal C}^*_a\,\bra{\psi'}\Phi_{\k\a}^{a}\,{\cal S}\ket{\psi}=0\;. 
\ee 
\eseq
These are linear constraints on the amplitudes involving the gauge
modes. 
As we are going to see, in relativistic Yang--Mills theory they
lead to the usual Ward identity implying that an 
amplitude vanishes when the
polarization vector of a gluon is replaced by its four-momentum
$k_\mu$. In general, they are more complicated and do not reduce to a
simple replacement of polarization vectors. Note, in particular, that 
in non-relativistic theories, the dispersion relations of gauge modes
entering (\ref{ampgm2}) need not be the same as those of the
physical particles.  

We can continue the process and add another combination of gauge modes
(\ref{BRSTaggen}) to a state already containing one such
combination. The ${\cal S}$-matrix element must again be zero due to
the identity (\ref{ampgm1}) and the nilpotency of $Q^{(2)}$. This
gives us
\be
\label{ampgm3}
\sum_{a,b}{\cal C}_a{\cal C}_b\,
\bra{\psi'}{\cal S}\,\Phi_{\k_1\a}^{a+}\Phi_{\k_2\beta}^{b+}\ket{\psi}=0\;,
\ee
and so on.

Another condition that the amplitudes between physical states must satisfy is
independence of the choice of gauge.\footnote{This
  condition can also be derived from the LSZ representation of the
  ${\cal S}$-matrix, see Appendix~\ref{app:Smatr}.} In concrete
calculations, this is easily verified by making sure that the gauge
parameters drop out from the answer. Since this condition is the same
in relativistic and non-relativistic theories, we are not going to
discuss it any further.


\subsection{Examples: Yang-Mills}

\subsubsection{Relativistic}
Let us first see how Eqs.~(\ref{ampgm2}) work in the standard case of the
relativistic Yang--Mills theory. For simplicity, we work in the
Feynman gauge, so the gauge-fixed Lagrangian reads,\footnote{The
  repeated Greek indices are summed with the Minkowski metric
  $\eta_{\mu\nu}={\rm diag}(-1,+1,+1,+1)$.}
\begin{equation}
\label{eq:YMquantum}
    \mathcal{L}_q^{\rm YM} =  -\frac{1}{4} F^{a}_{\mu \nu} F^{a}_{\mu \nu} -
    \frac{1}{2}  (\partial_{\mu} A_{\mu}^{a})^2 +
    \bar{c}^{a} \partial_{\mu} D_{\mu} c^{a} \, , 
\end{equation}
where 
\be
\label{YMdefs}
F_{\mu\nu}^a=\d_\mu A_\nu^a-\d_\nu A_\mu^a +g f^{abc} A_\mu^b
A_\nu^c~,~~~~~~
D_\mu c^a=\d_\mu c^a+g f^{abc}A_\mu^b c^c\;,
\ee
$g$ is the coupling constant, and $f^{abc}$ are the structure
constants of the gauge group. The quadratic kinetic term for the gauge
fields diagonalizes and they are straightforwardly quantized with the
result 
\bseq
\label{relYMmodes}
\begin{align}
&A^{a}_{\mu}(x) = 
\int \frac{d^3 k}{(2 \pi)^32\omega_\k} \,
A^{a}_{\mu\,\k} \,
\e^{- i \omega_{\mathbf{k}} t+i \mathbf{k} \mathbf{x}} + \text{h.c}
\;,\\
&[A_{\mu\,\k}^a,A_{\nu\,\k'}^{b+}]=2\omega_\k\eta_{\mu\nu}\delta^{ab}
(2\pi)^3\delta(\k-\k') \;,
\end{align}
\eseq
where $\omega_\k=k$.
The BRST transformation of the anti-ghost coincides, up to a sign,
with the gauge-fixing function,
\be
\label{relYMgtrans}
i[Q^{(2)},\bar c^a]_+=\s \bar c^a=-\d_\mu A_\mu^a\;,
\ee 
whence we read off
\be
\label{relYMgtrans1}
i[Q^{(2)},\bar c^{a+}_{\k}]_+=-ik_\mu A_{\mu\,\k}^{a+}\;.
\ee
Substituting this expression into Eqs.~(\ref{ampgm2}) we
find
\be
\label{relYMScons}
k_\mu\bra{\psi'}{\cal S}\,A_{\mu\,\k}^{a+}\ket{\psi}
=k_\mu\bra{\psi'}A_{\mu\,\k}^{a}\,{\cal S}\ket{\psi}=0\;.
\ee
On the other hand, the scattering amplitudes involving a physical gluon
with helicity $\pm 1$ in the initial or final state are given by 
\be
e^{(\pm 1)}_\mu\bra{\psi'}{\cal S}\,A_{\mu\,\k}^{a+}\ket{\psi}~,~~~~~~
e^{(\pm 1)*}_\mu\bra{\psi'}A_{\mu\,\k}^{a}\,{\cal S}\ket{\psi}\;,
\ee
where the transverse polarization vectors are defined as in
(\ref{polare}), with their temporal components set to zero. Thus, we
recover the standard Ward identity stating that 
the amplitudes in relativistic
Yang--Mills vanish whenever a gluon polarization vector is replaced by
$k_\mu$.

\subsubsection{Yang-Mills with Lifshitz scaling}

As a new application of the conditions (\ref{ampgm2}) we consider a
non-relativistic Yang--Mills theory with the Lagrangian
\begin{equation}
\label{YMLifL}
    \mathcal{L}^{\rm YM} = \frac{1}{2} F^{a}_{i 0}F^{a}_{i 0} -
    \frac{\kappa_1}{4} D_{i} F^{a}_{jk}\,D_{i} F^{a}_{jk}- \frac{\kappa_2}{2}
    D_{i} F^{a}_{ik}\,D_{j} F^{a}_{jk} - g \frac{\kappa_3}{3} \,
    f^{abc} F^{a}_{ij} F^{b}_{jk} F^{c}_{ki} \, , 
\end{equation}
where we use the notations (\ref{YMdefs}) and
$\kappa_{1,2,3}$ are new constant parameters. In what follows we will
denote the zeroth component of the gauge field by the calligraphic
letter, 
\be
{\cal A}^a\equiv A_0^a\;,
\ee 
to avoid confusion with the helicity $0$ polarization. The action
built from this Lagrangian is invariant under Lifshitz scaling
(\ref{eq:anis_scaling}) with $z=2$ in $(4+1)$-dimensional spacetime
with the following assignment of the scaling dimensions:
\be
\label{YMdims}
\dim {\cal A}^a=2~,~~~~~\dim A_i^a=1\;.
\ee 
When supplemented with a relevant operator $F_{ij}^aF_{ij}^a$, the model is
renormalizable. A similar model with $U(1)$ gauge group and fermionic
matter was studied in \cite{Iengo2010}.

We take the gauge fixing function and the operator $O^{-1}$ in the
gauge fixing term in the form consistent with the Lifshitz scaling, 
\be
\label{YMFO}
F^a=\dot{\cal A}^a+\xi\,\D\d_iA_i^a~,~~~~~~
O^{-1}_{ab}=\frac{\delta_{ab}}{\xi\D}\;.
\ee
Here $\xi$ is an arbitrary gauge fixing parameter. The tree-level
quantum Lagrangian then reads,
\begin{equation}
\label{qYMLif}
    \mathcal{L}_q^{\rm YM} 
= {\cal L}^{\rm YM}\!+
\frac{1}{2 \xi} \big(\dot{\cal A}^{a}\!+\! \xi \Delta \partial_i A^{a}_i
\big) 
\frac{1}{\Delta} 
\big(\dot{\cal A}^{a}\! +\! \xi \Delta \partial_j A^{a}_j \big) 
+\dot{\bar c}^a\big(\dot c^a\!+\!f^{abc}{\cal A}^b c^c\big)
+\xi\d_i\bar c^a\D\big(\d_ic^a\!+\!f^{abc}{A}^b_i c^c\big).
\end{equation}
The choice of the gauge ensures cancellation of the quadratic mixing terms
between ${\cal A}^a$ and~$A_i^a$.  
Diagonalization of the remaining quadratic Lagrangian is
straightforward and yields the general linear solution: 
\begin{subequations}
\label{modexpALif}
\begin{align}
&{\cal A}^{a}(\x,t) = 
\int \frac{d^4 k}{(2 \pi)^4}\, \frac{\sqrt{\xi}k}{2\omega_{\k 0}}\,
{\cal A}^{a}_{\mathbf{k}} \,
\e^{- i \omega_{\mathbf{k}0} t+i \mathbf{k} \mathbf{x}} + \text{h.c} \, ,\\
&A^{a}_{i}(\x,t) = 
\int \frac{d^4 k}{(2 \pi)^4} \sum_{\a=-1}^{+1}
\frac{e_{i}^{\a}(\k)}{2\omega_{\k\a}}\,
 A^{a}_{\mathbf{k}\a} \, 
\e^{- i \omega_{\mathbf{k}\a} t +i \mathbf{k} \mathbf{x}} + \text{h.c} \, ,\\
&c^{a}(\x,t) = \int \frac{d^4 k}{(2 \pi)^4} \,
\frac{1}{2\omega_{\k 0}}\,c^{a}_{\mathbf{k}}\, 
\e^{- i \omega_{\mathbf{k}0} t+i \mathbf{k} \mathbf{x}} + \text{h.c.} \, ,\\
&\bar{c}^{a}(\x,t) = 
\int\frac{d^4 k}{(2 \pi)^4} \,\frac{1}{2\omega_{\k 0}}\, 
\bar{c}^{a}_{\mathbf{k}}\, 
\e^{- i \omega_{\mathbf{k}0} t+i \mathbf{k} \mathbf{x}} - \text{h.c.}  \, ,
\end{align}
\end{subequations}
where the polarization vectors $e^\a_i(\k)$ are defined in
(\ref{polare}). The dispersion relations are different for the
transverse and longitudinal modes, as expected in theories without
Lorentz invariance:
\begin{equation}
    \omega_{\mathbf{k}1}^2 
= (\kappa_1 + \kappa_2) k^4~, 
~~~~~~~ \omega_{\mathbf{k}0}^2 = \xi k^4\;.
\end{equation}
Canonically quantizing the fields we obtain the commutation relations,  
\begin{subequations}
\label{YMlifcomms}
\begin{align}
\label{YMLifAtemp}
    &[{\cal A}^{a}_{\mathbf{k}}, {\cal A}^{b +}_{\mathbf{k}'}] 
= - 2 \omega_{\mathbf{k}0} \,\delta^{ab}  
\,  (2 \pi)^4 \, \delta(\mathbf{k}- \mathbf{k}') \, , \\
    &[A^{a}_{\mathbf{k}\a}, A^{b+}_{\mathbf{k}'\beta}] 
= 2 \omega_{\mathbf{k}1} \, \delta^{ab} \,\delta_{\a\beta}\,  (2
\pi)^4 \delta(\mathbf{k}- \mathbf{k}') \, , \\ 
    &[c^{a}_{\mathbf{k}}, \bar{c}^{b+}_{\mathbf{k}'}]_{+} 
=[\bar c^a_{\k},c^{b+}_{\k'}]_+
=- 2 \omega_{\mathbf{k}0} \, \delta^{ab}\,  (2 \pi)^4  
\delta(\mathbf{k}- \mathbf{k}')\;,
\end{align}
\end{subequations}
with all other (anti-)commutators vanishing.

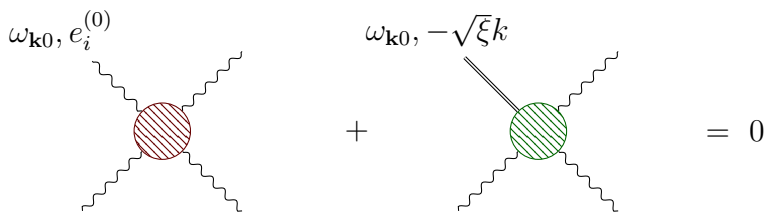
\begin{figure}[h]
\begin{center}
\begin{tikzpicture}
\begin{feynman}
\vertex (m1) [blob, draw=red!40!black, pattern color=red!40!black] {}; 
\vertex[above left=1.9cm of m1] (u1) {$\omega_{\k 0},e_i^{(0)}$};
\vertex[above right=1.5cm of m1] (u2) ;
\vertex[below left=1.5cm of m1] (d1) ; 
\vertex[below right=1.5cm of m1] (d2) ; 
\vertex[right=5cm of m1] (m2) [blob, draw=green!40!black, pattern
color=green!40!black] {};
\vertex[above left=1.9cm of m2] (u3) {$\omega_{\k 0},-\sqrt{\xi}k$};
\vertex[above right=1.5cm of m2] (u4) ; 
\vertex[below left=1.5cm of m2] (d3) ; 
\vertex[below right=1.5cm of m2] (d4) ;
\diagram* {
(u1) -- [photon] (m1) -- [photon] (d1),
(u2) -- [photon] (m1) -- [photon] (d2),
(u3) -- [double] (m2) -- [photon] (d3),
(u4) -- [photon] (m2) -- [photon] (d4)
};
\end{feynman}
\node[right=1.9cm of m1] (eq) {$+$};
\node[right=1.7cm of m2] {$= \ 0$};
\end{tikzpicture}
\end{center}
   \caption{Generalized Ward identity satisfied by the amplitudes in
     Yang--Mills theory 
with Lifshitz scaling. Wavy lines correspond to spatial gauge fields
$A_i^a$, whereas the straight double line --- to the temporal
component ${\cal A}^a$.}
   \label{fig:GaugeYM}
\end{figure}

According to the general rules, the BRST transformation of the
anti-ghost is
\begin{equation}
\label{YMQbarc}
i[Q^{(2)},\bar c^a]_+=
    \mathbf{s} \bar{c}^{a} = 
\frac{1}{\xi \, \Delta} \left(\dot{\cal A}^{a} 
+ \xi \Delta \partial_i A^{a}_i \right)\;.
\end{equation}
Comparing the Fourier decomposition of the left and right hand sides
we get,
\be
\label{YMQbarc1}
i[Q^{(2)},\bar c^{a+}_{\k}]_+=ik\, \big({\cal A}_\k^{a+}+A_{\k 0}^{a+}\big)\;.
\ee
Incidentally, this has the same form as in the relativistic case,
cf. Eq.~(\ref{relYMgtrans1}). Hence, the constraint (\ref{ampgm2})
becomes 
\be
\label{YMLifcond}
\bra{\psi'}{\cal S}\,A_{\k 0}^{a+}\ket{\psi}
+\bra{\psi'}{\cal S}\,{\cal A}_{\k}^{a+}\ket{\psi}=0\;,
\ee
and similarly for the outgoing mode. It can be represented
graphically as shown in Fig.~\ref{fig:GaugeYM}, where we explicitly
indicate the energy and polarization factors carried by the external legs.
Note that the factor for the ${\cal A}$-leg is negative due to the
minus sign in the commutator (\ref{YMLifAtemp}). We now observe an
important difference from the relativistic case. The verification of
the gauge invariance does not reduce to a mere substitution of the longitudinal
polarization in the external leg of a diagram for transverse gluons
--- the first diagram in the figure.
First, since the
dispersion relations of the longitudinal modes is different from that
of the transverse modes, the diagram 
must be re-evaluated
with a different incoming energy. Second, the interaction vertices for
the spatial and temporal parts of the gauge field are essentially
different, so the green blob in the second diagram is different from
the red blob and must be evaluated separately.
We have verified by an explicit calculation that the identity shown in
Fig.~\ref{fig:GaugeYM} holds for tree-level $2\to2$ amplitudes
in the theory (\ref{YMLifL}). 

\subsection{Application to Ho\v rava gravity}
\label{sec:GIHG}

We return to the projectable HG. All preliminary work has been already
done in Sec.~\ref{sec:BRSTHG}. We can directly use the BRST
transformation of the anti-ghost (\ref{newBRSTbarc}) which we write
explicitly:
\be
\label{BRSTbarcexpl}
i[Q^{(2)},\bar c_i]_+=-\frac{\sigma}{\D^2}\dot N_i
+\frac{\sigma\xi}{(1+\xi)\D^3}\d_i\d_j\dot N_j
+\frac{1}{2}(\d_j h_{ij}-\lambda \d_i h)\;.
\ee 
Expanding the left and right hand sides into Fourier modes according
to 
Eqs.~(\ref{fieldsdecomp}) we obtain simple relations
\bseq
\label{HGgaugemodes}
\begin{align}
&i[Q^{(2)},\bar c_{\k\a}^{+}]_+=\frac{ik}{\sqrt{2}}\,
(N_{\k\a}^++h_{\k\a}^+)\;,&& \a=\pm 1\;,\\
&i[Q^{(2)},\bar c_{\k\a}^{+}]_+=ik\sqrt{|1-\lambda|}\,
(N_{\k\a}^++h_{\k\a}^+)\;,&& \a=0\;.
\end{align}
\eseq
Substitution into Eqs.~(\ref{ampgm2}) yields the identities
\be
\label{HGgaugeid}
\bra{\psi'}{\cal S}\,h_{\k\a}^+\ket{\psi}+
\bra{\psi'}{\cal S}\,N_{\k\a}^+\ket{\psi}=0~,~~~~~\a=0,~\pm 1\;,
\ee
which are depicted graphically in Fig.~\ref{fig:GaugeHG}.
\begin{figure}[h]
\begin{center}
\begin{tikzpicture}
\begin{feynman}
\vertex (m1) [blob, draw=red!40!black, pattern color=red!40!black] {};
\vertex[above left=1.9cm of m1] (u1) {$\omega_{\k\a},\varepsilon^{\a}_{ij}$};
\vertex[above right=1.5cm of m1] (u2) ; 
\vertex[below left=1.5cm of m1] (d1) ; 
\vertex[below right=1.5cm of m1] (d2) ; 
\vertex[right=5cm of m1] (m2) [blob, draw=green!40!black, pattern
color=green!40!black] {}; 
\vertex[above left=1.9cm of m2] (u3) {$\omega_{\k\a},-\epsilon^{\a}_i$};
\vertex[above right=1.5cm of m2] (u4) ;
\vertex[below left=1.5cm of m2] (d3) ;
\vertex[below right=1.5cm of m2] (d4) ;

\diagram* {
(u1) -- [photon] (m1) -- [photon] (d1),
(u2) -- [photon] (m1) -- [photon] (d2),
(u3) -- [double] (m2) -- [photon] (d3),
(u4) -- [photon] (m2) -- [photon] (d4)
};
\end{feynman}
\node[right=1.9cm of m1] (eq) {$+$};
\node[right=1.7cm of m2] {$= \ 0~~~~~\text{for}~~\a=0,~\pm 1$};
\end{tikzpicture}
\end{center}
\caption{Generalized Ward identities for
the amplitudes in projectable Ho\v rava
  gravity. Wavy lines and the straight
  double line represent the spatial
  metric $h_{ij}$ and the shift $N_i$, respectively.}
\label{fig:GaugeHG}
\end{figure}
The polarization tensors corresponding to the external legs of the
diagrams in the figure are given in Eqs.~(\ref{polarN}),
(\ref{polarh}). Note that the shift polarization is multiplied by 
$(-1)$ due to the different signs in the commutators of the $h$ and
$N$ creation-annihilation operators, see Eqs.~(\ref{hcomm}),
(\ref{Ncomm}). 

We use the above identities to cross-check the validity of
our calculation of $2\to 2$ scattering amplitudes in the next section.

\section{Calculating the Amplitudes}
\label{sec:amplitude}

\subsection{Algorithm and overview of the result}
We have automated the computation of scattering amplitudes in HG using
the {\it xAct} package \cite{xAct,xPerm,xPert,xTras} 
for {\it Mathematica} \cite{Mathematica}. Our code \cite{github} 
starts by extracting propagators
and vertices from the action. For the propagators, we use the
gauge-fixed Lagrangian (\ref{L2q}). The gauge-fixing term is
quadratic and thus does not affect the vertices, which we obtain directly from
the original action (\ref{eq:HG_Action}) by taking variational
derivatives with respect to the metric perturbation $h_{ij}$ and the
shift $N_i$. Since we restrict to the tree level, we do not need the
propagators or vertices involving ghosts. Finally, the external lines
are determined from the mode decomposition of the fields
(\ref{fieldsdecomp}) and their commutators (\ref{modecomm}). More
details one the 
Feynman rules used in the calculation are given in Appendix~\ref{app:Feyn}.

We then follow the standard procedure to construct all diagrams
contributing to a given scattering process. For example, the
scattering amplitudes for two gravitons in the initial and final
states is given by the sum of the diagrams shown in
Fig.~\ref{fig:Scattering}. We treat all momenta and energies as
flowing into the diagram. 
The polarization tensors 
for incoming particles with negative energies are defined according to
\be
\label{polarout}
\ve_{ij}^\a(-\k,-\omega)=\ve_{ij}^\a(\k,\omega)=[\ve_{ij}^{-\a}(\k,\omega)]^*\;. 
\ee
This is consistent with the crossing rule that an incoming particle
is equivalent to an outgoing particle with opposite momentum and helicity.   
The amplitude ${\cal M}$ is defined in the
standard manner, as the ${\cal S}$-matrix element with unit operator
subtracted and the energy-momentum conserving $\delta$-function
stripped off,
\be
\label{Mdef}
{\cal S}=\mathbbm{1}+i{\cal M}(\k_I,\omega_I,\a_I) \,
(2\pi)^4\,\delta\Big(\sum_I\omega_I\Big)\, \delta\Big(\sum_{I}\k_I\Big)\;.
\ee  
The scattering of physical states corresponds to choosing the
helicities $\alpha_I$ in Fig.~\ref{fig:Scattering} equal to $\pm
2$ or $0'$. We have checked that such amplitudes, evaluated on-shell, 
are independent of the gauge parameters $\sigma$, $\xi$. 
In addition, we have evaluated the amplitudes with one gauge mode
having $\a=\pm 1$ or $0$, as well as the amplitudes with the shift in
the external line, and verified that on-shell they satisfy the
generalized Ward identity shown in Fig.~\ref{fig:GaugeHG}. 
Finally, we validated the code on the example
of general relativity and reproduced the standard results
\cite{Sannan1986}. The 
success of these tests makes us confident that the code works
correctly. 

\begin{figure}[t]
\begin{tikzpicture}
\begin{feynman}
\node(m1) {$i{\cal M}(\k_I,\omega_I,\a_I)$};
\node[left=2cm of m1] (a0);
\node[below=4cm of m1] (m0) {};
\vertex[right=2cm of m0] (m2);
\vertex[right=1.1cm of m2] (m3);
\node[above =1.2cm of m2] (a1);
\vertex[left=1.cm of a1] (u3) {$\mathbf{k}_2,\omega_2, \ve_2$};
\node[above =1.2cm of m3] (a2);
\vertex[right=1.cm of a2] (u4) {$\mathbf{k}_3,\omega_3, \ve_3$};
\node[below =1.2cm of m2] (a3);
\vertex[left=1.cm of a3] (d3) {$\mathbf{k}_1,\omega_1, \ve_1$};
\node[below =1.2cm of m3] (a4);
\vertex[right=1.cm of a4] (d4) {$\mathbf{k}_4,\omega_4, \ve_4$};
\vertex[right=4.8cm of m3] (m4);
\vertex[right=1.2cm of m4] (m5);
\node[above =1.2cm of m4] (b1);
\vertex[left=1.cm of b1] (u5) {$\mathbf{k}_2,\omega_2, \ve_2$};
\node[above =1.2cm of m5] (b2);
\vertex[right=1.cm of b2] (u6) {$\mathbf{k}_3,\omega_3, \ve_3$};
\node[below =1.2cm of m4] (b3);
\vertex[left=1.cm of b3] (d5) {$\mathbf{k}_1,\omega_1, \ve_1$};
\node[below =1.2cm of m5] (b4);
\vertex[right=1.cm of b4] (d6) {$\mathbf{k}_4,\omega_4, \ve_4$};
\vertex[right=5cm of m1] (m6);
\node[above=1.2cm of m6] (c1);
\vertex[left=1.3cm of c1] (u7) {$\mathbf{k}_2,\omega_2, \ve_2$};
\vertex[right=1.3cm of c1] (u8) {$\mathbf{k}_3,\omega_3, \ve_3$};
\node[below=1.2cm of m6] (c2);
\vertex[left=1.3cm of c2] (d7) {$\mathbf{k}_1,\omega_1,\ve_1$};
\vertex[right=1.3cm of c2] (d8) {$\mathbf{k}_4, \omega_4,\ve_4$};
\diagram* {
(u3) -- [photon] (m2) -- [photon] (d3),
(m2) -- [photon, edge label=$h$] (m3),
(u4) -- [photon] (m3) -- [photon] (d4),
(u5) -- [photon] (m4) -- [photon] (d5),
(m4) -- [double, edge label=$N$] (m5),
(u6) -- [photon] (m5) -- [photon] (d6),
(u7) -- [photon] (m6) -- [photon] (d7),
(u8) -- [photon] (m6) -- [photon] (d8)
};
\end{feynman}
\node[right=0cm of m1] (eq) {$=$};
\node[left=2.3cm of m2] {$+$};
\node[right=2.3cm of m5] {$+ \quad (t,u)$};
\node[right=2cm of m3] {$+$};
\end{tikzpicture}
    \caption{Feynman diagrams for 
$2 \rightarrow 2$ scattering of gravitons 
at tree level. Wavy lines represent an external leg or propagator of
the metric $h_{ij}$, and the double line is the propagator of the shift
$N_i$. 
All momenta and energies are incoming; 
$(t,u)$ stands for the
      diagrams with permutations $(\mathbf{k}_2,\omega_2, \ve_2) 
\leftrightarrow
      (\mathbf{k}_3,\omega_3,\ve_3)$ 
and $(\mathbf{k}_2,\omega_2, \ve_2) \leftrightarrow
      (\mathbf{k}_4,\omega_4, \ve_4)$.} 
    \label{fig:Scattering}
\end{figure}

The resulting expressions for the amplitudes are very long and are
available in the form of a {\it Mathematica} file
\cite{github}. Similar to general relativity \cite{Sannan1986}, they
can be cast into a sum of terms representing various contractions of
the polarization tensors with the external momenta, multiplied by
scalar functions of momenta and energies. However, 
the variety of structures in our case
is richer due to the presence of higher powers of momenta
(higher spatial derivatives) in
the vertices. In particular, we obtain terms containing six and eight
momenta contracted with the polarizations, such as e.g.,
\be
\label{nestruct}
(\k_3\ve_1\ve_2\k_4)(\k_1\ve_3\k_1)(\k_2\ve_4\k_2)~,~~~~~
(\k_2\ve_1\k_2)(\k_1\ve_2\k_1)(\k_4\ve_3\k_4)(\k_3\ve_4\k_3)\;,
\ee
where we have used condensed notations 
\be
\label{condnot}
(\k_1\ve_3\k_1)=k_1^i\,\ve_{3\,ij}\, k_1^j~,~~~~
(\k_1\ve_1\ve_2\k_4)=k_3^i\,\ve_{1\,ij}\,\ve_{2\,jk}\,k_4^k~,~~~~~\text{etc.}
\ee
We have not been able to reduce the 
structures (\ref{nestruct}) to those with
fewer momenta by using the momentum conservation or other
identities. 

The coefficient functions multiplying the aforementioned structures
depend on the scalar invariants of the momenta --- their absolute
values $k_I$, $I=1,2,3,4$, and scalar products. 
We express the latter
through the ``Mandelstam-like'' variables
\be
\label{STU}
S=(\k_1+\k_2)^2~,~~~~~
T=(\k_1+\k_3)^2~,~~~~~
U=(\k_1+\k_4)^2\;.
\ee
Note that as a consequence of momentum conservation these variables
obey the identity
\be
\label{STUid}
S+T+U=k_1^2+k_2^2+k_3^2+k_4^2\;.
\ee
The energy-conservation is implemented by using a
set of three independent combinations,
\be
\label{OmegaSTU}
\Omega_S=\omega_1+\omega_2~,~~~~~
\Omega_U=\omega_1+\omega_3~,~~~~~
\Omega_T=\omega_1+\omega_4\;.
\ee 
In this way we arrive at the coefficient functions depending on ten
variables $k_1$, $k_2$, $k_3$, $k_4$, $S$, $T$, $U$, 
$\Omega_S$, $\Omega_T$, $\Omega_U$ related
by the constraint (\ref{STUid}). We keep this form and simplify the
expressions as much as possible, without using the dispersion
relations until the very last step. The reason for this strategy is
twofold. First, it allows us to easily switch between physical and
gauge modes in order to verify the cancellation
(\ref{HGgaugeid}). Second, the dispersion relations introduce
non-analyticity (square-roots of the coefficients in
Eqs.~(\ref{omtt})--(\ref{oms})) which complicate the manipulation of
the formulas. The price to pay is that the off-shell amplitudes
preserve the dependence on the gauge parameters $\sigma$, $\xi$. This dependence
disappears once we put the amplitudes on-shell and assign
physical polarizations to the particles.

\subsection{Head-on collisions}
\label{sec:headon}

The expressions for the amplitudes greatly simplify in the special
case of head-on collisions when the momenta of two colliding particles
are opposite in direction and equal in magnitude.\footnote{In
  relativistic theories any collision can be brought to the head-on
  kinematics by a boost to the center-of-mass frame. This is not
  possible in HG.}
In more detail, we choose the particle momenta and energies to be:
\bseq
\begin{align}
\label{22mom}
&\k_1=\begin{pmatrix}
0\\0\\k
\end{pmatrix}\;,&&
\k_2=\begin{pmatrix}
0\\0\\-k
\end{pmatrix}\;,&&
\k_3=\begin{pmatrix}
-k'\sin\theta\\0\\-k'\cos\theta
\end{pmatrix}\;,&&
\k_4=\begin{pmatrix}
k'\sin\theta\\0\\k'\cos\theta
\end{pmatrix}\;,\\
\label{22eng}
&\omega_1=\sqrt{\nu_{(1)}}\,k^3\;,&&
\omega_2=\sqrt{\nu_{(2)}}\,k^3\;,&&
\omega_3=-\sqrt{\nu_{(3)}}\,k'^3\;,&&
\omega_4=-\sqrt{\nu_{(4)}}\,k'^3\;,
\end{align}
\eseq
where $\nu_{(I)}=\nu_5$ or $\nu_s$, depending on the type of the
physical graviton --- tensor or scalar. The final momentum $k'$ is
determined by the energy conservation,
\be
\label{engcons}
\big(\sqrt{\nu_{(1)}}+\sqrt{\nu_{(2)}}\big)\,k^3
=\big(\sqrt{\nu_{(3)}}+\sqrt{\nu_{(4)}}\big)\,k'^3\equiv E\;.
\ee  
Note that the physical momenta of the final particles $3$ and $4$ are
$-\k_3$ and $-\k_4$ and thus $\theta$ is the scattering angle defined
in the usual way as the angle between the directions of incoming
particle $1$ and outgoing particle $3$.

The amplitude depends on the polarizations of the particles 
$\alpha_I=\pm 2$ or $0'$ which we will write as 
$+, -, s$ for short.\footnote{A technical
  remark: The overall phases of the polarization vectors $e_i^{(\pm 1)}$
  defined in (\ref{polare}) and used to construct the graviton
  polarization tensors are ambiguous for particle $2$ moving in the
  direction opposite to the 3d axis. We set the phases to $0$ which
  renders all amplitudes real.}
We find it more convenient for the discussion of the physical
properties of the amplitudes in this section 
to label them
with the {\em physical} polarizations, i.e. upon performing
the crossing for final particles. In these notations, the amplitude
${\cal M}_{++,++}$ stands for elastic scattering of two right-handed
gravitons, the amplitude ${\cal M}_{++,+-}$ describes a process where
one right-handed graviton flips helicity, etc.

We find that the helicity amplitudes have the form,
\begin{equation}
\label{Mampl}
\mathcal{M}_{\a_1\a_2,\a_3\a_4} = G E^2
f_{\a_1\a_2,\a_3\a_4}\left(\cos\theta; u_s, v_a,\lambda\right)\;, 
\end{equation}
where 
\begin{equation}
\label{esscoup}
    u_{s} = \sqrt{\frac{\nu_{s}}{\nu_5}}~,~~~~~~~~ v_{a} =
    \frac{\nu_{a}}{\nu_5}~,~~ a=1,2,3\;,
\end{equation}
are the essential couplings of the theory introduced in
\cite{Barvinsky2023}. They are singled out by the requirement that
their RG running is independent of the gauge choice (this is not true
for $\nu_a$ individually). Note that the gravitational coupling $G$
multiplying the overall amplitude is not
essential, implying that its RG improvement depends on the gauge. This
is not a problem since the amplitude is not directly observable. We
will say more about this shortly. 

The functions $f_{\a_1\a_2,\a_3\a_4}$ describing the angular
dependence of the amplitudes are listed in
Appendix~\ref{app:amp}. They are rational functions of $\cos\theta$. 
Many of them have singularities in the forward scattering limit, as is
typical in theories with massless particles. The strongest singularity
is featured by elastic amplitudes with $\a_1=\a_3$, $\a_2=\a_4$ which
behave as $\sim \theta^{-6}$ at small $\theta$. On the other hand, the
helicity violating amplitude $f_{++,--}$ is regular at all
angles and is much simpler than the elastic amplitudes,
though, in contrast to general relativity, 
it does not vanish completely. 

Notably, when the dispersion relations of the transverse
traceless and scalar gravitons do not coincide ($u_s\neq 1$), the
amplitudes involving both types of particles have poles at non-zero
angles. They arise in $t$- and $u$- channels and are a
consequence of the fact that in HG a single graviton is kinematically
allowed to decay into a pair of gravitons with lower energies. Thus,
whenever, say, $\omega_3\neq \omega_1$, the propagator in the
$t$-channel diagram can go on-shell. For the head-on collisions this
is possible only if particles participating in the process are of
different types. 
For more general kinematics, we expect these resonant poles to
occur also in $2\to 2$ amplitudes for identical particles and in all
three $s,t,u$ channels.

One more peculiarity of amplitudes involving both tensors and scalars
can be illustrated on the example of $f_{++,+s}$. When $u_s\neq 1$,
this amplitude is finite in the forward and backward limits and in fact
vanishes in the way consistent with the conservation of angular
momentum. The incoming state has zero projection of the angular
momentum on the 3d axis. On the other hand, for the final state the
projection of the graviton spin becomes $+2$ or $-2$ for $\theta\to 0$
or $\pi$. This means that two units of angular momentum must be
carried away by the orbital wavefunction implying a $d$-wave
scattering. This leads to 
suppression
$\theta^2$ and $(\pi-\theta)^2$ in the two limits, 
respectively, which we indeed obtain from the direct computation,
cf. Eq.~(\ref{polyfact1}). By
contrast, in the case $u_s=1$ we recover the collinear singularities,
which are only partially compensated by the $d$-wave factors, 
see Eq.~(\ref{fppps}). Similar pattern
emerges for other amplitudes. More details on their angular dependence
can be found in Appendix~\ref{app:amp}. 

The quadratic growth of the amplitudes with energy is the same as in
general relativity where it is known to contradict the tree-level
unitarity. Nevertheless, it is consistent with unitarity in theories
with the Lifshitz scaling \cite{Blas2009Comment}. 
It is instructive to derive the cross section corresponding to the
amplitude (\ref{Mampl}). We define the cross section $\sigma$ in the standard
way, through the number of collisions happening in a unit of time and
volume in the intersection of two beams of particles with number
densities $n_1$, $n_2$: 
\begin{equation}
\label{ratedef}
    \frac{d N_{\text{coll}}}{dt \, dV} = \sigma\,
     n_{1} n_{2}  {\rm v}_{\text{rel}}\;,
\end{equation}
where 
\begin{equation}
\label{vreldef}
    {\rm v}_{\text{rel}} = |\mathbf{v}_1 - \mathbf{v}_2| = \Biggr| \frac{d
      \omega_1}{d \mathbf{k}_1} - \frac{d \omega_2}{d \mathbf{k}_2} \Biggr| 
\end{equation}
is the relative group velocity of colliding particles. Following the
usual steps, we obtain the standard expression
\begin{align}
\label{crosssec}
    \sigma = \frac{1}{4 \omega_1 \omega_2 {\rm v}_{\text{rel}}} \int 
\frac{d^3 k_3}{(2 \pi)^3 2 \omega_{3}} \frac{d^3 k_4}{(2 \pi)^3 2
  \omega_4} |\mathcal{M}|^2 
(2 \pi)^4 \delta\big(\sum\omega_I\big)   
    \,\delta\big(\sum\k_I\big) \;.
\end{align}
Let us for simplicity focus on the case when all particles
participating in the scattering are transverse traceless gravitons ---
the results for other cases are similar. Performing integration over
the phase space and expressing the energy and relative velocity
through the absolute value of graviton's momentum,
$E=2\sqrt{\nu_5}\,k^3$, ${\rm v}_{\rm rel}=6\sqrt{\nu_5}\,k^2$, we
arrive at the differential cross section 
\begin{equation}
\label{diffcrosssec}
    \frac{d\sigma_{\a_1\a_2,\a_3\a_4}}{\sin\theta \,d\theta} 
= \frac{G^2}{72 \pi\, \nu_5\, k^2} \left|f_{\a_1\a_2,\a_3\a_4}\right|^2\;.
\end{equation}
We observe that the cross section at fixed angle decreases as the
square of the inverse momentum (de Broglie wavelength squared) which
is a typical behavior in weakly coupled local theories compatible with
unitarity. On the other hand, the total cross section diverges at small angles
signaling the necessity of an infrared regulator. 

The cross section (\ref{diffcrosssec}) is proportional to the square
of the essential coupling \cite{Barvinsky2023}
\be
\label{esscoup1}
{\cal G}=\frac{G}{\sqrt{\nu_5}}\;.
\ee 
Also, as already noted, $f_{\a_1\a_2,\a_3\a_4}$ depends only on
essential couplings. This is reassuring. In contrast to the amplitude,
the cross section is a physical observable and its RG improvement must
be gauge invariant. We see that this in indeed the case.


\section{The limit $\lambda \rightarrow \infty$}
\label{sec:linf}

It was conjectured in \cite{Gumrukcuoglu2011_1} that projectable Ho\v
rava gravity can have a regular limit at $\lambda\to\infty$. This is
supported by the regularity of the dispersion relation for physical
transverse traceless and scalar gravitons, Eqs.~(\ref{omtt}),
(\ref{oms}), and by the regularity of the one-loop $\beta$-functions
for the essential couplings \cite{Barvinsky2021}. This
limit is interesting since it corresponds to a likely behavior of the
theory in the deep UV \cite{Gumrukcuoglu2011_1,Barvinsky2021}. In this
section we discuss evidence for its regularity from the scattering
amplitudes' perspective. We then prove the above conjecture by
recasting the $\lambda\to\infty$ theory in a manifestly regular form.

\subsection{Cancellation of enhanced terms in $\sigma,\xi$-gauge}
\label{sec:linfcancel}

A scrutiny of the expressions in Appendix~\ref{app:amp} for 
the head-on scattering
amplitudes between physical states shows that they are regular in the
limit (\ref{limit}). Using our symbolic code, we have
checked that this property holds also for arbitrary kinematics. This
is non-trivial. Indeed, interaction vertices contain contributions
proportional to $\lambda$. Thus, the amplitudes given by the diagrams
in Fig.~\ref{fig:Scattering} could, a priori, contain terms as large
as $O(\lambda^2)$. It is instructive to study how these large
contributions cancel.  

We start by observing that the polarizations of physical states are
traceless in the limit (\ref{limit}). This is, of course, always true
for the helicity $\pm 2$ gravitons, whereas for the scalar graviton we obtain
from Eqs.~(\ref{polarh}),
\be
\label{epsslim}
\ve_{ij}^{0'}(\k)=\sqrt{\frac{2}{3}}\big(\delta_{ij}
-3\hat k_i\hat k_j\big)+O(\lambda^{-1})\;. 
\ee
This removes many terms in the contraction of the interaction vertices
with the polarization tensors. Let us first consider the building
blocks involving a cubic vertex and two graviton
external legs. Using Eqs.~(\ref{Vhhh}), (\ref{VhhN}) we get,
\bseq
\begin{align}
\label{eeVh}
&\begin{tikzpicture}
\begin{feynman}
\vertex (c1) [dot];
\vertex[right=2cm of c1] (c2) {\(h_{mn}\)}; 
\vertex[below left=1.5cm of c1] (a2) {\(\ve_1\)};
\vertex[above left=1.5cm of c1] (a3) {\(\ve_2\)};
\diagram* {
(c1) -- [photon] (a2),
(a3) -- [photon] (c1),
(c1) -- [photon, edge label={\(\k_3,\omega_3\)}] (c2)
};
\node[label={[rotate=45,label distance=-0.1cm]\(\k_1,\omega_1\)},
below left=0.6cm of c1];
\node[label={[rotate=-45,label distance=-0.15cm]\(\k_2,\omega_2\)},
above left=0.6cm of c1];
\vertex[below=0cm of c2] (c3);
\node[right=0.8cm of c3] {\(\begin{aligned}
     =&~-i\frac{\lambda}{4}\,(\ve_1\ve_2)\,
(\omega_1+\omega_2)\,\omega_3\delta_{mn}+O(\lambda^0)\;,
\end{aligned}\)};
\end{feynman}
\end{tikzpicture}
\\
\label{eeVN}
&\begin{tikzpicture}
\begin{feynman}
\vertex (c1) [dot];
\vertex[right=2cm of c1] (c2) {\(N_m\)}; 
\vertex[below left=1.5cm of c1] (a2) {\(\ve_1\)};
\vertex[above left=1.5cm of c1] (a3) {\(\ve_2\)};
\diagram* {
(c1) -- [photon] (a2), 
(a3) -- [photon] (c1),
(c1) -- [double, edge label=\({\bf p}\)] (c2)
};
\node[label={[rotate=45,label distance=-0.1cm]\(\k_1,\omega_1\)},
below left=0.6cm of c1];
\node[label={[rotate=-45,label distance=-0.15cm]\(\k_2,\omega_2\)},
above left=0.6cm of c1];
\vertex[below=0cm of c2] (c3);
\node[right=0.8cm of c3] {\(\begin{aligned}
     =&~-i\frac{\lambda}{2}\,(\ve_1\ve_2)\,(\omega_1+\omega_2)\,p_m
     +O(\lambda^0)\;,
  \end{aligned}\)};
\end{feynman}
\end{tikzpicture}
\end{align}
\eseq
where $(\ve_1\ve_2)\equiv \ve_{1\,ij}\ve_{2\,ij}$ 
and we have taken into account the symmetry factors $3!$ and $2!$ for
the two diagrams, respectively, as well as a factor $\sqrt{G}$ for
each external leg.  

Next, the contraction of the graviton propagator (\ref{eq:props:2}) 
with $\delta_{mn}$ yields,
\be
\label{hhpropcontr}
\delta_{mn}\cdot 
\feynmandiagram[baseline=(a.base), horizontal= a to c]{
a [particle =\(h_{mn}\)]  -- [color=white] b -- [color=white] c
[particle=\(h_{pq}\)],  
a -- [photon, edge label={\(\k,\omega\)}]c,
};
=-\frac{4G}{3\lambda}{\cal P}_s\big(\delta_{pq}-3\hat k_p\hat k_q
\big)+O(\lambda^{-2})\;.
\ee
In deriving this expression 
we have used the limiting form of the longitudinal mode pole
factor,
\be
\label{P0limit}
{\cal P}_0=\frac{i}{\omega^2-\frac{(1-\lambda)(1+\xi)}{\sigma}k^6}
=i\frac{\sigma}{\lambda(1+\xi)k^6}+O(\lambda^{-2})\;.
\ee
Note that the first term in (\ref{hhpropcontr}) is again traceless and
vanishes when contracted with $\delta_{pq}$. Combining this with 
Eq.~(\ref{eeVh}) we conclude that for the physical states the diagram
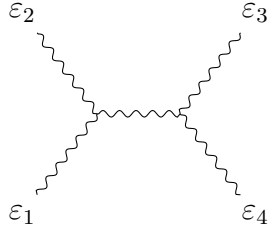
\[
\begin{tikzpicture}
\begin{feynman}
\node (m0) {};
\vertex[right=2cm of m0] (m2);
\vertex[right=1.1cm of m2] (m3);
\node[above =1.2cm of m2] (a1);
\vertex[left=1.cm of a1] (u3) {\(\ve_2\)};
\node[above =1.2cm of m3] (a2);
\vertex[right=1.cm of a2] (u4) {\(\ve_3\)};
\node[below =1.2cm of m2] (a3);
\vertex[left=1.cm of a3] (d3) {\(\ve_1\)};
\node[below =1.2cm of m3] (a4);
\vertex[right=1.cm of a4] (d4) {\(\ve_4\)};
\diagram*{
(u3) -- [photon] (m2) -- [photon] (d3),
(m2) -- [photon] (m3),
(u4) -- [photon] (m3) -- [photon] (d4),
};
\end{feynman}
\end{tikzpicture}
\]
is $O(\lambda^0)$, i.e. it is finite in the limit (\ref{limit}).

Consider now the diagram with the exchange of the shift. Here we have
from Eq.~(\ref{eq:props:1})
for the propagator:
\be
\label{NNpropcontr}
p_m\cdot
\feynmandiagram[baseline=(a.base), horizontal= a to c]{
a [particle =\(N_{m}\)] -- [double] b -- [double] c [particle=\(N_{n}\)],
a -- [double, edge label={\({\bf p}\)}] c,
};
=-i\frac{G}{\lambda}\, \frac{\hat p_n}{p}+O(\lambda^{-2})\;,
\ee
where we have again used the limiting form (\ref{P0limit}). Combining
with Eq.~(\ref{eeVN}), we find a $O(\lambda)$ contribution,
\be
\begin{tikzpicture}
\begin{feynman}
\node (m0) {};
\vertex[right=2cm of m0] (m2);
\vertex[right=1.1cm of m2] (m3);
\node[above =1.2cm of m2] (a1);
\vertex[left=1.cm of a1] (u3) {\(\ve_2\)};
\node[above =1.2cm of m3] (a2);
\vertex[right=1.cm of a2] (u4) {\(\ve_3\)};
\node[below =1.2cm of m2] (a3);
\vertex[left=1.cm of a3] (d3) {\(\ve_1\)};
\node[below =1.2cm of m3] (a4);
\vertex[right=1.cm of a4] (d4) {\(\ve_4\)};
\diagram*{
(u3) -- [photon] (m2) -- [photon] (d3),
(m2) -- [double] (m3),
(u4) -- [photon] (m3) -- [photon] (d4),
};
\vertex[below=0cm of m3] (c1);
\node[right=1.5cm of c1] {\(\begin{aligned}
     =&~-i\frac{G\lambda}{4}\,(\ve_1\ve_2)(\ve_3\ve_4)\,
(\omega_1+\omega_2)(\omega_3+\omega_4)
     +O(\lambda^0)\;.
  \end{aligned}\)};
\end{feynman}
\end{tikzpicture}
\ee
Note the minus sign in this expression which comes from the fact that
the momentum in the propagator is inflowing into one vertex and
outflowing from the other. Similar contributions with exchange of
particles $(2\leftrightarrow 3)$ and $(2\leftrightarrow 4)$ come from
the $t$ and $u$ channels. 

These $O(\lambda)$ contributions are precisely canceled by the
diagram with the 4-point vertex. Indeed, contraction with the
traceless polarizations leaves only terms in the third line
in Eq.~(\ref{Vhhhh}) which upon symmetrization read,
\be
\label{Vhhhhcontr}
\begin{tikzpicture}
\begin{feynman}
\vertex (c1) [dot];
\vertex[below left=1.5cm of c1] (a1) {\(\ve_1\)};
\vertex[above left=1.5cm of c1] (a2) {\(\ve_2\)};
\vertex[above right=1.5cm of c1] (a3) {\(\ve_3\)}; 
\vertex[below right=1.5cm of c1] (a4) {\(\ve_4\)}; 
\diagram* {
(a1) -- [photon] (c1), 
(a3)-- [photon] (c1),
(a2) -- [photon] (c1),
(a4) -- [photon](c1)
};
\node[label={[rotate=45,label distance=-0.1cm]\(\k_1,\omega_1\)}, below left=0.6cm of c1];
\node[label={[rotate=-45,label distance=-0.15cm]\(\k_2,\omega_2\)}, above left=0.6cm of c1];
\node[label={[rotate=45,label distance=-0.2cm]\(\k_3,\omega_3\)}, above right=0.7 of c1];
\node[label={[rotate=-45,label distance=-0.1cm]\(\k_4,\omega_4\)}, below right=0.8 of c1];
\vertex[below=0.8cm of c1] (c3);
\node[right=1.9cm of c3] {\(\begin{aligned}
     = i\frac{G\lambda}{4}\Big[&(\ve_1\ve_2)(\ve_3\ve_4)\,
(\omega_1+\omega_2)(\omega_3+\omega_4)\\
&+(\ve_1\ve_3)(\ve_2\ve_4)\,
(\omega_1+\omega_3)(\omega_2+\omega_4)\\
&+(\ve_1\ve_4)(\ve_3\ve_2)\,
(\omega_1+\omega_4)(\omega_3+\omega_2)\Big]
+O(\lambda^0)\;.
\end{aligned}\)};
\end{feynman}
\end{tikzpicture}
\ee  
Thus, we have confirmed explicitly the cancellation of dangerous
contributions to the amplitudes in the limit $\lambda\to\infty$. It
relies on a rather delicate interplay between the tracelessness of the
physical polarizations and the structure of the vertices and
propagators.


\subsection{Regular limit with an auxiliary field}
\label{sec:chi}

Encouraged by the previous results, we look for a way to cast the
action of HG in the form which would be manifestly regular at
$\lambda\to\infty$. This is indeed possible to do by integrating in an
auxiliary non-dynamical scalar field $\chi$ and rewriting the
$\lambda$-term in the Lagrangian as
\be
\label{Lchi}
-\frac{\lambda}{2G}\sqrt{\gamma}\, K^2\quad
\longrightarrow\quad \frac{\sqrt\gamma}{G}\bigg[-\chi
K+\frac{\chi^2}{2\lambda}\bigg]\;. 
\ee
Clearly, at finite $\lambda$ the two forms of the theory are
equivalent, since we can always integrate out $\chi$ and restore the
original action. On the other hand, in the new form we can easily take
the limit (\ref{limit}) and get for the action of HG,
\be
\label{Lchi1}
S\xrightarrow[\lambda \to \infty]{}
S'=\frac{1}{2G}\int d^3xdt\sqrt{\gamma}\,\big(K_{ij}K^{ij}-2\chi K
-{\cal V}\big)\;.
\ee
We see that the field $\chi$ takes the role of a Lagrange multiplier
constraining the extrinsic curvature to be traceless, $K=0$. 
Note that the new action is still invariant under Lifshitz scaling
(\ref{eq:anis_scaling}) if we assign $\dim \chi=3$.

Quantization of theories with Lagrange multipliers is in general
subtle. We need to make sure that the propagators of all the fields, including
$\chi$, are well-defined and the theory can be perturbatively
quantized. We also want to preserve renormalizability. For this, it
will suffice to have a gauge choice which renders all propagators {\em
regular} \cite{Anselmi:2008bq,Barvinsky:2015kil}. In real-time
signature adopted here the regularity condition is formulated as follows:
A propagator $\langle \Phi_1\Phi_2\rangle$ of two fields
$\Phi_1$, $\Phi_2$ with scaling dimensions $r_1$, $r_2$ is 
regular if it decomposes into a sum of terms of the form
\bseq
\be
\frac{P(\k,\omega)}{D(\k,\omega)}\;,
\ee
where $D$ is a product of monomials,
\be
D=\prod_{m=1}^M (A_m \omega^2-B_m k^6+i\epsilon)
\ee
\eseq
with strictly positive coefficients $A_m$, $B_m$, and $P(\k,\omega)$
is a polynomial of scaling degree less or equal $r_1+r_2+6(M-1)$.  

We cannot use the functions $F^i$ from Eq.~(\ref{eq:GFF}) for gauge
fixing since they contain terms proportional to $\lambda$ that
preclude setting $\lambda=\infty$. Then it appears impossible to
design a gauge that would eliminate the quadratic mixing between the scalar
parts of the metric $h_{ij}$, the shift $N_i$ and $\chi$. Thus, we
just pick up a 
gauge compatible with the scaling and disentangling at least the
helicity $\pm 1$ parts:
\begin{equation}
    \label{eq:chi-GFF}
\tilde F^i = \dot{N}^{i} + \frac{1}{2}{O}^{i j} \partial_k h^k_{j} \; ,
\end{equation}
with the same operator $O^{ij}$, as in Eq.~(\ref{eq:GFF}).
The propagators in this gauge are derived in
Appendix~\ref{app:chi}. We obtain many off-diagonal propagators
between $h_{ij}$, $N_i$ and $\chi$ which make practical calculations
rather cumbersome. Most importantly, however, all these propagators
are regular in the above sense guaranteeing the perturbative
renormalizability of the theory with the $\chi$-field. In particular,
this implies that no terms\footnote{Such
  terms would be irrelevant 
  by Lifshitz power counting.} 
with gradients or time derivatives of
the field $\chi$ are generated by quantum corrections and $\chi$ remains
non-dynamical.

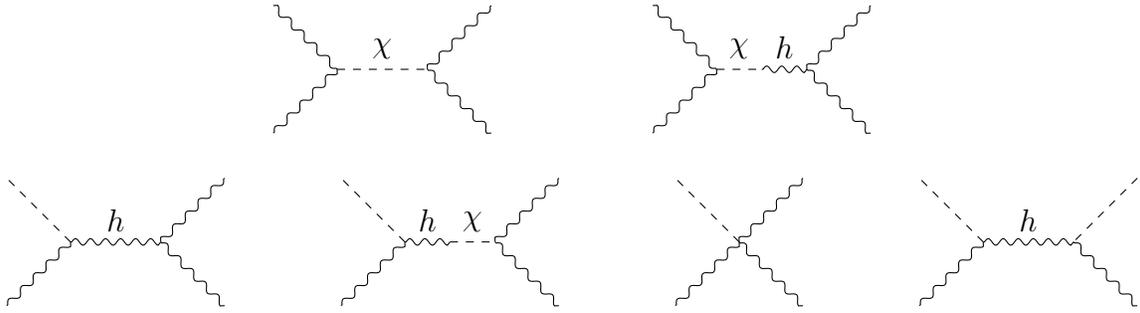
\begin{figure}[ht]
\begin{center}
\begin{tikzpicture}
\begin{feynman}
\vertex (m1);
\vertex[right=1.2cm of m1] (m2);
\vertex[below left=1.2cm of m1] (l1) {};
\vertex[above left =1.2cm of m1] (l2) {};
\vertex[above right=1.2cm of m2] (r1) {};
\vertex[below right=1.2cm of m2] (r2) {};
\diagram*{
(l1) --[photon] (m1)-- [photon] (l2),
(r1) --[photon] (m2)-- [photon] (r2),
(m1) -- [scalar, edge label=$\chi$] (m2)
};
\node[right=1.7cm of m2](n1){};
\vertex[right=2cm of n1] (m3);
\vertex[right=0.6cm of m3] (m4);
\vertex[right=0.6cm of m4] (m5);
\vertex[below left=1.2cm of m3] (l3) {};
\vertex[above left =1.2cm of m3] (l4) {};
\vertex[above right=1.2cm of m5] (r3) {};
\vertex[below right=1.2cm of m5] (r4) {};
\diagram*{
(l3) --[photon] (m3)-- [photon] (l4),
(r3) --[photon] (m5)-- [photon] (r4),
(m3) --[scalar, edge label=$\chi$](m4)-- [photon, edge label=$h$] (m5)
};
\end{feynman}
\end{tikzpicture}
\\
\begin{tikzpicture}
\begin{feynman}
\vertex (m3);
\vertex[right=0.6cm of m3] (m4);
\vertex[right=0.6cm of m4] (m5);
\vertex[below left=1.2cm of m3] (l3) {};
\vertex[above left =1.2cm of m3] (l4) {};
\vertex[above right=1.2cm of m5] (r3) {};
\vertex[below right=1.2cm of m5] (r4) {};
\diagram*{
(l3) --[photon] (m3)-- [scalar] (l4),
(r3) --[photon] (m5)-- [photon] (r4),
(m3) --[photon, edge label=$h$] (m5)
};
\node[right=1.4cm of m5](n2){};
\vertex[right=1.7cm of n2] (m6);
\vertex[right=0.6cm of m6] (m7);
\vertex[right=0.6cm of m7] (m8);
\vertex[below left=1.2cm of m6] (l5) {};
\vertex[above left =1.2cm of m6] (l6) {};
\vertex[above right=1.2cm of m8] (r5) {};
\vertex[below right=1.2cm of m8] (r6) {};
\diagram*{
(l5) --[photon] (m6)-- [scalar] (l6),
(r5) --[photon] (m8)-- [photon] (r6),
(m6) --[photon, edge label=$h$](m7)-- [scalar, edge label=$\chi$] (m8)
};
\node[right=1.4cm of m8](n3){};
\vertex[right=1.7cm of n3] (m9);
\vertex[below left=1.2cm of m9] (l7) {};
\vertex[above left =1.2cm of m9] (l8) {};
\vertex[above right=1.2cm of m9] (r7) {};
\vertex[below right=1.2cm of m9] (r8) {};
\diagram*{
(l7) --[photon] (m9)-- [scalar] (l8),
(r7) --[photon] (m9)-- [photon] (r8)
};
\node[right=1.4cm of m9](n4){};
\vertex[right=1.7cm of n4] (m10);
\vertex[right=0.6cm of m10] (m11);
\vertex[right=0.6cm of m11] (m12);
\vertex[below left=1.2cm of m10] (l9) {};
\vertex[above left =1.2cm of m10] (l10) {};
\vertex[above right=1.2cm of m12] (r9) {};
\vertex[below right=1.2cm of m12] (r10) {};
\diagram*{
(l9) --[photon] (m10)-- [scalar] (l10),
(r9) --[scalar] (m12)-- [photon] (r10),
(m10) --[photon, edge label=$h$] (m12)
};
\end{feynman}
\end{tikzpicture}
\end{center}
 \caption{Additional diagrams for the graviton
   scattering in the theory (\ref{Lchi1}) describing the
   $\lambda=\infty$ limit of projectable
   Ho\v rava gravity. The diagrams in the first row contribute to the
   amplitudes for the helicity $\pm 2$ states, and the diagrams in the
   second row must be further added for scattering of scalar
   gravitons. 
}
    \label{fig:chi-Scattering}
\end{figure}

To check the equivalence between the $\lambda\to\infty$ limit of the 
original formulation of
HG and the action (\ref{Lchi1}), 
we have computed the graviton scattering amplitudes directly with
the Feynman rules following from (\ref{Lchi1}). To avoid,
proliferation of diagrams, we fix one of the gauge
parameters,\footnote{This special choice spoils the regularity of the
  propagators in the above sense: The pole term corresponding to the
  helicity-$0$ gauge mode becomes $\tilde{\cal
    P}_0=i(\omega^2+i\epsilon)^{-1}$, i.e. it does not depend on the
  spatial momentum. This, however, is not a problem for the tree-level
  calculation where no possible divergences associated with this
  behavior can arise.}
$\xi=-1$. This eliminates the off-diagonal propagators involving the
shift $N_i$, as well as the overlap of the shift with the scalar
graviton state (see Appendix~\ref{app:chi}). On the other hand, the
mixing between the metric and $\chi$ still remains, implying that we
need to include diagrams with internal, and for scalar gravitons --
external, $\chi$-lines. This gives us the set of new diagrams shown in 
Fig.~\ref{fig:chi-Scattering} which must be added to those of 
Fig.~\ref{fig:Scattering}, with all possible permutations of the
external states. Note that the $h^3$, $h^4$ and $h^2N$ vertices for
this new calculation can be obtained from the expressions used in
Sec.~\ref{sec:amplitude} by simply dropping the parts containing
$\lambda$. At the same time we have new cubic and quartic vertices
with a $\chi$-line giving rise to diagrams in
Fig.~\ref{fig:chi-Scattering}.    

We have evaluated the amplitudes for the physical transverse-traceless
and scalar gravitons in the $\chi$-theory using our code and found that they
exactly coincide with the $\lambda\to\infty$
limit of the amplitudes computed with the original HG
action. 
This
confirms that the action (\ref{Lchi1}) 
correctly captures the
dynamics of HG at $\lambda\to\infty$. 

All in all, we conclude that the limit (\ref{limit}) of
projectable HG is regular and is described by the action
(\ref{Lchi1}).


\section{Conclusions}
\label{sec:discussion}
In this paper we computed tree-level scattering amplitudes in
projectable HG in $(3+1)$ dimensions. For this purpose, we developed a
symbolic computer code which can be found at \cite{github}. We focused
on the high-energy behavior of the theory keeping only marginal
interactions with respect to Lifshitz scaling with $z=3$.

We started by deriving the Ward identities for the amplitudes which we
used to cross-check our computation. Our approach is based on the BRST
quantization and is not restricted to HG. We illustrated it on the
case of a Yang--Mills theory with Lifshitz scaling. To the best of our
knowledge, this is the first derivation of Ward identities in
non-relativistic gauge theories.

We next discussed the general structure of the HG scattering
amplitudes and presented explicit results for the case of head-on
collisions, i.e. collisions with vanishing total momentum. The
amplitudes have peculiar dependence on the scattering angle. Their
dependence on the collision energy is compatible with tree-level
unitarity. In particular, the differential cross section decreases as
the square of the colliding particles' momentum, as it should be for a
theory weakly coupled in UV. 

We found that the amplitudes remain finite in the limit when the
coupling constant $\lambda$ in the kinetic term of the Lagrangian is
taken to infinity. We have further reformulated the action of the
theory in the form which is manifestly regular at $\lambda\to\infty$
and checked that it reproduces the same scattering amplitudes. This
establishes the $\lambda\to\infty$ limit as a viable location for
asymptotically free UV fixed points \cite{Barvinsky2021}.

Our research opens several directions. The tree amplitudes that we
computed have analytic properties quite similar to those in
relativistic theories: They have poles corresponding to physical
particles in the internal propagators, feature soft and collinear
singularities, etc. It would be interesting to understand if these
properties can be exploited in adapting to HG the powerful on-shell
methods developed for relativistic gauge theories and gravity
\cite{Elvang:2013cua}. An obvious missing ingredient is the
spinor-helicity formalism which relies on Lorentz invariance. Whether
an adequate substitute for it exists in non-relativistic theories is
an open question.

Another possible extension of our is the study of amplitudes
beyond tree level. On top of the usual issues associated with infrared
divergences, which are also present in relativistic context, such
study will have to face several new challenges. To see them consider a single
tensor or scalar graviton with the dispersion relation (\ref{omtt}) or
(\ref{oms}). Energy and momentum conservation allow it to decay into
two or more gravitons of lower energy. This implies absence of any
stable asymptotic states, thus undermining the standard assumptions
used in the definition of the ${\cal S}$-matrix. Hopefully, this
problem can be overcome by adapting the methods used in relativistic
theories to describe scattering of metastable particles. Another
peculiarity of HG gravity and non-relativistic theories in general is
that the parameters entering into particles' dispersion relations
receive loop corrections and exhibit RG running. The definition of the
asymptotic states must take these corrections into account order by
order in the loop expansion, which further challenges the standard
construction of the ${\cal S}$-matrix.   

Having established good behavior of the projectable HG in UV, our work
motivates revisiting its low-energy properties. It is known
\cite{Koyama:2009hc,Blas2010} that Minkowski background in this theory
suffers from a tachyon-like instability associated with the scalar
graviton mode. It is important to understand the fate of this
instability. Can it lead to a new phase of the theory which could be
phenomenologically viable? We plan to address this question in future.

Finally, it will be interesting to apply the amplitude-based approach
developed in this work to the non-projectable version of HG where it
can provide a valuable information about the UV properties of the
theory.

\subsection*{Acknowledgments}

We thank Andrei Barvinsky, Diego Blas, Alexander Kurov, Maxim Pospelov 
and Oriol Pujolas
for useful discussions. 
The work is supported by
the 
Natural Sciences and Engineering Research Council (NSERC) of Canada.
Research at Perimeter Institute is supported in part by the Government
of Canada through the Department of Innovation, Science and Economic
Development Canada and by the Province of Ontario through the Ministry
of Colleges and Universities.

\appendix

\renewcommand{\thesection}{\Alph{section}}
\renewcommand{\thesubsection}{\Alph{section}.\arabic{subsection}}
\renewcommand{\theequation}{\Alph{section}.\arabic{equation}}


\section{Helicity decomposition}
\label{app:decomp}

In this Appendix we diagonalize the quadratic Lagrangian (\ref{L2q})
and summarize the relations obeyed by particle creation--annihilation
operators. We start by splitting the fields into tensor, vector and
scalar parts, 
\begin{subequations} 
\label{eq:mode_decomp}
\begin{align}
    h_{ij} = \zeta_{ij} +\partial_{i} v_{j}
    + \partial_{j} v_i
+ \left(\delta_{ij} -
      \frac{\partial_i \partial_j}{\Delta} \right) \psi +
    \frac{\partial_i \partial_j}{\Delta} E\;, \\
    N_i = u_i+\partial_i B~,~~~~ 
      c_i = w_i+ \partial_i C~,~~~~
  \bar{c}_i = \bar{w}_i +\partial_i \bar{C}  \, ,    
\end{align}
\end{subequations}
where the components satisfy
\be
\label{tcond}
\d_i\zeta_{ij}=\zeta_{ii}=\d_i v_i=\d_i u_i=\d_i w_i=\d_i \bar w_i=0\;.
\ee
The Lagrangian separates into contributions of different sectors:
\bseq
\label{L2parts}
\begin{align}
\label{L2t}
\LL_q^{(2t)}=\frac{1}{2G}\bigg\{&\frac{\dot\zeta_{ij}^2}{4}
+\frac{\nu_5}{4}\zeta_{ij}\D^3\zeta_{ij}\bigg\},\\
\label{L2v}
\LL_q^{(2v)}=\frac{1}{2G}\bigg\{&-\frac{1}{2}\dot v_i\D\dot v_i
-\frac{1}{4\sigma}v_i\D^4 v_i
-\dot u_i\frac{\sigma}{\D^2} \dot u_i
-\frac{1}{2}u_i\D u_i
+2\dot{\bar w}_i\dot w_i
+\frac{1}{\sigma}\bar w_i\D^3 w_i\bigg\},\\
\LL_q^{(2s)}=\frac{1}{2G}\bigg\{&\frac{1-2\lambda}{2}\dot\psi^2
-\lambda\dot E\dot\psi+\frac{1-\lambda}{4}\dot E^2
+\bigg(\frac{8\nu_4+3\nu_5}{2}
+\frac{\lambda^2(1+\xi)}{\sigma}\bigg)\psi\D^3\psi\notag\\
&-\frac{\lambda(1-\lambda)(1+\xi)}{\sigma}E\D^3\psi
+\frac{(1-\lambda)^2(1+\xi)}{4\sigma}E\D^3E\notag\\
&+\dot B\frac{\sigma}{(1+\xi)\D}\dot B+(1-\lambda) B\D^2B
-2\dot{\bar C}\D\dot C-\frac{2(1-\lambda)(1+\xi)}{\sigma}
\bar C\D^4 C\bigg\}.
\label{L2s}
\end{align}
\eseq
The scalar part still contains mixing between the $\psi$ and $E$
components, which is removed by the change of variables,
\begin{equation} 
\label{eq:fieldS}
  E\mapsto  \tilde E = E - \frac{2 \lambda}{1- \lambda} \psi \, .
\end{equation}
The final Lagrangian in this sector reads,
\be
\label{L2sfin}
\LL_q^{(2\psi\tilde E)}=\frac{1}{2G}
\bigg\{\frac{1-3\lambda}{2(1-\lambda)}\dot\psi^2
+\frac{8\nu_4+3\nu_5}{2}\psi\D^3\psi
+\frac{1-\lambda}{4}\dot{\tilde E}^2
+\frac{(1-\lambda)^2(1+\xi)}{4\sigma}\tilde E\D^3 \tilde E\bigg\}. 
\ee
Note that the positivity of the kinetic term for the gauge invariant
scalar $\psi$ requires $\lambda$ to be outside the range
$1/3\leq\lambda\leq 1$. From Eqs.~(\ref{L2parts}), (\ref{L2sfin}) we
read off the dispersion relations (\ref{omtt}), (\ref{omga}),
(\ref{oms}) 
quoted in the main text.

Collecting the helicity modes together, we obtain the expressions for
the local fields which we write in the form,
\bseq
\label{fieldsdecomp}
\begin{align}
\label{hdecomp}
&h_{ij}(\x,t)=\sqrt{G} \int \frac{d^3k}{(2\pi)^3}\sum_{\a}
\frac{\varepsilon_{ij}^\a(\k)}{2\omega_{\k\a}}\,h_{\k\a}\,\e^{-i\omega_{\k\a}t+i\k\x}
+\text{h.c.}\;,
\\
\label{Ndecomp}
&N_{i}(\x,t)=\sqrt{G} \int \frac{d^3k}{(2\pi)^3}\sum_{\a}
\frac{\epsilon_{i}^\a(\k)}{2\omega_{\k\a}}\,N_{\k\a}\,\e^{-i\omega_{\k\a}t+i\k\x}
+\text{h.c.}\;,
\\
\label{cdecomp}
&c_{i}(\x,t)=\sqrt{G} \int \frac{d^3k}{(2\pi)^3}\sum_{\a}
\frac{e_i^\a(\k)}{2\omega_{\k\a}}\,c_{\k\a}\,\e^{-i\omega_{\k\a}t+i\k\x}
+\text{h.c.}\;,
\\
\label{barcdecomp}
&\bar c_{i}(\x,t)=\sqrt{G} \int \frac{d^3k}{(2\pi)^3}\sum_{\a}
\frac{e_{i}^\a(\k)}{2\omega_{\k\a}}\,\bar c_{\k\a}\,\e^{-i\omega_{\k\a}t+i\k\x}
-\text{h.c.}\;,
\end{align}
\eseq
where the sum runs over the helicities $\a$ contained in the
corresponding field (see Eq.~(\ref{modeset})). Note that the ghosts
$c_i$ are taken to be Hermitian, whereas the anti-ghosts $\bar c_i$
are anti-Hermitian. The former property is needed for the Hermiticity
of the BRST operator, whereas the latter then follows from the
Hermiticity of the Lagrangian.

We normalize the mode coefficients in such a way that upon
quantization they become the annihilation--creation operators with the
commutation relations:
\bseq
\label{modecomm}
\begin{align}
\label{hcomm}
&[h_{\k\a},h_{\k'\b}^+]=2\omega_{\k\a}\,\delta_{\a\b}\,(2\pi)^3\delta(\k-\k')
[\sign(1-\lambda)]^{\delta_{\a 0}}\;,\\
\label{Ncomm}
&[N_{\k\a},N_{\k'\b}^+]=-2\omega_{\k\a}\,\delta_{\a\b}\,(2\pi)^3\delta(\k-\k')
[\sign(1-\lambda)]^{\delta_{\a 0}}\;,\\
\label{ccomm}
&[c_{\k\a}, \bar c_{\k'\b}^+]_+=[\bar c_{\k\a}, c_{\k'\b}^+]_+=
-2\omega_{\k\a}\,\delta_{\a\b}\,(2\pi)^3\delta(\k-\k')\;.
\end{align}
\eseq
Two comments are in order. First note that we use the ``relativistic''
normalization including a factor $2\omega$ for the operators and
corresponding scattering states. Though in our case it is not
connected with Lorentz invariance, it is still convenient since it
results in dimensionless $2\to 2$ scattering amplitudes. Second, the
helicity $\pm 1$ modes of the shift $N_i$ clearly have negative
norm. In the helicity $0$ sector the situation is subtler. Here the
negative-norm state is in $N_i$ or $h_{ij}$, depending on whether
$\lambda$ is less or bigger than $1$, as reflected by the last factor
in Eqs.~(\ref{hcomm}), (\ref{Ncomm}).

It remains to specify the polarization vectors and tensors entering
Eqs.~(\ref{modecomm}). Let us start with the ghosts. Their
polarization vectors are given by the standard orthonormal triad which
for the momentum with polar and azimuthal angles $\theta$, $\phi$
has the form,
\be
\label{polare}
e_i^{(0)}\equiv \hat k_i=
\begin{pmatrix}
\sin\theta\cos\phi\\
\sin\theta\sin\phi\\
\cos\theta
\end{pmatrix}~,~~~~~
e_i^{(\pm 1)}=\mp\frac{\e^{\pm i\phi}}{\sqrt{2}}
\begin{pmatrix}
\cos\theta\cos\phi\mp i\sin\phi\\
\cos\theta\sin\phi\pm i\cos\phi\\
-\sin\theta
\end{pmatrix}~.
\ee
The polarizations in $N_i$ differ by normalizations that can be read
out the Lagrangians (\ref{L2v}), (\ref{L2s}):
\be
\label{polarN}
\epsilon_i^{(\pm 1)}=\frac{k^2}{\sqrt{\sigma}}\,e_i^{(\pm 1)}~,~~~~~
\epsilon_i^{(0)}=k^2\sqrt{\frac{|1+\xi|}{\sigma}}\, \hat k_i\;.
\ee
Finally, the polarization tensors in $h_{ij}$ are constructed from the
triad as follows:
\bseq
\label{polarh}
\begin{gather}
\ve_{ij}^{(\pm2)}=2e_i^{(\pm 1)}e_j^{(\pm 1)}~,~~~~
\ve_{ij}^{(\pm1)}=\sqrt{2}\left(\e_i^{(\pm 1)}\hat k_j
+\hat k_i e_j^{(\pm 1)}\right)\;,\\
\ve_{ij}^{(0)}=\frac{2}{\sqrt{|1-\lambda|}}\hat k_i\hat k_j~,~~~~
\ve_{ij}^{(0')}=\sqrt{\frac{2(1-\lambda)}{1-3\lambda}}
\left(\delta_{ij}-\frac{1-3\lambda}{1-\lambda}\hat k_i\hat k_j\right).
\end{gather}
\eseq

\section{BRST-Invariance of the ${\cal S}$-Matrix}
\label{app:Smatr}

In this Appendix we review the derivation of Eq.~(\ref{eq:sym_Smatr})
stating that the ${\cal S}$-matrix of a gauge theory commutes with the
{\em asymptotic} quadratic BRST operator $Q^{(2)}$. We follow
Ref.~\cite{Becchi:1996yh} generalizing the analysis to an abstract gauge theory
which need not enjoy Lorentz invariance. We adopt the conventions and
notations of \cite{Barvinsky:2017zlx} (except for (anti-)ghosts 
which we denote with $c$, instead of $\omega$).

Consider a gauge theory with local gauge-invariant action
$S$ built out of gauge and matter fields $\varphi^{a}$, where the label 
$a$ collectively denotes all field indices and coordinates.
The fields linearly transform
under the action of the gauge group via 
\begin{equation}
    \delta_{\ve} \varphi^{a} = \ve^{\alpha}(P^{a}_{\ \alpha}
    + R^{a}_{\ b \alpha} \varphi^{b}) \, , 
\end{equation}
where $\ve^\a$ is the transformation parameter.
The gauge fields are supplemented by the Faddeev--Popov ghosts $c^\a$,
anti-ghosts $\bar c_\a$, and the Nakanishi--Lautrup field $b_\a$,
related by the BRST transformations,
\begin{equation}
\label{eq:BRST_trans_gen}
    \mathbf{s} \varphi^{a} = c^\a(P^{a}_{\ \alpha} + R^{a}_{\ b \alpha}
    \varphi^{b}) \, ,~~~~~ 
    \mathbf{s} c^{\alpha} = \frac{1}{2} C^{\alpha}_{\ \beta \gamma}
    c^{\beta} c^{\gamma} \, , ~~~~
    \mathbf{s} \bar{c}_{\alpha} = b_{\alpha} \, ,~~~~
    \mathbf{s} b_{\alpha} = 0 \; ,
\end{equation}
where $C^{\a}_{\ \beta\gamma}$ are the structure constants of the
gauge group. 
Implementing the BRST quantization procedure we obtain the quantum tree-level
action $S_{q}$ invariant under \eqref{eq:BRST_trans_gen},
\begin{equation}
\label{eq:quant_act_gen}
    S_{q} = S[\varphi] + b_{\alpha} \chi^{\alpha}_{a} \varphi^{a} -
    \frac{1}{2} b_{\alpha} O^{\alpha \beta} b_{\beta} -
    \bar{c}_{\alpha} \chi^{\alpha}_{a}(P^{a}_{\ \beta} +
    R^{a}_{\ b\beta} \varphi^{b}) c^{\beta} \;, 
\end{equation}
where we have chosen linear gauge-fixing functions
$\chi^{\alpha}_{a}\varphi^{a}$. Note that since the transformations
(\ref{eq:BRST_trans_gen}) are non-linear, the conserved BRST charge
$Q$ generating them in the Heisenberg picture is non-linear as well. 
However, instead of pursuing the operator quantization, we use
the path integral approach.  
 
We define the
generating functional with sources for all the fields and their BRST
variations:  
\begin{equation}
\label{BRSTpartfunc}
    Z[J, \bar\xi, {\xi},y, \gamma, \zeta] = \int \! D \Phi^A \,
    \exp\Big\{i\big(S_{q}[\varphi, c, \bar{c}, b] + J_{a} \varphi^{a} +
      \bar{\xi}_{\alpha} c^{\alpha} + \xi^{\alpha}
      \bar{c}_{\alpha} + y^{\alpha} b_{\alpha} + \gamma_{a}
      \mathbf{s} \varphi^{a} + \zeta_{\alpha} \mathbf{s} c^{\alpha}
    \big)\Big\} \,, 
\end{equation}
where $\Phi^A$ stands collectively for all the fields $\varphi^a$,
$c^\a$, $\bar{c}_\a$ and $b_\a$. 
We further define the partition function 
(generating functional for the connected diagrams):
\begin{equation}
    W = - i \log Z \, ,
\end{equation}
and its Legendre transform --- the effective action 
\begin{equation}
\label{GWLegendre}
    \Gamma\big[\langle\varphi\rangle, \langle c\rangle ,
    \langle\bar{c}\rangle ,  \langle b\rangle , \gamma, \zeta\big] = W - J_{a}
    \langle \varphi^{a}\rangle - \bar{\xi}_{\alpha} \langle
    c^{\alpha}\rangle -\xi^{\alpha} \langle c_{\alpha}\rangle
    - y^{\alpha} \langle b_{\alpha}\rangle \;,
\end{equation}
with the quantities in angular brackets denoting the {\em mean}
fields. By definition, the latter are variational derivatives of $W$
with respect to the sources,\footnote{We define the derivatives with
  respect to anti-commuting variables as acting from the left,
  i.e. the differential of a function $f(\theta)$ of a Grassmann
  variable $\theta$ is $df=d\theta \, f'(\theta)$.}
\begin{equation}
\label{eq:Leg_cur_def}
  \langle\varphi^{a}\rangle =  \frac{\delta W}{\delta J_{a}} \, , ~~~~
\langle c^{\alpha}\rangle =
\frac{\delta W}{\delta \bar{\xi}_{\alpha}} \, , ~~~~
 \langle \bar{c}_{\alpha}\rangle =  \frac{\delta W}{\delta \xi^{\alpha}} \, , ~~~~
 \langle b_{\alpha} \rangle=  \frac{\delta W}{\delta y^{\alpha}} \, .   
\end{equation}
Note that at tree-level the effective action is
\be
\Gamma^{\rm tree}=S_q+\gamma_a {\s \varphi^a}+\zeta_\a \s c^\a\;. 
\ee 
The relation (\ref{GWLegendre}) implies the equality of the
variational derivatives, 
\begin{gather}
\label{eq:anti-field_der}
    \frac{\delta \Gamma}{\delta \gamma_{a}} = \frac{\delta W}{\delta
      \gamma_{a}} \, . 
\end{gather}
Importantly, the partition function satisfies the identities (see e.g
\cite{Barvinsky:2017zlx} for the derivation),
\bseq
\begin{align}
\label{eq:ST}
  &{\bf D}W\equiv  
\left(-J_{a} \frac{\delta}{\delta \gamma_{a}} + \bar{\xi}_{\alpha}
      \frac{\delta}{\delta \zeta_{\alpha}} + \xi^{\alpha}
      \frac{\delta}{\delta y^{\alpha}} \right) W 
    = 0\;,\\ 
\label{eq:y_der_Gamma}
&\left(\chi^{\alpha}_{a} \frac{\delta}{\delta J_{a}} -
      O^{\alpha \beta} \frac{\delta}{\delta y^{\beta}} +
      y^{\alpha} \right) W = 0 \;.
\end{align}
\eseq
The first equation here is the Slavnov--Taylor identity following from
the BRST symmetry (\ref{eq:BRST_trans_gen}), whereas the second is the
equation of motion for the Nakanishi--Lautrup field.

We now use the Lehmann--Symanzik--Zimmermann (LSZ) reduction 
(see \cite{Collins:2019ozc} for a recent
discussion) to define the ${\cal S}$-matrix from the correlation
functions. In a compact form, it can be written as 
(see e.g. \cite{Itzykson1980})
\begin{equation}
\label{SLSZ}
    {\cal S} = \, : \exp{ \left(- {\Phi}^{A}_{\text{as}}\,
        \mathcal{K}_{AB}\, \frac{\delta}{\delta
          \mathcal{J}_{B}} \right)}: \, Z[{\cal J}]
    \bigr|_{\mathcal{J}=0}\equiv {\bf K}\,Z[{\cal J}]\bigr|_{\mathcal{J}=0}\;. 
\end{equation}
Here $\Phi_{\text{as}}^A=\{\varphi_{\text{as}}^a,c_{\text{as}}^\a,\bar
c_{\text{as}\, \a}\}$ are the asymptotic gauge and (anti-)ghost field
operators,\footnote{
We consider the asymptotic states as being generated by the free
fields. This may not be
true for a variety of reasons, such as infrared divergences or
particle instability, see
discussion in Sec.~\ref{sec:discussion}. We proceed under the assumption
that these issues can be properly handled on the case-by-case basis.

In principle, one could also introduce the
  asymptotic Nakanishi--Lautrup field, but we choose not to do it
  since $b_\a$ is not an independent variable on-shell, being
  expressed through the gauge-fixing function.} 
and ${\cal J}_A=\{J_a,\bar\xi_\a,\xi^\a,\gamma_a,\zeta_\a,y^\a\}$ are
the corresponding currents supplemented with the BRST sources.
Colon around the
exponent stand for the normal ordering with respect to particle
creation-annihilation operators contained inside
$\Phi_{\text{as}}^A$. The differential operator ${\cal K}_{AB}$ is taken
from the wave equations satisfied by the asymptotic fields, 
\begin{equation}
    \mathcal{K}_{AB} \Phi^{B}_{\text{as}} = 0 \,. 
\end{equation}
Despite these equations, the exponent in (\ref{SLSZ}) is non-trivial
because the operator ${\cal K}_{AB}$ in it acts to the right and cancels with
the on-shell poles of the Green's functions produced by the
variational derivatives with respect to the currents. The vertical
line with subscript ``${\cal J}=0$'' means that all sources must be set to
zero {\em after} taking the variational derivatives.

In the second equality in (\ref{SLSZ}) we have introduced the notation
${\bf K}$ for the exponential factor acting on $Z[{\cal J}]$.
This object is a ``double operator'': it is a variational
operator acting on functionals of the currents, and
a quantum-mechanical operator in the asymptotic Fock space. We observe
that 
\begin{equation}
\label{eq:ST_S-matrix}
  [\mathbf{K},\mathbf{\mathbf{D}}]\, W[{\cal J}]
  \bigr|_{\mathcal{J}=0}
=\mathbf{K} \, \mathbf{D} \, W[J] \bigr|_{\mathcal{J}=0}
- \mathbf{D} \, \mathbf{K} \, W[J] \bigr|_{\mathcal{J}=0}=0\, .
\end{equation}
Indeed, the first term vanishes due to the Slavnov--Taylor identity
(\ref{eq:ST}), whereas the second term is zero because $\mathbf{D}$ is
proportional to the sources. Evaluating $[\mathbf{K},\mathbf{D}]$
on the l.h.s. as the commutator of two variational operators we
obtain,\footnote{Note the different signs of the ghost and the
  anti-ghost terms stemming from their anti-commutativity: $\bar c_\a
  ({\cal K}^c)^\a_{\ \beta}c^\beta =- c^\beta 
({\cal K}^c)^\a_{\ \beta} \bar c_\a$.} 
\begin{equation}
\label{commKD}
    [\mathbf{K},\mathbf{D}]\, W[{\cal J}] \bigr|_{\mathcal{J}=0} = \,:
    \mathbf{K}\cdot \left( \varphi^{a}_{\text{as}}\, \mathcal{K}^{\varphi}_{ab}\,
      \frac{\delta}{\delta \gamma_{b}} - \bar{c}_{\text{as}\,\alpha}\,
      (\mathcal{K}^{c})^{\alpha}_{\ \beta}\, \frac{\delta}{\delta
        \zeta_{\beta}} + c^{\alpha}_{\text{as}}\, (\mathcal{K}^{c})_{\
        \alpha}^{\beta}\, \frac{\delta}{\delta y^{\beta}}\right):\,
    W[\mathcal{J}]  \bigr|_{\mathcal{J}=0}\;. 
\end{equation}

Let us discuss the terms in brackets one by one, starting from the
last. Using the relation \eqref{eq:y_der_Gamma} it can be transformed
as 
\begin{equation}
    \frac{\delta W}{\delta y^{\beta}}  = O^{-1}_{\beta \alpha}
    \chi^{\alpha}_{a} \frac{\delta W}{\delta J_{a}} +
    O^{-1}_{\beta \alpha} y^{\alpha}\, . 
\end{equation}
The second term on the right hand side does not contribute because
upon acting with $\mathbf{K}$ it either leaves something proportional
to $y^{\a}$ which is zero when we take currents to be zero, or, if
the derivatives from $\mathbf{K}$ hit $y^\a$ instead of the generating
functional, we are not getting poles from the Green's functions to
compensate the action of $(\mathcal{K}^{c})_{\ \alpha}^{\beta}$. 

The second term in (\ref{commKD}) amounts to
\begin{equation}
    \frac{\delta W}{\delta \zeta_{\beta}} = \langle \mathbf{s}
    c^{\beta} \rangle = \left\langle \frac{1}{2} C^{\beta}_{\ \gamma
      \delta} c^{\gamma} c^{\delta} + \dots \right\rangle 
\end{equation}
with the dots representing 
corrections coming from renormalization. The 
diagrams contributing to this matrix element 
do not have poles since there are no one-particle states
with ghost number 2.
Hence they vanish once
we act on them by ${\cal K}^c$ and restrict on-shell.

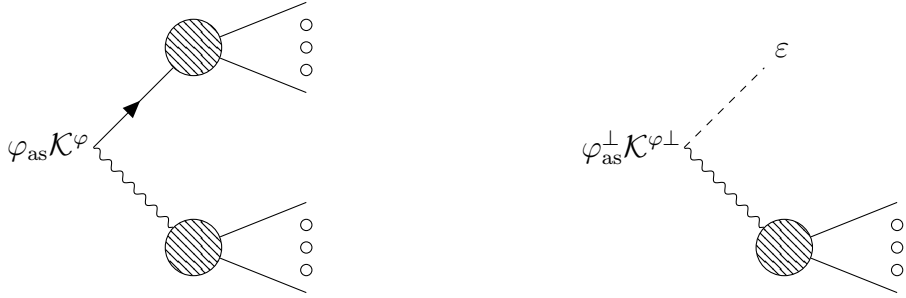
\begin{figure}[ht]
\begin{center}
\begin{tikzpicture}
\begin{feynman}
\vertex (m1) {$\varphi_{\rm as}{\cal K}^{\varphi}$};
\vertex[right=0.6cm of m1] (m2);
\vertex[above right=1.5cm of m2] (l2) [blob, draw=black!40!black,
pattern color=black!40!black] {};  
\vertex[below right=1.5cm of m2] (l3) [blob, draw=black!40!black,
pattern color=black!40!black] {}; 
\vertex[right=1.5cm of l2, empty dot] (r1) {};
\vertex[above=0.3cm of r1, empty dot] (ra1) {};
\vertex[above=0.6cm of r1] (ra3) ;
\vertex[below=0.3cm of r1, empty dot] (rb1) {};
\vertex[below=0.6cm of r1] (rb3) ;
\vertex[right=1.5cm of l3, empty dot] (q1) {};
 \vertex[above=0.3cm of q1, empty dot] (qa1) {};
 \vertex[above=0.6cm of q1] (qa3) ;
 \vertex[below=0.3cm of q1, empty dot] (qb1) {};
 \vertex[below=0.6cm of q1] (qb3) ;
\diagram*{
(m2) --[fermion] (l2),
(m2) --[photon] (l3),
(l2) --(ra3),
(l2) --(rb3),
(l3) --(qa3),
(l3) --(qb3),
};
\end{feynman}
\end{tikzpicture}
\qquad\qquad\qquad\qquad
\begin{tikzpicture}
\begin{feynman}
\vertex (m1) {$\varphi_{\rm as}^\perp{\cal K}^{\varphi\perp}$};
\vertex[right=0.7cm of m1] (m2);
\vertex[above right=1.5cm of m2] (l2) {$\ve$};  
\vertex[below right=1.5cm of m2] (l3) [blob, draw=black!40!black,
pattern color=black!40!black] {}; 
\vertex[right=1.5cm of l3, empty dot] (q1) {};
 \vertex[above=0.3cm of q1, empty dot] (qa1) {};
 \vertex[above=0.6cm of q1] (qa3) ;
 \vertex[below=0.3cm of q1, empty dot] (qb1) {};
 \vertex[below=0.6cm of q1] (qb3) ;
\diagram*{
(m2) --[scalar] (l2),
(m2) --[photon] (l3),
(l3) --(qa3),
(l3) --(qb3),
};
\end{feynman}
\end{tikzpicture}
\end{center}
 \caption{Diagrams arising from the second terms in
   Eq.~(\ref{antifield_der_long}) (left) and Eq.~(\ref{deltaSphys})
   (right). They do not have on-shell poles to cancel the vanishing
   vertex factor $\varphi_{\rm as}{\cal K}^\varphi$.}
    \label{fig:discon}
\end{figure}

The first term in Eq.~(\ref{commKD}) requires a bit more work. Using
Eqs.~(\ref{eq:Leg_cur_def}), (\ref{eq:anti-field_der}) 
we can write a Taylor expansion,
\begin{align}
    \frac{\delta W}{\delta \gamma_{b}} = \frac{\delta \Gamma}{\delta
      \gamma_{b}} &=  \frac{\delta^2 \Gamma}{\delta
      \langle c^{\alpha}\rangle\delta
      \gamma_{b}}\bigg|_{\langle\Phi\rangle=0} 
\langle c^{\alpha}\rangle + 
\frac{\delta^3 \Gamma}{\delta\langle\varphi^a\rangle\delta
      \langle c^{\alpha}\rangle\delta
      \gamma_{b}}\bigg|_{\langle\Phi\rangle=0} 
\langle c^{\alpha}\rangle \langle\varphi^a\rangle+
\dots \notag\\
& =  \frac{\delta^2 \Gamma}{\delta
      \langle c^{\alpha}\rangle\delta
      \gamma_{b}}\bigg|_{\langle\Phi\rangle=0} 
\frac{\delta W}{\delta \bar\xi_\a} + 
\frac{\delta^3 \Gamma}{\delta\langle\varphi^a\rangle\delta
      \langle c^{\alpha}\rangle\delta
      \gamma_{b}}\bigg|_{\langle\Phi\rangle=0} 
\frac{\delta W}{\delta \bar\xi_\a}\frac{\delta W}{\delta J_a}+
\dots 
\, .
\label{antifield_der_long}
\end{align}
The expansion starts with the term linear in the ghost field since the
l.h.s. has unit ghost number\footnote{The source $\gamma_a$ has ghost
  number $-1$ from the way it enters the partition function (\ref{BRSTpartfunc})
  in combination with the BRST variation of the gauge field.} 
and thus vanishes at $\langle
c^\a\rangle=0$. 
The second and subsequent terms lead to the diagrams of the form shown
on the left of Fig.~\ref{fig:discon} which do not have poles. Thus, the only pole
contribution comes from the first term. We notice that at tree level
the second variational derivative entering it coincides with the
generator of linear gauge transformations,   
\begin{equation}
     \frac{\delta^2 \Gamma}{\delta \langle c^{\alpha}\rangle \delta
       \gamma_{b} }\bigg|_{\langle\Phi\rangle=0} = P^{b}_{\ \alpha}\;.
\end{equation}
In fact, this relation remains valid also after taking into account
loop corrections, with $P^b_\a$ understood as the generator acting on
properly normalized asymptotic fields \cite{Becchi:1996yh}. 

We further
have the identity,
\begin{equation}
\label{Ktrans}
    \mathcal{K}^{\varphi}_{ab} P^b_{\ \a} = (\mathcal{K}_{ab}^{\varphi \perp} +
     \chi^{\beta}_{a} O^{-1}_{\beta\gamma}
     \chi^{\gamma}_{b})P^b_{\ \a}
=\chi^{\beta}_{a} O^{-1}_{\beta\gamma}
     \chi^{\gamma}_{b} P^b_{\ \a}\;,
\end{equation}
where we have split the wave operator for the asymptotic gauge fields
into the `transverse' and `longitudinal' parts and used that the former is
gauge invariant. By `transverse' part here we mean the operator coming
from the original action $S$, whereas the `longitudinal' part 
arises upon eliminating from the action (\ref{eq:quant_act_gen}) the
non-dynamical field $b_\a$. 
Finally, we recall the structure of the ghost wave
operator which is again read off from (\ref{eq:quant_act_gen}),
\be
\label{Kghost}
({\cal K}^c)^\a_{\ \beta}=-\chi^\a_a P^a_{\ \beta}\;.
\ee
Combining together the above results gives,
\begin{equation}
    [\mathbf{K},\mathbf{D}]\, W[J] \bigr|_{\mathcal{J}=0} =\,
: \mathbf{K}\cdot \left(-\varphi^{a}_{\text{as}} \chi_a^\a
      O^{-1}_{\a\beta}\,({\cal K}^c)^\beta_{\ \gamma}\frac{\delta}{\delta\bar\xi_\gamma}
      - c^{\alpha}_{\text{as}}P^a_{\ \a}\, {\cal K}^\varphi_{ab}
      \frac{\delta}{\delta J_{b}}\right):\, W[\mathcal{J}] \bigr|_{\mathcal{J}=0} \;.
\end{equation}
We recognize here the linear BRST variations of the asymptotic fields
generated by $Q^{(2)}$,
\begin{align}
&i[Q^{(2)},\bar{c}_{\text{as}\,\alpha}]_{+} = 
  O^{-1}_{\alpha \beta} \chi^{\beta}_{a}
    \varphi^{a}_{\text{as}}\;,
    &&i[Q^{(2)},\varphi^{a}_{\text{as}}] = P^{a}_{\ \alpha} c^{\alpha}_{\text{as}}\;.
\end{align}
Recall also that the linear BRST variation of the ghost field
vanishes, $i[Q^{(2)},c^\a_{\text{as}}]_+=0$. This allows us to write
\begin{equation}
\label{eq:repr_ST_Q}
    [\mathbf{K},\mathbf{D}]\, W[J] \bigr|_{\mathcal{J}=0} =
    i[Q^{(2)},\mathbf{K}]\, W[J] \bigr|_{\mathcal{J}=0} \, ,
\end{equation}
where on the r.h.s. we have the commutator of operators acting on the
asymptotic Fock space. 
Together with Eq.~(\ref{eq:ST_S-matrix}) and the definition of the
${\cal S}$-matrix (\ref{SLSZ}) it implies Eq.~(\ref{eq:sym_Smatr}). 

For completeness, let us also show that the elements of the ${\cal
  S}$-matrix (\ref{SLSZ}) between the states containing only physical
particles do not depend on the choice of gauge. The physical particle
states are interpolated by `transverse' components of the asymptotic
fields satisfying $\chi^\a_a\varphi^{a\perp}_{\text{as}}=0$. Thus, the
restriction of the ${\cal S}$-matrix to the physical states can be
written as
\be
\label{Sphys}
{\cal S}^{\rm phys}=\, :\exp\left(-\varphi^{a\perp}_{\rm as}
{\cal K}^{\varphi\perp}_{ab}\frac{\delta}{\delta J_b}\right):\,Z[{\cal
J}]\bigr|_{{\cal J}=0}\;.
\ee
An infinitesimal change of the gauge-fixing functions $\delta \chi_a^\a$ can be
compensated by a properly chosen gauge transformation $\delta_\ve
\varphi^a$ of the integration variables in the path integral
(\ref{BRSTpartfunc}), so 
that we have,
\be
\label{deltaZ}
\delta Z[{\cal J}] =\langle iJ_a\delta_\ve \varphi^a\rangle Z[{\cal J}]
=J_a\ve^\a \left(iP^a_{\ \a}+R^a_{\ b\a}\frac{\delta}{\delta J_b}\right)
Z[{\cal J}]\;.
\ee
Substituting this into Eq.~(\ref{Sphys}) we obtain,
\be
\label{deltaSphys}
\delta {\cal S}^{\rm phys}=\,:{\bf K}\cdot
\left(-i\varphi_{\rm as}^{a\perp}{\cal K}^{\varphi\perp}_{ab} 
\ve^\a P_{\ \a}^b
- \varphi_{\rm as}^{a\perp}{\cal K}^{\varphi\perp}_{ab} 
\ve^\a R^b_{\ c\a}\frac{\delta }{\delta J_c}\right):\, Z[{\cal
J}]\bigr|_{{\cal J}=0}\;.
\ee
The first term in brackets vanishes due to the gauge invariance of
${\cal K}^{\varphi\perp}_{ab}$, whereas the second term leads to the
diagrams shown on the right of Fig.~\ref{fig:discon} and does
not have on-shell poles. This implies $\delta{\cal S}^{\rm phys}=0$,
as expected.

\section{Feynman Rules in $\sigma,\xi$-gauge}

\label{app:Feyn}

Here we summarize the Feynman rules used in the
computation of graviton $2\to2$ scattering amplitudes 
in the gauge of Sec.~\ref{sec:BRSTHG}. We also include the ingredients 
entering diagrams with an external shift $N_i$ which are used for
verification of the gauge consistency relation~(\ref{HGgaugeid}). \\   

\noindent{\bf External lines:}
\bseq
\label{extlegs}
\begin{align}
&\feynmandiagram[baseline=(a.base), horizontal= a to c]{
a [particle =\(N_{i}\)] -- [double] b -- [double] c,
a -- [double, edge label={\(\k,\omega\)}] c,
};\quad= -\sqrt{G}\,\epsilon^\a_{i}(\k,\omega)\;,\\
&\feynmandiagram[baseline=(a.base), horizontal= a to c]{
a [particle =\(h_{ij}\)] -- [color=white] b -- [color=white] c,
a -- [photon, edge label={\(\k,\omega\)}] c,
};\quad= \sqrt{G}\,\ve^\a_{ij}(\k,\omega)\;,
\label{extlegsh}
\end{align}
\eseq
with
\be
\epsilon^\a_{i}(\k,\omega)=
\begin{cases}
\epsilon^\a_{i}(\k)\,,&\omega>0\\
-\epsilon^\a_{i}(-\k)\,,&\omega<0
\end{cases}\quad,
\qquad\qquad\quad
\ve^\a_{ij}(\k,\omega)=
\begin{cases}
\ve^\a_{ij}(\k)\,,&\omega>0\\
\ve^\a_{ij}(-\k)\,,&\omega<0
\end{cases}
\ee
The positive-frequency polarization factors 
$\epsilon^\a_{i}(\k)$, $\ve^\a_{ij}(\k)$
are 
given by Eqs.~(\ref{polarN}),
(\ref{polarh}). 
Note that we treat all momenta and energies as flowing into the
diagram.\\

\noindent{\bf Propagators:}
\begin{subequations}
\label{eq:props}
\begin{align}
\label{eq:props:1}
\feynmandiagram[baseline=(a.base), horizontal= a to c]{
a [particle =\(N_{i}\)] -- [double] b -- [double] c [particle=\(N_{j}\)],
a -- [double, edge label={\(\k,\omega\)}] c,
};
=& ~-G\bigg[\frac{k^4}{
  \sigma} (\delta_{ij} -\hat{k}_i \hat{k}_j) \mathcal{P}_{1} +
\frac{(1+\xi)k^4}{\sigma} \hat{k}_i \hat{k}_j \mathcal{P}_0\bigg]\,,
\\ 
\label{eq:props:2}
\feynmandiagram[baseline=(a.base), horizontal= a to c]{
a [particle =\(h_{ij}\)]  -- [color=white] b -- [color=white] c
[particle=\(h_{kl}\)],  
a -- [photon, edge label={\(\k,\omega\)}]c,
};
=&~ 2G\bigg\{
(\delta_{i k}\delta_{j l} + \delta_{i l} \delta_{j k})
\mathcal{P}_{tt} - \delta_{i j}\delta_{kl} \left[
  \mathcal{P}_{tt}- \frac{1-\lambda}{1-3\lambda} \mathcal{P}_s
\right] \nonumber \\   
   & \qquad- (\delta_{i k}\hat{k}_{j}\hat{k}_{l} 
    + \delta_{i l}\hat{k}_{j}\hat{k}_{k} + \delta_{j
      k}\hat{k}_{i}\hat{k}_{l} + \delta_{j l}\hat{k}_{i}\hat{k}_{k})
    \bigr[\mathcal{P}_{tt}-\mathcal{P}_{1} \bigr] \nonumber \\ 
 &\qquad   + (\delta_{i j} \hat{k}_{k} \hat{k}_{l} + \delta_{k l}
    \hat{k}_{i}\hat{k}_{j} )
    \bigr[\mathcal{P}_{tt}-\mathcal{P}_{s} \bigr] \nonumber \\ 
 & \qquad
 +\hat{k}_{i}\hat{k}_{j}\hat{k}_{k}\hat{k}_{l}\left[\mathcal{P}_{tt} 
      + \frac{1-3 \lambda}{1-\lambda} \mathcal{P}_{s} - 4
      \mathcal{P}_{1} +
      \frac{2\mathcal{P}_0}{1-\lambda} \right] \bigg\}\,.
\end{align} 
\end{subequations}
Here $\hat \k$ is the unit vector along the momentum, and the pole
factors are
\bseq
\label{proppoles}
\begin{align}
    \mathcal{P}_1 &= \frac{i}{\omega^2 - \omega^2_{1}(k)+i\epsilon}~,~~~~~~~
\mathcal{P}_0 =\frac{i}{\omega^2 - \omega^2_0(k)+i\epsilon} \\
\mathcal{P}_{tt}& =\frac{i}{\omega^2 -  \omega^2_{tt}(k)+i\epsilon} 
~,~~~~~~~
    \mathcal{P}_s= \frac{i}{\omega^2 - \omega^2_{s}(k)+i\epsilon}\;,
\end{align}
\eseq
with the dispersion relations (\ref{omtt})---(\ref{oms}). The
Euclidean version of these 
propagators was derived in \cite{Barvinsky:2015kil}.\\

\noindent{\bf Vertices.} In our calculation we use the following
vertices:
\bseq
\label{Vertices}
\begin{gather}
\label{Vgggg}
\begin{tikzpicture}
\begin{feynman}
\vertex (c1) [dot];
\vertex[right=1.2cm of c1] (c2) {\(h_{mn}\)}; 
\vertex[below left=1.2cm of c1] (a2) {\(h_{ij}\)};
\vertex[above left=1.2cm of c1] (a3) {\(h_{kl}\)};
\diagram* {
(c1) -- [photon] (a2),
(a3) -- [photon] (c1),
(c1) -- [photon] (c2)
};
\end{feynman}
\end{tikzpicture}
\qquad~~
\begin{tikzpicture}
\begin{feynman}
\vertex (c1) [dot];
\vertex[right=1.2cm of c1] (c2) {\(N_m\)}; 
\vertex[below left=1.2cm of c1] (a2) {\(h_{ij}\)};
\vertex[above left=1.2cm of c1] (a3) {\(h_{kl}\)};
\diagram* {
(c1) -- [photon] (a2), 
(a3) -- [photon] (c1),
(c1) -- [double] (c2)
};
\end{feynman}
\end{tikzpicture}
\qquad~~
\begin{tikzpicture}
\begin{feynman}
\vertex (c1) [dot];
\vertex[below left=1.2cm of c1] (a1) {\(h_{ij}\)};
\vertex[above left=1.2cm of c1] (a2) {\(h_{kl}\)};
\vertex[above right=1.2cm of c1] (a3) {\(h_{mn}\)}; 
\vertex[below right=1.2cm of c1] (a4) {\(h_{pq}\)}; 
\diagram* {
(a1) -- [photon] (c1), 
(a3)-- [photon] (c1),
(a2) -- [photon] (c1),
(a4) -- [photon](c1)
};
\end{feynman}
\end{tikzpicture}
\\
\begin{tikzpicture}
\begin{feynman}
\vertex (c1) [dot];
\vertex[right=1.2cm of c1] (c2) {\(N_l\)}; 
\vertex[below left=1.2cm of c1] (a2) {\(h_{ij}\)};
\vertex[above left=1.2cm of c1] (a3) {\(N_k\)};
\diagram* {
(c1) -- [photon] (a2), 
(a3) -- [double] (c1), 
(c1) -- [double] (c2),
};
\end{feynman}
\end{tikzpicture}
\qquad\qquad
\begin{tikzpicture}
\begin{feynman}
\vertex (c1) [dot];
\vertex[below left=1.2cm of c1] (a1) {\(h_{ij}\)};
\vertex[above right=1.2cm of c1] (a2) {\(h_{kl}\)};
\vertex[below right=1.2cm of c1] (a3) {\(h_{mn}\)}; 
\vertex[above left=1.2cm of c1] (a4) {\(N_p\)}; 
\diagram* {
(a1) -- [photon] (c1), 
(a3)-- [photon] (c1),
(a2) -- [photon] (c1),
(a4) -- [double](c1)
};
\end{feynman}
\end{tikzpicture}
\end{gather}
\eseq
The vertices in the first line enter the graviton scattering
amplitude, see Fig.~\ref{fig:Scattering}, whereas the vertices in the
second line are used to verify
the identity 
(\ref{HGgaugeid}).

The full 
expressions for the vertices are lengthy
and not illuminating. We present explicitly only the parts of (\ref{Vgggg})
which are proportional to the coupling constant $\lambda$ and
could lead to large contributions to the graviton amplitudes in the limit
$\lambda\to \infty$. These are used in the proof of
Sec.~\ref{sec:linfcancel} 
that the divergent contributions actually cancel.  
\bseq
\label{vertices}
\begin{align}
\label{Vhhh}
&\begin{tikzpicture}
\begin{feynman}
\vertex (c1) [dot];
\vertex[right=1.5cm of c1] (c2) {\(h_{mn}\)}; 
\vertex[below left=1.5cm of c1] (a2) {\(h_{ij}\)};
\vertex[above left=1.5cm of c1] (a3) {\(h_{kl}\)};
\diagram* {
(c1) -- [photon] (a2),
(a3) -- [photon] (c1),
(c1) -- [photon, edge label={\(\k_3,\omega_3\)}] (c2)
};
\node[label={[rotate=45,label distance=-0.1cm]\(\k_1,\omega_1\)},
below left=0.6cm of c1];
\node[label={[rotate=-45,label distance=-0.15cm]\(\k_2,\omega_2\)},
above left=0.6cm of c1];
\vertex[below=1.1cm of c2] (c3);
\node[right=0.8cm of c3] {\(\begin{aligned}
     =  i\frac{\lambda}{48G}\Big[&\delta_{ij}\delta_{kl}\delta_{mn}
(\omega_1\omega_2+\omega_2\omega_3+\omega_3\omega_1)\\
&-\delta_{ij}(\delta_{km}\delta_{ln}+\delta_{kn}\delta_{lm})
\omega_1(\omega_2+\omega_3)\\
&-\delta_{kl}(\delta_{im}\delta_{jn}+\delta_{in}\delta_{jm})
\omega_2(\omega_3+\omega_1)\\
&-\delta_{mn}(\delta_{ik}\delta_{jl}+\delta_{il}\delta_{jk})
\omega_3(\omega_1+\omega_2)\Big]+O(\lambda^0)
\end{aligned}\)};
\end{feynman}
\end{tikzpicture}
\\
\label{VhhN}
&\begin{tikzpicture}
\begin{feynman}
\vertex (c1) [dot];
\vertex[right=1.5cm of c1] (c2) {\(N_m\)}; 
\vertex[below left=1.5cm of c1] (a2) {\(h_{ij}\)};
\vertex[above left=1.5cm of c1] (a3) {\(h_{kl}\)};
\diagram* {
(c1) -- [photon] (a2), 
(a3) -- [photon] (c1),
(c1) -- [double, edge label=\({\bf p}\)] (c2)
};
\node[label={[rotate=45,label distance=-0.1cm]\(\k_1,\omega_1\)},
below left=0.6cm of c1];
\node[label={[rotate=-45,label distance=-0.15cm]\(\k_2,\omega_2\)},
above left=0.6cm of c1];
\vertex[below=0.7cm of c2] (c3);
\node[right=0.8cm of c3] {\(\begin{aligned}
     =
     -i\frac{\lambda}{8G}
\Big[&\delta_{ij}\delta_{kl}(\omega_1k_{1m}+\omega_2k_{2m}) 
+(\delta_{ik}\delta_{jl}+\delta_{jk}\delta_{il})(\omega_1+\omega_2)p_m\\
&-\delta_{ij}\omega_1(\delta_{lm} k_{1k}
+\delta_{km} k_{1l})
-\delta_{kl}\omega_2(\delta_{jm} k_{2i}
+\delta_{im} k_{2j})\Big]\\
&+O(\lambda^0)\;,
  \end{aligned}\)};
\end{feynman}
\end{tikzpicture}
\\
\label{Vhhhh}
&\begin{tikzpicture}
\begin{feynman}
\vertex (c1) [dot];
\vertex[below left=1.5cm of c1] (a1) {\(h_{ij}\)};
\vertex[above left=1.5cm of c1] (a2) {\(h_{kl}\)};
\vertex[above right=1.5cm of c1] (a3) {\(h_{mn}\)}; 
\vertex[below right=1.5cm of c1] (a4) {\(h_{pq}\)}; 
\diagram* {
(a1) -- [photon] (c1), 
(a3)-- [photon] (c1),
(a2) -- [photon] (c1),
(a4) -- [photon](c1)
};
\node[label={[rotate=45,label distance=-0.1cm]\(\k_1,\omega_1\)}, below left=0.6cm of c1];
\node[label={[rotate=-45,label distance=-0.15cm]\(\k_2,\omega_2\)}, above left=0.6cm of c1];
\node[label={[rotate=45,label distance=-0.2cm]\(\k_3,\omega_3\)}, above right=0.7 of c1];
\node[label={[rotate=-45,label distance=-0.1cm]\(\k_4,\omega_4\)}, below right=0.8 of c1];
\vertex[below=1.5cm of c1] (c3);
\node[right=1.9cm of c3] {\(\begin{aligned}
     = i\frac{\lambda}{64G}\,{\rm sym}\,\Big\{\omega_1\omega_2\Big[
&2\delta_{ij}(\delta_{kq}\delta_{lm}\delta_{np}\!+\!\delta_{lq}\delta_{km}\delta_{np}
\!+\!\delta_{kq}\delta_{ln}\delta_{mp}\!+\!\delta_{lq}\delta_{kn}\delta_{mp}\\
&+\!\delta_{kp}\delta_{lm}\delta_{nq}\!+\!\delta_{lp}\delta_{km}\delta_{nq}
\!+\!\delta_{kp}\delta_{ln}\delta_{mq}\!+\!\delta_{lp}\delta_{kn}\delta_{mq})\\
&+\!2(\delta_{im}\delta_{jn}\!+\!\delta_{in}\delta_{jm})
(\delta_{kp}\delta_{lq}\!+\!\delta_{kq}\delta_{lp})\\
&-\!4\delta_{ij}\delta_{mn}(\delta_{kp}\delta_{lq}\!+\!\delta_{kq}\delta_{lp})
\!-\!\delta_{ij}\delta_{kl}(\delta_{mp}\delta_{nq}\!+\!\delta_{mq}\delta_{np})\\
&+\delta_{ij}\delta_{kl}\delta_{mn}\delta_{pq}\Big]\Big\}+O(\lambda^0)\;.
\end{aligned}\)};
\end{feynman}
\end{tikzpicture}
\end{align}
\eseq
In the last expression `sym' stands for symmetrization over
the graviton lines.

\section{Angular dependence of head-on amplitudes}
\label{app:amp}

Throughout this Appendix we denote $x=\cos\theta$. The subscripts
$+,-,s$ stand for the $\pm 2$, and $0'$-helicity gravitons. 
Here we use the {\em physical} helicities to label the incoming and outgoing
particles: For example, the subscript $++,++$ means that both
gravitons in the initial and final states are right-handed.
For the
relation of the angular
functions $f_{\a_1\a_2,\a_3\a_4}$ to the full amplitude 
see Eq.~(\ref{Mampl}).

\subsection{Processes without scalar gravitons}

Using the notation $\hat u_s^2=\frac{1-3\lambda}{1-\lambda} u_s^2
=\frac{8\nu_4}{\nu_5}+3$, we have:
\begin{align}
   f_{++,++}= &f_{--,--}\notag=\\
=&\frac{1}{512 \hat u_s^2 (1-x^2)^3}
\bigg[x^8\Big(-161 - 320 v_{2}^2 + v_{2} (464 - 720 v_{3}) + 39
   \hat{u}_{s}^2 - 9 v_{3}^2 (45 - 11 \hat{u}_{s}^2) \notag \\
 &+ 6 v_{3} (87 - 85 \hat{u}_{s}^2)\Big) 
 +4 x^6 \Big(231 + 443 \hat{u}_{s}^2 - 72 v_{3}^2 \hat{u}_{s}^2 - 16
   v_{2} (21 - 8 \hat{u}_{s}^2) \notag\\
&+ 6 v_{3} (63 - 53 \hat{u}_{s}^2)\Big) 
   +2 x^4 \Big(-287 + 448 v_{2}^2 - 4783 \hat{u}_{s}^2 - 16 v_{2} (49 -
   63 v_{3} + 48 \hat{u}_{s}^2) \notag\\
&+ 63 v_{3}^2 (9 + \hat{u}_{s}^2) - 6
   v_{3} (147 + 295 \hat{u}_{s}^2)\Big) -4 x^2 \Big(581 + 128 v_{2}^2
   - 6343 \hat{u}_{s}^2 \notag\\
&- 16 v_{2} (35 - 18 v_{3} + 24 \hat{u}_{s}^2) + 54 v_{3}^2 (3 -
\hat{u}_{s}^2) - 6 v_{3} (105 + 269 \hat{u}_{s}^2)\Big)
-169 - 64 v_{2}^2 \notag\\
&- 19921 \hat{u}_{s}^2 - 9 v_{3}^2 (9 + 17
\hat{u}_{s}^2) + 16 v_{2} (13 - 9 v_{3} + 32 \hat{u}_{s}^2) + 6 v_{3}
(39 - 613 \hat{u}_{s}^2) 
\bigg]\;;
\end{align}
\begin{align}
    f_{++,+-}=&f_{--,-+}=f_{+-,++} = f_{-+,--}=\notag\\
 =& \frac{1}{512 \hat{u}_{s}^2 (1-x^2)}\bigg[
x^4 \Big(133 + 64 v_{2}^2 - 16 v_{2} (13 - 12 v_{3}) - 243
\hat{u}_{s}^2 + 9 v_{3}^2 (15 - \hat{u}_{s}^2)\notag\\
& - 12 v_{3} (23 - 13 \hat{u}_{s}^2)\Big)
    -2x^2 \Big(211 + 64 v_{2}^2 - 16 v_{2} (7 - 9 v_{3}) - 285
    \hat{u}_{s}^2 + 9 v_{3}^2 (9 + \hat{u}_{s}^2) \notag\\
&- 12 v_{3} (15 - 11 \hat{u}_{s}^2)\Big) 
    +64 v_{2}^2 - 16 v_{2} (1 - 6 v_{3}) + 27 v_{3}^2 (1 +
    \hat{u}_{s}^2) + 12 v_{3} (5 + 21 \hat{u}_{s}^2)\notag\\ 
&- 11 (13 + 69
    \hat{u}_{s}^2) 
\bigg]\;;
\end{align}
\begin{align}
   f_{++,--} =&f_{--,++}=\notag\\
=&\frac{1 }{512 \hat{u}_{s}^2}\bigg[3 x^2 
  \Big(-35 + 64 v_{2}^2 - 16 v_{2} (1 - 7 v_{3}) + 501 \hat{u}_{s}^2
  + 3 v_{3}^2 (15 - \hat{u}_{s}^2) \notag \\
&+ 2 v_{3} (5 - 79
  \hat{u}_{s}^2)\Big)
+121 + 64 v_{2}^2 - 1375 \hat{u}_{s}^2 + 9 v_{3}^2 (1 +
   \hat{u}_{s}^2) + 66 v_{3} (1 + 13 \hat{u}_{s}^2) \notag\\
&+ 16 v_{2} (11 + 3
   v_{3} + 32 \hat{u}_{s}^2)
\bigg]\;;
\end{align}
\begin{align}
    f_{+-,+-} = &\frac{1+x}{512 \hat{u}_{s}^2
      (1-x)^3}
\bigg[-x^4\Big(64 v_{2}^2 - 16 v_{2} (13 - 12 v_{3}) +
    27 v_{3}^2 (5 + \hat{u}_{s}^2) - 12 v_{3} (23 + 15 \hat{u}_{s}^2)\notag\\
   & + 7 (19 + 59 \hat{u}_{s}^2)\Big)
-6 x^3 (4 - v_{3}) \Big(16 v_{2} + 3 v_{3} (7 + 3 \hat{u}_{s}^2) -
    4 (5 + 7 \hat{u}_{s}^2)\Big) \nonumber \\ 
    &+2x^2 \Big(-221 + 64 v_{2}^2 - 16 v_{2} (7 - 9 v_{3}) - 205
    \hat{u}_{s}^2 + 18 v_{3}^2 (3 - \hat{u}_{s}^2) + 12 v_{3} (3 + 13
    \hat{u}_{s}^2)\Big) \nonumber \\ 
    &+6x (4 - v_{3}) \Big(16 v_{2} + 3 v_{3} (3 - \hat{u}_{s}^2) +
    4 (1 - \hat{u}_{s}^2)\Big) \nonumber \\ 
    &-145 - 64 v_{2}^2 + v_{2} (16 - 96 v_{3}) + 103 \hat{u}_{s}^2 +
    v_{3} (84 - 60 \hat{u}_{s}^2) - 9 v_{3}^2 (5 + \hat{u}_{s}^2) 
\bigg]\;.
\end{align}

\subsection{Processes with one scalar graviton}

Due to different dispersion relations of scalar and tensor modes the 
structure of the amplitudes involving both types of particles
is more complicated. Let us consider the
case when the scalar graviton is in the final state. Then the momentum
of outgoing particles is related to the incoming momentum $k$ as
\be
\label{alpha1}
k'=\varkappa k~,~~~~~~~~\varkappa=\left(\frac{2}{1+u_s}\right)^{1/3}\;.
\ee
Using this notation, we can write
\bseq
\begin{align}
   f_{\a_1\a_2,\a_3 s} = \sqrt{\frac{2 (1- \lambda)}{1 - 3
        \lambda}}\frac{P_{\a_1\a_2,\a_3 s}(x)}{g_{1}(x)}\;,~~~~~\a_I=+,-\;,
\end{align}
\eseq
where
\begin{align}
\label{denom1}
    g_1(x) &= \bigl((1 - 2 x \varkappa + \varkappa^2)^3 -  (1 - u_{s}
    \varkappa^3)^2\bigr) \bigl(u_{s}^2 (1 - 2 x \varkappa + \varkappa^2)^3 -
    (1 - u_{s} \varkappa^3)^2\bigr) \nonumber \\
    &\times \bigl((1 + 2 x \varkappa + \varkappa^2)^3 -  (1 - u_{s}
    \varkappa^3)^2\bigr) \bigl(u_{s}^2 (1 + 2 x \varkappa + \varkappa^2)^3 -
    (1 - u_{s} \varkappa^3)^2\bigr) \;, 
\end{align}
and $P_{\a_1\a_2,\a_3 s}(x)$ are polynomials of $14$th degree in $x$
that are too cumbersome to present explicitly. Note that for $u_s\neq
1$ the denominator
(\ref{denom1}) has roots at non-zero scattering angles. As discussed in
Sec.~\ref{sec:headon}, this corresponds to resonant poles in the
amplitude due to on-shell graviton decays. On the other hand, $g_1(x)$ is
regular in the forward and  backward limits $x=\pm 1$. In fact, the amplitude
vanishes in these limits since the polynomials in the numerator can be 
factorized as
\be
\label{polyfact1}
P_{++,+s}=(1-x^2) \tilde P_{++,+s}~,~~~~
P_{++,-s}=(1-x^2) \tilde P_{++,-s}~,~~~~
P_{+-,-s}=(1-x)^3(1+x) \tilde P_{+-,-s}\;,
\ee
and similarly for the channels obtained by parity and time inversion.
This is consistent with conservation of angular momentum (see
Sec.~\ref{sec:headon}). 

The amplitudes greatly simplify if the dispersion relations of the
tensor and scalar gravitons coincide: $u_s=1$, $\varkappa=1$. Then we have:
\begin{align}
\label{fppps}
    f_{++,+s}=& f_{--,-s}=f_{+s,++}=f_{-s,--}=\notag\\
=&\frac{1}{128\sqrt{2(1-\lambda)(1 - 3 \lambda)^3}(1-x^2)^2}
    \bigg[x^6 \Bigl(81 + 80 v_{2}^2 (1 - \lambda)^2 - 245 \lambda +
    230 \lambda^2 \notag\\
&+ 18 v_{3}^2 (5\! - \!8 \lambda\! +\! 3 \lambda^2) 
- 3 v_{3} (31\! -\! 21 \lambda \!- \!8 \lambda^2) - 4 v_{2} (1 -
    \lambda) \bigl(49 \!-\! 80 \lambda \!-\! v_{3} (48\! -\! 51
    \lambda)\bigr)\Bigr) \nonumber \\ 
    &- x^4 \Bigl(447 + 240 v_{2}^2 (1 - \lambda)^2 - 1175 \lambda +
    402 \lambda^2 + 18 v_{3}^2 (17 - 38 \lambda + 21 \lambda^2)
    \nonumber \\  
    &- 3 v_{3} (111 - 165 \lambda + 76 \lambda^2) - 4 v_{2} (1 -
    \lambda) \bigl(85 - 60 \lambda - 3 v_{3} (48 - 59
    \lambda)\bigr)\Bigr) \nonumber \\ 
    &+ x^2 \Bigl(-1221 + 112 v_{2}^2 (1 - \lambda)^2 + 5729 \lambda -
    5870 \lambda^2 + 54 v_{3}^2 (5 - 16 \lambda + 11 \lambda^2)
    \nonumber \\ 
    &+ 3 v_{3} (159 - 821 \lambda + 608 \lambda^2) + 4 v_{2} (1 -
    \lambda) \bigl(121 - 536 \lambda + 3 v_{3} (32 - 59
    \lambda)\bigr)\Bigr) \nonumber \\ 
    &+ \Bigl(1587 + 48 v_{2}^2 (1 - \lambda)^2 - 6851 \lambda + 6426
    \lambda^2 - 4 v_{2} (1 - \lambda) \bigl(253 - (708 + 51 v_{3})
    \lambda\bigr) \nonumber \\ 
    &- 54 v_{3}^2 (1 - 6 \lambda + 5 \lambda^2) - 3 v_{3} (335 - 1253
    \lambda + 884 \lambda^2)\Bigr) \bigg]\;;
\end{align}
\begin{align}
    f_{++,-s} =& f_{--,+s}=f_{-s,++}=f_{+s,--}=\notag\\
=&\frac{1}{128\sqrt{2(1-\lambda)(1 - 3 \lambda)^3}}\bigg[
  x^2 \Bigl(299 + 18 v_{3}^2 (1 - \lambda) + 48 v_{2}^2 (1
    - \lambda)^2 - 1247 \lambda + 882 \lambda^2 \nonumber \\ 
    &- 3 v_{3} (7 \!-\! 49 \lambda \!+\! 40 \lambda^2) + 4 v_{2} (1\! -\!
    \lambda) \bigl(17\! -\! 48 \lambda \!+\! 3 v_{3} (6 \!-\! 5
    \lambda)\bigr)\Bigr) -299 - 48 v_{2}^2 (1 \!-\! \lambda)^2\notag\\
& + 1211 \lambda - 36 v_{3}^2
    (1\! -\! \lambda) \lambda - 810 \lambda^2 
- 4 v_{2} (1\! -\! \lambda) \bigl(35 + 3 v_{3} (4\! -\! \lambda) - 84
    \lambda\bigr) \notag\\
&- 3 v_{3} (11 - 9 \lambda + 4 \lambda^2) 
\bigg]\;;
\end{align}
\begin{align}
    f_{+-,-s}=& f_{-+,+s}=f_{-s,+-}=f_{+s,-+}=\notag\\
=&\frac{1}{128\sqrt{2(1-\lambda)(1 - 3 \lambda)^3}(1+x)^2}\bigg[
 x^4 \Bigl(187 + 16 v_{2}^2 (1 - \lambda)^2 - 651 \lambda +
    466 \lambda^2 \notag\\
&+ 27 v_{3}^2 (2\! - \!5 \lambda \!+\! 3 \lambda^2)
   - 6 v_{3} (34\! - \! 105 \lambda \!+\! 72 \lambda^2) - 4 v_{2} (1 \!-\!
    \lambda) \bigl(33\! -\! 64 \lambda\! - \! 3 v_{3} (5\! -\! 6
    \lambda)\bigr)\Bigr) \nonumber \\ 
    &+3 x^3 (4 - v_{3}) \Big(18 - 4 v_{2} (1 - \lambda) - 56
    \lambda + 36 \lambda^2 - 3 v_{3} (3 - 7 \lambda + 4
    \lambda^2)\Big) \nonumber \\ 
    &- x^2 \Bigl(64 v_{2}^2 (1 - \lambda)^2 + 9 v_{3}^2 (13 - 31
    \lambda + 18 \lambda^2) + 2 (25 - 69 \lambda + 62 \lambda^2)
    \nonumber \\ 
    &- 6 v_{3} (47 - 140 \lambda + 96 \lambda^2) - 4 v_{2} (1 -
    \lambda) \bigl(78 - 160 \lambda - 9 v_{3} (5 - 6
    \lambda)\bigr)\Bigr) \nonumber \\ 
    &+3x (4 - v_{3}) \Bigl(-18 + 64 \lambda - 52 \lambda^2 + 4 v_{2}
    (5 - 13 \lambda + 8 \lambda^2) + 3 v_{3} (7 - 19 \lambda + 12
    \lambda^2)\Bigr) \nonumber \\ 
    &-137 + 48 v_{2}^2 (1 - \lambda)^2 + 561 \lambda - 438 \lambda^2
    + 9 v_{3}^2 (3 - 4 \lambda + \lambda^2) \nonumber \\ 
    &- 6 v_{3} (1 + 5 \lambda - 8 \lambda^2) - 12 v_{2} (1 -
    \lambda) \bigl(7 - 16 \lambda - v_{3} (6 - 4 \lambda)\bigr)\bigg] \;.
\end{align}

\subsection{Processes with two scalar gravitons}

\subsubsection{Two scalars in the final state}
Here the relation between the outgoing and incoming momenta is
\be
\label{alpha2}
k'=\varkappa k~,~~~~~~\varkappa=u_s^{-1/3}\;,
\ee
and the angular functions have the form 
\be
f_{\a_1\a_2,ss}=\frac{2(1-\lambda)}{(1-3\lambda)}
\frac{P_{\a_1\a_2,ss}(x)}{g_2(x)}~,~~~~\a_I=+,-\;,  
\ee
with the denominator
\be
\label{denom2}
g_{2}(x)=\bigl(1 + (2 - 4 x^2) \varkappa^2 + \varkappa^4\bigr)^6\;.
\ee
We
observe that this denominator does not have any zeros for $u_s\neq 1$.
The absence of resonances is explained as follows. 
For the processes at hand 
the energies of initial and final particles are the same. So 
the energy
flowing in the propagators of intermediate states in $t-$ and
$u-$channels vanishes, whereas the momentum does not. 
For the $s$-channel the situation is opposite.
Thus these
propagators 
never become on-shell. 
For the case of different helicities $\a_1\neq
\a_2$ the amplitude vanishes in the collinear limits since
\be
P_{+-,ss}=(1-x^2)^2 \tilde P_{+-,ss}\;,
\ee
as required by the angular momentum conservation.

The $14$th order polynomials $P_{\a_1\a_2,ss}(x)$ are again too
lengthy in general. We explicitly present the amplitude for the case $u_s=1$:
\begin{align}
    f_{++,ss} \!=&f_{--,ss}=f_{ss,++}=f_{ss,--}=\notag\\
=&\frac{1}{128(1\!-\!\lambda)^2(1\! -\! 3 \lambda)^2(1\!-\!x^2)}\bigg[
 \!-\!x^4 \Bigl(\!87\! -\! 527 \lambda \!+\! 887 \lambda^2 \!\!-\! 421
    \lambda^3 \!\!-\! 42 \lambda^4\! \!-\! 9 v_{3}^2 (1\! -\! \lambda)^3
    (2 \!-\! 3\lambda) \notag\\ 
&+ 12 v_{3} (1 - \lambda)^2 (8 - 24 \lambda + 17 \lambda^2)+ 8 v_{2} (1 - \lambda)^3 \bigl(20 - 51 \lambda - v_{3} (3 - 6
    \lambda)\bigr)\Bigr)\notag\\
& +x^2 (1 - \lambda) \Bigl(270 + 36 v_{3} - 207
    v_{3}^2 - 1416 \lambda - 36 v_{3} \lambda + 900 v_{3}^2 \lambda + 1874 \lambda^2
    + 216 v_{3} \lambda^2 \notag\\
&- 1179 v_{3}^2 \lambda^2 - 548 \lambda^3 -
    216 v_{3} \lambda^3 + 486 v_{3}^2 \lambda^3 - 1536 v_{1} (1 -
    \lambda)^2 (1 - 3 \lambda) \notag\\ 
&- 16 v_{2}^2 (1 - \lambda)^2 (9 - 22 \lambda) - 8 v_{2} (1 -
    \lambda) \bigl(26 - 122 \lambda + 70 \lambda^2 + 9 v_{3} (5 - 17
    \lambda + 12 \lambda^2)\bigr)\Bigr) \nonumber \\ 
    &-183 + 60 v_{3} + 225 v_{3}^2 + 1159 \lambda - 360 v_{3} \lambda
    - 1206 v_{3}^2 \lambda - 2387 \lambda^2 + 432 v_{3} \lambda^2 +
    2268 v_{3}^2 \lambda^2 \notag\\
&+ 1953 \lambda^3\!-\! 24 v_{3} \lambda^3\! -\! 1818 v_{3}^2
\lambda^3\! -\! 558 \lambda^4 \!-\!
    108 v_{3} \lambda^4 \!+\! 531 v_{3}^2 \lambda^4 \!+\! 1536 v_{1} (1\! -\!
    \lambda)^3 (1\! -\! 3 \lambda) \nonumber \\ 
    &+ 16 v_{2}^2 (1 \!-\! \lambda)^3 (13 \!-\! 30 \lambda) + 8 v_{2}
    (1 \!-\!
    \lambda)^2 \bigl(46 \!-\! 185 \lambda\! +\! 105 \lambda^2\! +\! 18 v_{3} (3 \!-\!
    10 \lambda \!+\! 7 \lambda^2)\bigr)
\bigg] \;;
\end{align}
\begin{align}
    f_{+-,ss} =&f_{ss,+-}=\notag\\
=& \frac{-1}{128(1\!-\!\lambda)^2(1\! -\! 3 \lambda)^2(1\!-\!x^2)}\bigg[
x^4 \Bigl(273 - 1633 \lambda + 3433 \lambda^2 - 3179
    \lambda^3 +\ 1122 \lambda^4 \notag\\
&+ 9 v_{3}^2 (1 - \lambda)^3 (4 - 9
    \lambda) - 12 v_{3} (1 - \lambda)^2 (22 - 80 \lambda
    + 59 \lambda^2) \notag\\
&- 8 v_{2} (1\! -\! \lambda)^3 \bigl(20\! -\! 51 \lambda \!-\! v_{3} (3\! -\! 6
    \lambda)\bigr)\Bigr) 
- x^2 (1 - \lambda) \Bigl(242 - 1240 \lambda + 2126 \lambda^2 -
    1372 \lambda^3\notag\\
& + 16 v_{2}^2 (1\! -\! \lambda)^2 (1\! +\! 2 \lambda)\! +\! 9
    v_{3}^2 (1 \!-\! \lambda)^2 (7 \!-\! 6 \lambda) \!-\! 8 v_{2} (1\! -\! \lambda)
    \bigl(46\! -\! 9 v_{3} (1 \!-\! \lambda)\! -\! 150 
    \lambda \!+\! 98 \lambda^2\bigr) \notag\\
&+ 12 v_{3} (-35 + 151 \lambda - 194
    \lambda^2 + 78 \lambda^3)\Bigr) 
-31 + 151 \lambda - 51 \lambda^2 - 367 \lambda^3 + 282 \lambda^4\notag\\
& + 9 v_{3}^2 (1 - \lambda)^3 (7 - 5 \lambda) + 16 v_{2}^2 (1 -
    \lambda)^3 (5 - 6 \lambda) - 12 v_{3} (1 - \lambda)^3 (13 - 27
    \lambda) \notag\\
&+ 8 v_{2} (1 - \lambda)^2 \bigl(-26 + 18 v_{3} (1 -
    \lambda)^2 + 87 \lambda - 63 
    \lambda^2\bigr) \bigg]\;.
\end{align}

\subsubsection{Tensor -- scalar scattering}
In this case the absolute value of the initial and final momenta is
the same, $k'=k$. The amplitude has the form,
\begin{equation}
   f_{\a_1 s,\a_2 s} = \frac{2 (1- \lambda)}{(1 - 3 \lambda)}
   \frac{P_{\a_1 s,\a_2 s}(x)}{g_3(x)}~,~~~~~~\a_I=+,-\;, 
\end{equation}
with
\begin{equation}
    g_3(x)=64 u_{s}^2 (1 - x)^3 \big((1-u_s)^2-8(1+x)^3)\big) 
\bigl((1-u_s)^2- 8u_{s}^2(1+x)^3\bigr)\;, 
\end{equation}
and $P_{\a_1 s,\a_2 s}(x)$ an 11th order polynomial. The amplitude has
resonant poles at non-zero values of $x$ and diverges in the forward
limit. The divergence is alleviated if the helicities of tensor
gravitons in the initial and 
final states are different,
\be
\label{polfact2}
P_{\a_1 s,\a_2 s}=(1-x)^2 \tilde P_{\a_1 s,\a_2 s}~~~\text{for}~~\a_1\neq \a_3\;,
\ee 
consistently with the angular momentum conservation. For  
the same initial and final helicities and $u_s\neq 1$ 
the amplitude vanishes in the
backward limit:
\be
\label{polfact2}
P_{\a_1 s,\a_2 s}=(1+x)^2 \tilde P_{\a_1 s,\a_2 s}~~~\text{for}~~\a_1=\a_3\;.
\ee

The case of identical tensor and scalar dispersion relations,
$u_{s}=1$, 
leads to:
\begin{align}
    f_{+s,+s} =& f_{-s,-s}=\notag\\
=&\frac{-1}{256(1-\lambda)^2(1-3\lambda)^2(1-x)^3(1+x)}\bigg[
 x^6 \Bigl(64 v_{2}^2 (1 - \lambda)^4 + 27 v_{3}^2 (1 - \lambda)^3 (4
 - 5 \lambda) \notag\\
&- 3 v_{3} (119\! -\! 605 \lambda \!+\! 1081 \lambda^2\! -\! 819
 \lambda^3\! +\! 224 \lambda^4) + 4 (93 \!-\! 533 \lambda \!+\! 1052 \lambda^2\! -\!
 878 \lambda^3 \!+\! 262 \lambda^4)\notag\\
& + 4 v_{2} (1 - \lambda)^2 \bigl(3
 v_{3} (15 - 34 \lambda + 19 \lambda^2) - 4 (21 - 67 \lambda + 43
 \lambda^2)\bigr)\Bigr) \nonumber \\ 
    &+ x^5 \Bigl(-342 + 198 v_{3} + 153 v_{3}^2 - 192 v_{1}(1 - \lambda)^4 (7 - 8
    v_{2} - 9 v_{3})  + 1618 \lambda - 1014 v_{3}
    \lambda \nonumber \\ 
    &- 657 v_{3}^2 \lambda - 2450 \lambda^2 + 1854 v_{3} \lambda^2 +
    1053 v_{3}^2 \lambda^2 + 1178 \lambda^3 - 1434 v_{3} \lambda^3 -
    747 v_{3}^2 \lambda^3 \nonumber \\ 
    &+ 60 \lambda^4 + 396 v_{3} \lambda^4 + 198 v_{3}^2 \lambda^4 + 16
    v_{2}^2 (1 - \lambda)^3 (31 - 30 \lambda) \nonumber \\ 
    &+ 8 v_{2} (1 - \lambda)^2 \bigl(-20 + 8 \lambda + 18 \lambda^2 +
    3 v_{3} (29 - 59 \lambda + 30 \lambda^2)\bigr)\Bigr) \nonumber \\ 
    &+x^4 \Bigl(-738 + 1461 v_{3} - 441 v_{3}^2 - 192 v_{1} (1 - 8
    v_{2} - 9 v_{3}) (1 - \lambda)^4 + 4086 \lambda - 8187 v_{3}
    \lambda\notag\\
& + 2394 v_{3}^2 \lambda - 7522 \lambda^2 + 15903 v_{3}
    \lambda^2 - 4536 v_{3}^2 \lambda^2 + 6178 \lambda^3 - 13029 v_{3} \lambda^3
    + 3654 v_{3}^2 \lambda^3 \notag\\
&- 2084 \lambda^4 + 3852 v_{3} \lambda^4 -
    1071 v_{3}^2 \lambda^4 + 16 v_{2}^2 (1 - \lambda)^3 (9 + 2
    \lambda) - 4 v_{2} (1 -\lambda)^2 (-324 + 69 v_{3} \notag\\
&+ 1156 \lambda
    - 348 v_{3} \lambda -  
    848 \lambda^2 + 279 v_{3} \lambda^2)\Bigr) 
+2 x^3 (1 - \lambda) \Bigl(770 - 918 v_{3} + 117 v_{3}^2\notag\\
& + 192 v_{1} (13 - 8 v_{2} - 9 v_{3}) (1 - \lambda)^3 - 3708 \lambda
    + 4704 v_{3} \lambda - 684 v_{3}^2 \lambda + 5250 \lambda^2 -
    6990 v_{3} \lambda^2 \notag\\
&+ 1017 v_{3}^2 \lambda^2 - 2132 \lambda^3 + 3180 v_{3} \lambda^3 -
450 v_{3}^2 \lambda^3 - 16 v_{2}^2 (1 - \lambda)^2 (21 - 10 \lambda)\notag\\
&+ 8 v_{2} (1 \!-\! \lambda) \bigl(28 - 264 \lambda + 262 \lambda^2 + 3
v_{3} (11 \!-\! \lambda \!-\! 10 \lambda^2)\bigr)\Bigr) 
    +x^2 \Bigl(\!-1072 \!+\! 453 v_{3} \!+\! 90 v_{3}^2 \notag\\
&+ 384 v_{1}(1 - \lambda)^4 (7 - 8 v_{2} - 9 v_{3}) + 7688 \lambda - 2223 v_{3}
    \lambda - 855 v_{3}^2 \lambda 
- 18724 \lambda^2 + 3843 v_{3} \lambda^2 \notag\\
&+ 2025 v_{3}^2 \lambda^2
    + 17692 \lambda^3 - 2913 v_{3} \lambda^3 - 1845 v_{3}^2 \lambda^3
    - 5504 \lambda^4 + 840 v_{3} \lambda^4 + 585 v_{3}^2 \lambda^4
    \notag\\
& - 32 v_{2}^2 (1 \!-\!
    \lambda)^3 (21 \!-\! 16 \lambda) + 4 v_{2} (1\! -\! \lambda)^2 \bigl(276
    \!-\! 700 \lambda \!+\! 404 \lambda^2 \!-\! 3 v_{3} (55 \!-\! 54 \lambda \!-\!
    \lambda^2)\bigr)\Bigr) \nonumber \\ 
    &+ x \Bigl(-910 + 1638 v_{3} - 531 v_{3}^2 - 192 v_{1}(1 - \lambda)^4 (19 - 8
    v_{2} - 9 v_{3})  + 5322 \lambda - 10422 v_{3}
    \lambda \notag\\
&+ 2979 v_{3}^2 \lambda - 10634 \lambda^2 + 22302 v_{3}
    \lambda^2 - 5751 v_{3}^2 \lambda^2 + 8882 \lambda^3 - 19866 v_{3} \lambda^3
    + 4689 v_{3}^2 \lambda^3 \notag\\
&- 2724 \lambda^4 + 6348 v_{3} \lambda^4
    - 1386 v_{3}^2 \lambda^4 - 16 v_{2}^2 (1 - \lambda)^3 (5 - 42
    \lambda) + 8 v_{2} (1 - \lambda)^2 \bigl(76 - 568 \lambda \notag\\
&+ 570
    \lambda^2 \!-\! 3 v_{3} (23\! -\! 105 \lambda \!+\! 82 \lambda^2)\bigr)\Bigr)
- 192 v_{1} (1 \!-\! \lambda)^4 (13 \!-\! 8 v_{2} \!-\! 9 v_{3})
   \! +\! 1726 \!-\! 1557 v_{3} \notag\\
&+ 387 v_{3}^2  - 11658 \lambda + 8787 v_{3} \lambda -
    1800 v_{3}^2 \lambda + 26998 \lambda^2
- 17271 v_{3} \lambda^2 + 3078 v_{3}^2 \lambda^2 \notag\\
&- 25446 \lambda^3 + 14445 v_{3} \lambda^3 - 2304 v_{3}^2 \lambda^3 
+ 16 v_{2}^2 (1 - \lambda)^3 (45 - 62 \lambda)
+ 8396 \lambda^4 - 4404 v_{3} \lambda^4 \notag\\
&+ 639 v_{3}^2 \lambda^4  - 4 v_{2} (1 - \lambda)^2 (516 -
    285 v_{3} + 1652 \lambda - 696 v_{3} \lambda - 1208 \lambda^2 +
    411 v_{3} \lambda^2)\bigg] \;;
\end{align}
\begin{align}
    f_{+s,-s} = &f_{-s,+s}=\notag\\
=&\frac{1}{256(1-\lambda)^2(1-3\lambda)^2(1-x^2)}\bigg[
 - x^4 \Bigl(64 v_{2}^2 (1 - \lambda)^4 + 81 v_{3}^2 (1 - \lambda)^3
 \lambda \notag\\
&+ 4 (2 - 30 \lambda + 67 \lambda^2 - 51 \lambda^3 + 16
 \lambda^4) + 3 v_{3} (81 - 459 \lambda + 871 \lambda^2 - 685 \lambda^3 + 192
    \lambda^4) \notag\\
&+ 4 v_{2} (1 - \lambda)^2 \bigl(3 v_{3} (7 - 10
    \lambda + 3 \lambda^2) + 4 (13 - 51 \lambda + 35 \lambda^2)\bigr)\Bigr)
    \nonumber \\ 
    &- x^3 \Bigl(78 - 108 v_{3} + 279 v_{3}^2 - 192 v_{1} (1 - \lambda)^4 (7 - 8
    v_{2} - 9 v_{3})  - 194 \lambda + 240 v_{3}
    \lambda \notag\\
&- 1089 v_{3}^2 \lambda \!-\! 694 \lambda^2 \!+\! 1593 v_{3}^2
    \lambda^2 \!+\! 1878 \lambda^3 \!-\! 312 v_{3} \lambda^3 \!-\! 1035 v_{3}^2
    \lambda^3 \!-\! 1036 \lambda^4 \!+\! 
    180 v_{3} \lambda^4 \notag\\
&+ 252 v_{3}^2 \lambda^4 + 16 v_{2}^2 (1 -
    \lambda)^3 (39 - 38 \lambda) + 8 v_{2} (1 - \lambda)^2 \bigl(-48
    + 64 \lambda - 22 \lambda^2 \notag\\
& + 3 v_{3} (40 \!-\! 79 \lambda \!+\! 39 \lambda^2)\bigr)\Bigr) 
+ x^2 (1\! -\! \lambda) \Bigl(10 \!+\! 54 v_{3}\! - \! 81 v_{3}^2
\!-\!  192 v_{1}(1 \!-\! \lambda)^3
    (11 \!+\! 8 v_{2} \!+\! 3 v_{3}) \notag\\
& - 84 \lambda - 528 v_{3}
    \lambda + 252 v_{3}^2 \lambda - 46 \lambda^2 + 918 v_{3} \lambda^2 - 261 v_{3}^2 \lambda^2 + 44
    \lambda^3 - 444 v_{3} \lambda^3 + 90 v_{3}^2 \lambda^3 \nonumber
    \\ 
    &+ 16 v_{2}^2 (1 - \lambda)^2 (-29 + 26 \lambda) - 8 v_{2} (1 -
    \lambda) \bigl(56 - 72 \lambda + 26 \lambda^2 + v_{3} (39 - 69
    \lambda + 30 \lambda^2)\bigr)\Bigr) \nonumber \\ 
    &+ x \Bigl(294 - 180 v_{3} + 315 v_{3}^2 - 192 v_{1} (7 - 8 v_{2}
    - 9 v_{3}) (1 - \lambda)^4 - 1706 \lambda + 792 v_{3} \lambda
    \notag\\
&- 1269 v_{3}^2 \lambda 
+ 2994 \lambda^2 - 1416 v_{3} \lambda^2 + 1917 v_{3}^2 \lambda^2
    - 1842 \lambda^3 + 1152 v_{3} \lambda^3 - 1287 v_{3}^2 \lambda^3
    \notag\\
& + 292 \lambda^4 - 348 v_{3} \lambda^4 + 324 v_{3}^2 \lambda^4 + 16 v_{2}^2 (1 -
    \lambda)^3 (43 - 46 \lambda) + 8 v_{2} (1 - \lambda)^2 \bigl(-48
    + 72 \lambda \nonumber \\ 
    &- 38 \lambda^2 + 3 v_{3} (44 - 91 \lambda + 47
    \lambda^2)\bigr)\Bigr) -218 + 192 v_{1}
    (1 - \lambda)^4 
    (11 + 8 v_{2} + 3  v_{3}) 
 + 261 v_{3} \notag\\
& + 117 v_{3}^2 + 1486 \lambda - 1251
    v_{3} \lambda - 432  v_{3}^2 \lambda - 3426 \lambda^2 + 2199 v_{3}
    \lambda^2 + 594 v_{3}^2 \lambda^2 + 3330 \lambda^3 \notag\\
&- 1677 v_{3} \lambda^3 -
    360 v_{3}^2 \lambda^3 - 1156 \lambda^4 + 468 v_{3} \lambda^4
   + 81 v_{3}^2 \lambda^4 + 16 v_{2}^2 (1 - \lambda)^3 (37 - 38
    \lambda) \notag\\
&+ 4 v_{2} (1 - \lambda)^2 \bigl(3 v_{3} (41 - 80 \lambda
    + 39 \lambda^2) + 4 (41 - 83 \lambda + 40 \lambda^2)\bigr)\bigg]\;.
\end{align}


\subsection{Processes with three and four scalar gravitons}

For scattering with three scalar gravitons --- one in the beginning
and two in the end --- the final and initial momenta are related by
\be
\label{alpha3}
k'=\varkappa k~,~~~~~~\varkappa=\left(\frac{1+u_s}{2u_s}\right)^{1/3}\;.
\ee
The angular dependence of the amplitude reads,
\begin{equation}
     f_{\a s,ss} = \left(\frac{2 (1- \lambda)}{1 - 3
         \lambda}\right)^{3/2}  \frac{(1-x^2)
P_{\a s,ss}(x)}{g_{4}(x)}\;,~~~~~~\a=+,-\;, 
\end{equation}
where
\begin{align}
\label{denom4}
    g_{4}(x) &= \bigl((1 + 2 x \varkappa +\varkappa^2)^3 -  (1 -  u_{s}
    \varkappa^3)^2\bigr) \bigl(u_{s}^2 (1 + 2 x \varkappa + \varkappa^2)^3 -
    (1 -  u_{s} \varkappa^3)^2\bigr) \nonumber \\ 
    & \times \bigl((1 - 2 x \varkappa + \varkappa^2)^3 -  (u_{s} -  u_{s}
    \varkappa^3)^2\bigr) \bigl(u_{s}^2 (1 - 2 x \varkappa + \varkappa^2)^3 -
    (u_{s} -  u_{s} \varkappa^3)^2\bigr)\;, 
\end{align}
and $P_{\a s,ss}(x)$ is a 12th order polynomial. For $u_s\neq 1$ the
denominator has zeros at non-zero angles, and the amplitude vanishes
in the forward and backward limits.

For $u_{s} = 1$ the amplitude simplifies (though it still remains
quite lengthy):
\begin{align}
    f_{+s,ss}=& f_{-s,ss}=f_{ss,+s}=f_{ss,-s}=\notag\\
    =&\frac{1}{64\sqrt{2(1\!-\!\lambda)^3(1\!-\!3\lambda)^5}(1\!-\!x^2)^2}\bigg[ 
  x^6 (1\! -\! \lambda) \Bigl(119 \!-\! 16 v_{2}^2 (1\! -\!
    \lambda)^3 \!-\! 582 \lambda \!+\! 825 \lambda^2 \!-\! 422 \lambda^3 \nonumber
    \\ 
    &+ 9 v_{3}^2 (1 \!-\! \lambda)^2 (1\! -\! 6 \lambda)\! -\! 6 v_{3} (30\! -\!
    147 \lambda \!+\! 193 \lambda^2 \!-\! 74 \lambda^3) \!-\! 4 v_{2} (1 \!-\!
    \lambda) \bigl(35 \!-\! 139 \lambda \!+\! 92 \lambda^2\notag\\
& - v_{3} (3 + 3 \lambda - 6 \lambda^2)\bigr)\Bigr)
    +x^4 \Bigl(-275 + 102 v_{3} + 90 v_{3}^2 - 192 v_{1}(1 - \lambda)^3 \bigl(15 - 4
    v_{2} (1 - \lambda) \notag\\
&- 6 v_{3} (1 - \lambda) - 29 \lambda\bigr) + 1717 \lambda - 966 v_{3}
\lambda - 234 v_{3}^2 \lambda - 3811 \lambda^2 + 2286 v_{3} \lambda^2
+ 162 v_{3}^2 \lambda^2 \notag\\
&+ 3799 \lambda^3 \!-\! 2058 v_{3} \lambda^3 \!+\! 18 v_{3}^2 \lambda^3 \!-\!
    1422 \lambda^4 \!+\! 636 v_{3} \lambda^4 
\!-\! 36 v_{3}^2 \lambda^4 \!+\! 16 v_{2}^2 (1 \!-\! \lambda)^3 (17 \!-\! 15
    \lambda) \notag\\
&- 4 v_{2} (1 - \lambda) \bigl(161 - 554 \lambda + 653
    \lambda^2 - 264 \lambda^3 - 6 v_{3} (1 - \lambda)^2 (19 + 13
    \lambda)\bigr)\Bigr) \nonumber \\ 
    &- x^2 \Bigl(167 - 480 v_{3} + 387 v_{3}^2 - 384 v_{1}(1 -
    \lambda)^3 \bigl(15 -
    8 v_{2} (1 - \lambda) - 9 v_{3} (1 - \lambda) - 31 \lambda\bigr)\notag\\
   & - 1381 \lambda + 2214 v_{3} \lambda - 1539 v_{3}^2 \lambda + 3615 \lambda^2 -
    3540 v_{3} \lambda^2 + 2295 v_{3}^2 \lambda^2 - 3319 \lambda^3
    \nonumber \\ 
    &+ 2322 v_{3} \lambda^3 - 1521 v_{3}^2 \lambda^3 + 902 \lambda^4 -
    516 v_{3} \lambda^4 + 378 v_{3}^2 \lambda^4 + 16 v_{2}^2 (1 -
    \lambda)^3 (67 - 71 \lambda) \nonumber \\ 
    &- 4 v_{2} (1 - \lambda) \bigl(451 - 1790 \lambda + 2183
    \lambda^2 - 836 \lambda^3 - 3 v_{3} (1 - \lambda)^2 (129 - 134
    \lambda)\bigr)\Bigr)\notag\\
&+323 - 546 v_{3} + 288 v_{3}^2 - 576 v_{1}(1 - \lambda)^3 \bigl(5 - 4 v_{2} (1 -
\lambda) - 4 v_{3} (1 - \lambda) - 11 \lambda\bigr) \notag\\
&- 2493 \lambda \!+\! 3126 v_{3} \lambda\! -\! 1224 v_{3}^2 \lambda \!+\! 6595
\lambda^2\! -\! 6234 v_{3} \lambda^2 \!+\! 1944 v_{3}^2 \lambda^2 \!-\! 6927
\lambda^3
\!+\! 5226 v_{3} \lambda^3 \notag\\
&- 1368 v_{3}^2 \lambda^3 + 2478
\lambda^4 - 1572 v_{3} \lambda^4 + 360 v_{3}^2 \lambda^4 + 48 v_{2}^2
(1 - \lambda)^3 (17 - 19 \lambda) \notag\\
&- 12 v_{2} (1 - \lambda)
\bigl(101 - 450 \lambda + 609 \lambda^2 - 256 \lambda^3 - 2 v_{3} (1
- \lambda)^2 (46 - 53 \lambda)\bigr) 
 \bigg]\;.
\end{align} 

Finally, we consider the amplitude with four scalar gravitons. The
kinematics in this case is simple, $k'=k$. Still, the amplitude is
rather lengthy since it involves all vertices in an intricate way. In
general it has the form, 
\begin{equation}
     f_{ss,ss} = \left(\frac{2 (1- \lambda)}{1 - 3
         \lambda}\right)^{2}  \frac{P_{ss,ss}(x)}{(1-x^2)^3}\;, 
\end{equation}
where $P_{ss,ss}(x)$ is an even polynomial of degree $8$. The
amplitude has only forward singularities. 
For $u_s=1$
we have:
\begin{align}
    f_{ss,ss} =& \frac{1}{64(1-\lambda)^2(1-3\lambda)^3(1-x^2)^3} 
\bigg[
 x^8 (1 - \lambda)^2 \Bigl(185 - 32 v_{2}^2 (1 -
    \lambda)^3 - 992 \lambda + 1525 \lambda^2 \notag \\
&- 838 \lambda^3 - 9 v_{3}^2 (1 - \lambda)^2 (1 + 6 \lambda) 
- 8 v_{2} (1 - \lambda) \bigl(15 + 6 v_{3} (1 - \lambda) - 68
    \lambda + 41 \lambda^2\bigr)\notag\\
& + 12 v_{3} (-15 + 80 \lambda - 103
    \lambda^2 + 36 \lambda^3)\Bigr) 
+x^6 \Bigl(-500 \!+\! 408 v_{3} \!+\! 171 v_{3}^2 \!+\! 3592 \lambda \!-\! 3624
    v_{3} \lambda \notag\\
&- 954 v_{3}^2 \lambda - 9152 \lambda^2 + 11280 v_{3}
    \lambda^2 + 2106 v_{3}^2 \lambda^2 + 10584 \lambda^3 - 16344 v_{3}
    \lambda^3 - 2304 v_{3}^2 \lambda^3 \notag\\
&- 5340 \lambda^4 \!+\! 11304 v_{3}
    \lambda^4 \!+\! 1251 v_{3}^2 \lambda^4 \!+\! 768 \lambda^5 
\!-\! 3024 v_{3} \lambda^5 \!-\! 270 v_{3}^2 \lambda^5 \!+\! 192 v_{2}^2 (1\!-\!
     \lambda)^4 (1 \!-\! 2 \lambda)\notag\\
& - 384 v_{1} (1 - \lambda)^3 (1 + 2
    \lambda - 15 \lambda^2) - 48 v_{2} (1 - \lambda)^2 \bigl(-2 + 34
    \lambda - 98 \lambda^2 
    + 70 \lambda^3 \notag\\
& - v_{3} (1 - \lambda)^2 (8 - 15
    \lambda)\bigr)\Bigr) 
+x^4 \Bigl(458 + 585 v_{3}^2 + 73728 v_{1}^2 (1 - \lambda)^5 -
    3844 \lambda + 792 v_{3} \lambda \notag\\
& - 2574 v_{3}^2 \lambda + 12580
    \lambda^2 - 3840 v_{3} \lambda^2 
+ 4446 v_{3}^2 \lambda^2 - 20984 \lambda^3 + 7128 v_{3} \lambda^3
    - 3744 v_{3}^2 \lambda^3 \notag\\
&+ 17554 \lambda^4 - 5904 v_{3} \lambda^4
    + 1521 v_{3}^2 \lambda^4 - 5684 \lambda^5 
+ 1824 v_{3} \lambda^5 + 64 v_{2}^2 (1 -
    \lambda)^4 (124 - 119 \lambda) \notag\\
&- 234 v_{3}^2 \lambda^5 + 384 v_{1} (1 - \lambda)^3
    \bigl(-3 + 128 v_{2} (1 - \lambda)^2 + 42 v_{3} (1 - \lambda)^2
   + 30 \lambda - 55 \lambda^2\bigr) \notag\\
&- 16 v_{2} (1 - \lambda)^2
    \bigl(3 - 151 \lambda + 469 \lambda^2 - 341 \lambda^3 + 15 v_{3}
    (1 - \lambda)^2 (-20 + 17 \lambda)\bigr)\Bigr) \nonumber \\ 
    &+x^2 \Bigl(-300 + 936 v_{3} - 1647 v_{3}^2 - 147456 v_{1}^2 (1 -
    \lambda)^5 + 2744 \lambda - 6936 v_{3} \lambda + 8082 v_{3}^2
    \lambda \notag\\
& - 9120 + 18288 v_{3} \lambda^2 - 15858 v_{3}^2 \lambda^2 + 13992
    \lambda^3 - 22728 v_{3} \lambda^3 + 15552 v_{3}^2 \lambda^3 -
    10532 \lambda^4 \nonumber \\ 
    &+ 13656 v_{3} \lambda^4 \!-\! 7623 v_{3}^2 \lambda^4 \!+\! 3200 \lambda^5
    \!-\! 3216 v_{3} \lambda^5 \!+\! 1494 v_{3}^2 \lambda^5 
\!-\! 64 v_{2}^2 (1 \!-\! \lambda)^4 (255 \!-\! 254 \lambda) \notag\\
&- 384 v_{1}
    (1 - \lambda)^3 \bigl(-33 + 256 v_{2} (1 - \lambda)^2 + 84 v_{3}
    (1 - \lambda)^2 + 150 \lambda - 137 \lambda^2\bigr) \nonumber \\ 
    &+ 16 v_{2} (1 - \lambda)^2 \bigl(234 - 1290 \lambda + 1978
    \lambda^2 - 926 \lambda^3 - 15 v_{3} (1 - \lambda)^2 (44 - 43
    \lambda)\bigr)\Bigr) \nonumber \\ 
    &+733 - 1164 v_{3} + 900 v_{3}^2 + 73728 v_{1}^2 (1 - \lambda)^5
    - 6890 \lambda + 8448 v_{3} \lambda - 4536 v_{3}^2 \lambda + 23886
    \lambda^2 \notag\\
& - 22392 v_{3} \lambda^2 + 9144 v_{3}^2 \lambda^2 - 37880 \lambda^3 + 28080 v_{3}
    \lambda^3 - 9216 v_{3}^2 \lambda^3 + 27949 \lambda^4 - 16956 v_{3}
    \lambda^4 \notag\\
&+ 4644 v_{3}^2 \lambda^4 - 7814 \lambda^5 
+ 3984 v_{3} \lambda^5 - 936 v_{3}^2 \lambda^5 + 32 v_{2}^2 (1 -
    \lambda)^4 (257 - 259 \lambda) \notag\\
&+ 384 v_{1} (1 - \lambda)^3
    \bigl(-29 + 128 v_{2} (1 - \lambda)^2 
+ 42 v_{3} (1 - \lambda)^2 + 122 \lambda - 97 \lambda^2\bigr) \notag\\
&- 8 v_{2} (1 - \lambda)^2 \bigl(459 - 2399 \lambda + 3497 \lambda^2
    - 1549 \lambda^3 - 6 v_{3} (1 - \lambda)^2 (113 - 115
    \lambda)\bigr) \bigg]\;.
\end{align}

\section{Modes and propagators 
with auxiliary field}
\label{app:chi}

The tensor and vector parts of the quadratic Lagrangian following from
the action (\ref{Lchi1}) with the gauge fixing (\ref{eq:chi-GFF}) are
the same as in the original HG action, see Eqs.~(\ref{L2t}),
(\ref{L2v}). The difference, however, occurs in the scalar
sector. Using the same decomposition as in Eqs.~(\ref{eq:mode_decomp})
we obtain,
\be
\label{L2sprim}
\begin{split}
{\tilde{\cal L}_g^{(2s)}}=\frac{1}{2G}\bigg\{&\frac{\dot\psi^2}{2}
+\frac{\dot E^2}{4}-2\chi\dot\psi-\chi\dot E
+\frac{8\nu_4+3\nu_5}{2}\psi\D^3\psi
+\frac{1+\xi}{4\sigma}E\D^3 E\\
&+\dot B\frac{\sigma}{(1+\xi)\D}\dot B+B\D^2 B+2\chi\D B
-2\dot{\bar C}\D\dot C-\frac{2(1+\xi)}{\sigma}\bar C\D^4 C\bigg\}.
\end{split}
\ee 
The ghost part, of course, decouples and leads to a simple propagator
which, combined with the vector contribution, gives
\be
\label{chi-ccprop}
\feynmandiagram[baseline=(a.base), horizontal= a to c]{
a [particle =\(c_i\)]  -- [color=white] b -- [color=white] c
[particle=\(\bar c_j\)],  
a -- [fermion, edge label={}]c,
};
=G\Big[\delta_{ij} {\cal P}_1+\hat k_i\hat k_j (\tilde\P_0-\P_1)\Big]\;,
\ee
where $\P_1$ is given in (\ref{proppoles}) and
\be
\label{newproppole}
\tilde\P_0=\frac{i}{\omega^2-\tilde\nu_0 k^6+i\epsilon}~,~~~~~~
\tilde\nu_0\equiv\frac{1+\xi}{\sigma}\;.
\ee
The other components $\psi$, $E$, $B$, $\chi$ all mix with each
other. To find their propagators, we switch to the Fourier space and
invert the mixing matrix. Combining with the propagators of tensor and
vector components we arrive at, 
\bseq
\label{newprops}
\begin{align}
\label{chichiprop}
\feynmandiagram[baseline=(a.base), horizontal= a to c]{
a [particle =\(\chi\)]  -- [color=white] b -- [color=white] c
[particle=\(\chi\)],  
a -- [scalar, edge label={}]c,
};
&= G\bigg\{-\frac{i}{3} + \frac{2\nu_{s}}{3} k^6
   \mathcal{P}_{s}\bigg\}, \\ 
\label{chiNprop}
\feynmandiagram[baseline=(a.base), horizontal= a to c]{
a [particle =\(\chi\)]  -- [scalar] b -- [double] c
[particle=\(N_i\)],
};
&= iG\frac{\tilde\nu_0k^5 \hat k_i}{3 (\nu_s-\tilde\nu_0)} 
\bigg\{2\nu_{s} \mathcal{P}_s - (3\nu_s-\tilde\nu_0) \tilde{\mathcal{P}}_0\bigg\},
   \\ 
\label{chihprop}
\feynmandiagram[baseline=(a.base), horizontal= a to c]{
a [particle =\(\chi\)]  -- [scalar] b -- [photon] c
[particle=\(h_{ij}\)],  
};
&= iG \frac{2\omega}{3} \bigg\{-\delta_{ij}\mathcal{P}_{s}
 +\hat{k}_{i} \hat{k}_{j} \frac{3\nu_s-\tilde\nu_0}{\nu_s-\tilde\nu_0}
\big[\P_s-\tilde\P_0\big]\bigg\},
     \\ 
\label{newNNprop}
\feynmandiagram[baseline=(a.base), horizontal= a to c]{
a [particle =\(N_i\)]  -- [color=white] b -- [color=white] c
[particle=\(N_j\)],  
a -- [double, edge label={}]c,
};
& = -G\bigg\{ \!\frac{k^4}{ \sigma} (\delta_{ij}\! -\!
   \hat{k}_i \hat{k}_j) \mathcal{P}_{1} 
\!-\! \frac{\tilde\nu_0 k^4\hat{k}_i \hat{k}_j}{3(\nu_s\!\!-\!\tilde\nu_0)} 
\bigg[\frac{2\tilde\nu_0\nu_s}{\nu_s\!\!-\!\tilde\nu_0}\P_s
\!-\!\frac{2\tilde\nu_0^2}{\nu_s\!\!-\!\tilde\nu_0}\tilde\P_0
\!+\!i(3\nu_s\!\!-\!\tilde\nu_0)\omega^2\tilde\P_0^2\bigg]\!\bigg\},
\\ 
\label{Nhprop}
\feynmandiagram[baseline=(a.base), horizontal= a to c]{
a [particle =\(N_i\)]  -- [double] b -- [photon] c
[particle=\(h_{jk}\)],  
};
&=  \frac{2G\omega\tilde\nu_0}{3(\nu_s\!-\!\tilde\nu_0)k}\bigg\{
\!\!\!-\!\hat k_i\delta_{jk}\big[\P_s\!-\!\tilde\P_0\big]
\!+\!\hat k_i\hat k_j\hat k_k
(3\nu_s\!-\!\tilde\nu_0)
\bigg[\frac{\P_s\!-\!\tilde\P_0}{\nu_s\!-\!\tilde\nu_0}
\!+\!i k^6\tilde\P_0^2\bigg]\!\bigg\},
   \\
\label{newhhprop}
\feynmandiagram[baseline=(a.base), horizontal= a to c]{
a [particle =\(h_{ij}\)]  -- [color=white] b -- [color=white] c
[particle=\(h_{kl}\)],  
a--[photon]c,
};
&= 2G\bigg\{(\delta_{i k}\delta_{j l} +
   \delta_{i l} \delta_{j k}) \mathcal{P}_{tt} - \delta_{i
     j}\delta_{kl} \left[ \mathcal{P}_{tt} -\frac{\P_s}{3}\right]
   \nonumber \\  
&\qquad~~~~~
- (\delta_{i k}\hat{k}_{j}\hat{k}_{l} 
    + \delta_{i l}\hat{k}_{j}\hat{k}_{k} + \delta_{j
      k}\hat{k}_{i}\hat{k}_{l} + \delta_{j l}\hat{k}_{i}\hat{k}_{k})
    \big[\mathcal{P}_{tt}-\mathcal{P}_{1} \big] \nonumber \\
    &\qquad~~~~~
+ (\delta_{i j} \hat{k}_{k} \hat{k}_{l} + \delta_{k l}\hat{k}_{i}\hat{k}_{j})
     \bigg[\mathcal{P}_{tt}-\frac{3\nu_s-\tilde\nu_0}{3(\nu_s-\tilde\nu_0)}
\mathcal{P}_{s} + \frac{2\tilde\nu_0}{3(\nu_s-\tilde\nu_0)} 
\tilde{\mathcal{P}}_0 \bigg]
    \nonumber \\ 
 &\qquad~~~~~
+\hat{k}_{i}\hat{k}_{j}\hat{k}_{k}\hat{k}_{l}\bigg[
\mathcal{P}_{tt}
      + \frac{(3\nu_s-\tilde\nu_0)^2}{3(\nu_s-\tilde\nu_0)^2} \mathcal{P}_{s} 
- 4 \mathcal{P}_{1}\notag\\
 &\qquad\qquad\qquad~~~~~~~~
- \frac{4\tilde\nu_0(3\nu_s-2\tilde\nu_0)}{3(\nu_s-\tilde\nu_0)^2}
\tilde{\mathcal{P}}_0 
     + i \frac{2\tilde\nu_0(3 \nu_s-\tilde\nu_0)k^6}{3(\nu_s-\tilde\nu_0)}
      \tilde{\mathcal{P}}_0^2  \bigg]\bigg\}.
\end{align}
\eseq
Here $\P_{tt}$ and $\P_s$ are the same as in (\ref{proppoles}), and
$\nu_s$ is given by the $\lambda\to\infty$ limit of
Eq.~(\ref{oms}), $\nu_s=(8/3)\nu_4+\nu_5$. 

We observe that all propagators (\ref{chi-ccprop}), (\ref{newprops})
are regular in the sense defined in Sec.~\ref{sec:chi}. This follows
from three properties. First, the pole factors $\P_{tt}$, $\P_s$,
$\P_1$, $\tilde\P_0$ are regular. Second, the propagators scale
homogeneously under the Lifshitz transformations, in the way
compatible with the scaling dimensions of the corresponding
fields. And third, the inverse powers of the spatial momentum $k$
contained in the unit vector $\hat k_i$ cancel when we bring the
combinations in the square brackets to the common denominator. As
shown in Ref.~\cite{Barvinsky:2015kil}, the regularity of the
propagators is sufficient for the renormalizability of the theory.

Let us also note the presence of double poles $\tilde\P_0^2$. They
signal presence of a linearly growing gauge mode, similarly as it
happens in the Maxwell theory in general covariant gauges (see
e.g. Sec.~18 of \cite{Strathdee1994}). 
 
Mixing between different components in the Lagrangian (\ref{L2sprim})
implies that the scalar graviton state has overlap not only with the
metric $h_{ij}$, but also the shift $N_i$ and the field $\chi$. To see
this, let us write the eigenmode equations following from
(\ref{L2sprim}):
\bseq
\begin{align}
&\omega^2\psi-2i\omega\chi-3\nu_sk^6\psi=0\;,\\
&\omega^2 E-2i\omega\chi-\tilde\nu_0k^6E=0\;,\\
&-\omega^2B+\tilde\nu_0k^6B-\tilde\nu_0k^4\chi=0\;,\\
&2i\omega\psi+i\omega E-2k^2 B=0\;.
\end{align}
\eseq
Substituting here the dispersion relation of the scalar graviton,
$\omega=\sqrt{\nu_s}\, k^3$, and expressing all fields in terms of
$\psi$ we obtain,
\be
E=-\frac{2\nu_s}{\nu_s-\tilde\nu_0}\psi\;,~~~~~~
B=-\frac{i\tilde\nu_0\sqrt{\nu_s}\,k}{\nu_s-\tilde\nu_0}\psi\;,~~~~~~
\chi=i\sqrt{\nu_s}\,k^3\psi\;.
\ee
So indeed, for the scalar graviton eigenmode all fields are in general
non-vanishing. 

The situation simplifies considerably if we choose the gauge $\xi=-1$,
entailing $\tilde\nu_0=0$. Then the admixture of the scalar graviton
to the shift vanishes, which also eliminates the mixed propagators
(\ref{chiNprop}), (\ref{Nhprop}). The normalization of the scalar
graviton mode is deduced by imposing the canonical commutations
relations on $\psi$ and its conjugate momentum
\[
\pi_\psi=\frac{\dot\psi-2\chi}{2G}\;.
\] 
Collecting everything together, we
find the scalar graviton contribution to the metric and the field
$\chi$ in the $\xi=-1$ gauge,
\bseq
\begin{align}
\label{hdecompnew}
&h_{ij}(\x,t)\ni \sqrt{G}\int\frac{d^3k}{(2\pi)^3 2\omega_s}
\,\ve^{(0')}_{ij}\,h_{\k 0'}\,\e^{-i\omega_s
  t+i\k\x}+\text{h.c.}~,~~~~~~~
\ve_{ij}^{(0')}=\sqrt{\frac{2}{3}}\big(\delta_{ij}-3\hat k_i\hat
k_j\big)\;,\\
&\chi(\x,t)=\sqrt{G}\int\frac{d^3k}{(2\pi)^3 2\omega_s}\,
i\omega_s\sqrt{\frac{2}{3}}\,h_{\k 0'}\,\e^{-i\omega_s
  t+i\k\x}+\text{h.c.}\;,
\end{align}
\eseq
where $h_{\k 0'}$ is the scalar graviton annihilation operator
satisfying 
\be
[h_{\k 0'},h^+_{\k' 0'}]=2\omega_s\, (2\pi)^3\delta(\k-\k')\;.
\ee
This provides us with the expressions for the external lines of the
scalar diagrams for the scalar graviton scattering. The form of the
$h$-line is unchanged, see Eq.~(\ref{extlegsh}), with the polarization
tensor from (\ref{hdecompnew}). Whereas the $\chi$-line reads,
\be
\feynmandiagram[baseline=(a.base), horizontal= a to c]{
a [particle =\(\chi\)] -- [color=white] b -- [color=white] c,
a -- [scalar, edge label={}] c,
};\quad= i\sqrt{\frac{2G}{3}}\,\omega\;.
\ee

\bibliographystyle{JHEP}
\bibliography{library.bib}

\providecommand{\href}[2]{#2}\begingroup\raggedright\begin{thebibliography}{10}

\bibitem{Horava2009}
P.~Horava, {\it {Quantum Gravity at a Lifshitz Point}},  {\em Phys. Rev. D}
  {\bf 79} (2009) 084008, [\href{http://arxiv.org/abs/0901.3775}{{\tt
  arXiv:0901.3775}}].

\bibitem{Blas2010}
D.~Blas, O.~Pujolas, and S.~Sibiryakov, {\it {Models of non-relativistic
  quantum gravity: The Good, the bad and the healthy}},  {\em JHEP} {\bf 04}
  (2011) 018, [\href{http://arxiv.org/abs/1007.3503}{{\tt arXiv:1007.3503}}].

\bibitem{Mukohyama2010}
S.~Mukohyama, {\it {Horava-Lifshitz Cosmology: A Review}},  {\em Class. Quant.
  Grav.} {\bf 27} (2010) 223101, [\href{http://arxiv.org/abs/1007.5199}{{\tt
  arXiv:1007.5199}}].

\bibitem{Sotiriou2010}
T.~P. Sotiriou, {\it {Horava-Lifshitz gravity: a status report}},  {\em J.
  Phys. Conf. Ser.} {\bf 283} (2011) 012034,
  [\href{http://arxiv.org/abs/1010.3218}{{\tt arXiv:1010.3218}}].

\bibitem{Wang2017}
A.~Wang, {\it {Ho\v{r}ava gravity at a Lifshitz point: A progress report}},
  {\em Int. J. Mod. Phys. D} {\bf 26} (2017), no.~07 1730014,
  [\href{http://arxiv.org/abs/1701.06087}{{\tt arXiv:1701.06087}}].

\bibitem{Barvinsky2023}
A.~O. Barvinsky, {\it {Ho\v{r}ava models as palladium of unitarity and
  renormalizability in quantum gravity}},
  \href{http://arxiv.org/abs/2301.13580}{{\tt arXiv:2301.13580}}.

\bibitem{Stelle:1976gc}
K.~S. Stelle, {\it {Renormalization of Higher Derivative Quantum Gravity}},
  {\em Phys. Rev. D} {\bf 16} (1977) 953--969.

\bibitem{Stelle:1977ry}
K.~S. Stelle, {\it {Classical Gravity with Higher Derivatives}},  {\em Gen.
  Rel. Grav.} {\bf 9} (1978) 353--371.

\bibitem{Salvio2014}
A.~Salvio and A.~Strumia, {\it {Agravity}},  {\em JHEP} {\bf 06} (2014) 080,
  [\href{http://arxiv.org/abs/1403.4226}{{\tt arXiv:1403.4226}}].

\bibitem{Einhorn2014}
M.~B. Einhorn and D.~R.~T. Jones, {\it {Naturalness and Dimensional
  Transmutation in Classically Scale-Invariant Gravity}},  {\em JHEP} {\bf 03}
  (2015) 047, [\href{http://arxiv.org/abs/1410.8513}{{\tt arXiv:1410.8513}}].

\bibitem{Salvio:2015gsi}
A.~Salvio and A.~Strumia, {\it {Quantum mechanics of 4-derivative theories}},
  {\em Eur. Phys. J. C} {\bf 76} (2016), no.~4 227,
  [\href{http://arxiv.org/abs/1512.01237}{{\tt arXiv:1512.01237}}].

\bibitem{Strumia:2017dvt}
A.~Strumia, {\it {Interpretation of quantum mechanics with indefinite norm}},
  {\em MDPI Physics} {\bf 1} (2019), no.~1 17--32,
  [\href{http://arxiv.org/abs/1709.04925}{{\tt arXiv:1709.04925}}].

\bibitem{Blas2009}
D.~Blas, O.~Pujolas, and S.~Sibiryakov, {\it {Consistent Extension of Horava
  Gravity}},  {\em Phys. Rev. Lett.} {\bf 104} (2010) 181302,
  [\href{http://arxiv.org/abs/0909.3525}{{\tt arXiv:0909.3525}}].

\bibitem{EmirGumrukcuoglu:2017cfa}
A.~Emir~G\"umr\"uk\c{c}\"uo\u{g}lu, M.~Saravani, and T.~P. Sotiriou, {\it
  {Ho\v{r}ava gravity after GW170817}},  {\em Phys. Rev. D} {\bf 97} (2018),
  no.~2 024032, [\href{http://arxiv.org/abs/1711.08845}{{\tt
  arXiv:1711.08845}}].

\bibitem{Bellorin:2022qeu}
J.~Bellorin, C.~Borquez, and B.~Droguett, {\it {Cancellation of divergences in
  the nonprojectable Ho\v{r}ava theory}},  {\em Phys. Rev. D} {\bf 106} (2022),
  no.~4 044055, [\href{http://arxiv.org/abs/2207.08938}{{\tt
  arXiv:2207.08938}}].

\bibitem{Bellorin:2022efu}
J.~Bellorin, C.~Borquez, and B.~Droguett, {\it {BRST symmetry and unitarity of
  the Ho\v{r}ava theory}},  {\em Phys. Rev. D} {\bf 107} (2023), no.~4 044059,
  [\href{http://arxiv.org/abs/2212.14079}{{\tt arXiv:2212.14079}}].

\bibitem{Barvinsky:2015kil}
A.~O. Barvinsky, D.~Blas, M.~Herrero-Valea, S.~M. Sibiryakov, and C.~F.
  Steinwachs, {\it {Renormalization of Ho\v{r}ava gravity}},  {\em Phys. Rev.
  D} {\bf 93} (2016), no.~6 064022,
  [\href{http://arxiv.org/abs/1512.02250}{{\tt arXiv:1512.02250}}].

\bibitem{Barvinsky:2017zlx}
A.~O. Barvinsky, D.~Blas, M.~Herrero-Valea, S.~M. Sibiryakov, and C.~F.
  Steinwachs, {\it {Renormalization of gauge theories in the background-field
  approach}},  {\em JHEP} {\bf 07} (2018) 035,
  [\href{http://arxiv.org/abs/1705.03480}{{\tt arXiv:1705.03480}}].

\bibitem{Barvinsky2019}
A.~O. Barvinsky, M.~Herrero-Valea, and S.~M. Sibiryakov, {\it {Towards the
  renormalization group flow of Horava gravity in $(3+1)$ dimensions}},  {\em
  Phys. Rev. D} {\bf 100} (2019), no.~2 026012,
  [\href{http://arxiv.org/abs/1905.03798}{{\tt arXiv:1905.03798}}].

\bibitem{Barvinsky2021}
A.~O. Barvinsky, A.~V. Kurov, and S.~M. Sibiryakov, {\it {Beta functions of
  (3+1)-dimensional projectable Ho\v{r}ava gravity}},  {\em Phys. Rev. D} {\bf
  105} (2022), no.~4 044009, [\href{http://arxiv.org/abs/2110.14688}{{\tt
  arXiv:2110.14688}}].

\bibitem{Gumrukcuoglu2011_1}
A.~E. Gumrukcuoglu and S.~Mukohyama, {\it {Horava-Lifshitz gravity with
  $\lambda\to\infty$}},  {\em Phys. Rev. D} {\bf 83} (2011) 124033,
  [\href{http://arxiv.org/abs/1104.2087}{{\tt arXiv:1104.2087}}].

\bibitem{Frenkel2020}
A.~Frenkel, P.~Horava, and S.~Randall, {\it {Perelman's Ricci Flow in
  Topological Quantum Gravity}},  \href{http://arxiv.org/abs/2011.11914}{{\tt
  arXiv:2011.11914}}.

\bibitem{xAct}
J.~M. Martin-Garcia, ``{xAct: Efficient tensor computer algebra for the Wolfram
  Language}.'' \url{http://www.xact.es/}.

\bibitem{xPerm}
J.~M. Martin-Garcia, {\it {xPerm: fast index canonicalization for tensor
  computer algebra}},  {\em Comput. Phys. Commun.} {\bf 179} (2008) 597--603,
  [\href{http://arxiv.org/abs/0803.0862}{{\tt arXiv:0803.0862}}].

\bibitem{xPert}
D.~Brizuela, J.~M. Martin-Garcia, and G.~A. Mena~Marugan, {\it {xPert: Computer
  algebra for metric perturbation theory}},  {\em Gen. Rel. Grav.} {\bf 41}
  (2009) 2415--2431, [\href{http://arxiv.org/abs/0807.0824}{{\tt
  arXiv:0807.0824}}].

\bibitem{xTras}
T.~Nutma, {\it {xTras : A field-theory inspired xAct package for mathematica}},
   {\em Comput. Phys. Commun.} {\bf 185} (2014) 1719--1738,
  [\href{http://arxiv.org/abs/1308.3493}{{\tt arXiv:1308.3493}}].

\bibitem{Mathematica}
{Wolfram Research{,} Inc.}, ``{Mathematica, {V}ersion 12.2}.''
\newblock Champaign, IL, 2020.

\bibitem{github}
J.~I. Radkovski. \url{https://github.com/JIRadkovski}.

\bibitem{Koyama:2009hc}
K.~Koyama and F.~Arroja, {\it {Pathological behaviour of the scalar graviton in
  Ho\v{r}ava-Lifshitz gravity}},  {\em JHEP} {\bf 03} (2010) 061,
  [\href{http://arxiv.org/abs/0910.1998}{{\tt arXiv:0910.1998}}].

\bibitem{Izumi2011}
K.~Izumi and S.~Mukohyama, {\it {Nonlinear superhorizon perturbations in
  Horava-Lifshitz gravity}},  {\em Phys. Rev. D} {\bf 84} (2011) 064025,
  [\href{http://arxiv.org/abs/1105.0246}{{\tt arXiv:1105.0246}}].

\bibitem{Gumrukcuoglu2011}
A.~E. Gumrukcuoglu, S.~Mukohyama, and A.~Wang, {\it {General relativity limit
  of Horava-Lifshitz gravity with a scalar field in gradient expansion}},  {\em
  Phys. Rev. D} {\bf 85} (2012) 064042,
  [\href{http://arxiv.org/abs/1109.2609}{{\tt arXiv:1109.2609}}].

\bibitem{Becchi:1975nq}
C.~Becchi, A.~Rouet, and R.~Stora, {\it {Renormalization of Gauge Theories}},
  {\em Annals Phys.} {\bf 98} (1976) 287--321.

\bibitem{Tyutin:1975qk}
I.~V. Tyutin, {\it {Gauge Invariance in Field Theory and Statistical Physics in
  Operator Formalism}},  {\em LEBEDEV-75-39} (1975)
  [\href{http://arxiv.org/abs/0812.0580}{{\tt arXiv:0812.0580}}].

\bibitem{Weinberg:1996kr}
S.~Weinberg, {\em {The quantum theory of fields. Vol. 2: Modern applications}}.
\newblock Cambridge University Press, 2013.

\bibitem{Kugo:1977yx}
T.~Kugo and I.~Ojima, {\it {Manifestly Covariant Canonical Formulation of
  Yang-Mills Field Theories. 1}},  {\em Prog. Theor. Phys.} {\bf 60} (1978)
  1869.

\bibitem{Kugo:1979gm}
T.~Kugo and I.~Ojima, {\it {Local Covariant Operator Formalism of Nonabelian
  Gauge Theories and Quark Confinement Problem}},  {\em Prog. Theor. Phys.
  Suppl.} {\bf 66} (1979) 1--130.

\bibitem{Becchi:1996yh}
C.~Becchi, {\it {Introduction to BRS symmetry}},
  \href{http://arxiv.org/abs/hep-th/9607181}{{\tt hep-th/9607181}}.

\bibitem{Iengo2010}
R.~Iengo and M.~Serone, {\it {A Simple UV-Completion of QED in 5D}},  {\em
  Phys. Rev. D} {\bf 81} (2010) 125005,
  [\href{http://arxiv.org/abs/1003.4430}{{\tt arXiv:1003.4430}}].

\bibitem{Sannan1986}
S.~Sannan, {\it {Gravity as the Limit of the Type {II} Superstring Theory}},
  {\em Phys. Rev. D} {\bf 34} (1986) 1749.

\bibitem{Blas2009Comment}
D.~Blas, O.~Pujolas, and S.~Sibiryakov, {\it {Comment on `Strong coupling in
  extended Horava-Lifshitz gravity'}},  {\em Phys. Lett. B} {\bf 688} (2010)
  350--355, [\href{http://arxiv.org/abs/0912.0550}{{\tt arXiv:0912.0550}}].

\bibitem{Anselmi:2008bq}
D.~Anselmi, {\it {Weighted power counting and Lorentz violating gauge theories.
  I. General properties}},  {\em Annals Phys.} {\bf 324} (2009) 874--896,
  [\href{http://arxiv.org/abs/0808.3470}{{\tt arXiv:0808.3470}}].

\bibitem{Elvang:2013cua}
H.~Elvang and Y.-t. Huang, {\it {Scattering Amplitudes}},
  \href{http://arxiv.org/abs/1308.1697}{{\tt arXiv:1308.1697}}.

\bibitem{Collins:2019ozc}
J.~Collins, {\it {A new approach to the LSZ reduction formula}},
  \href{http://arxiv.org/abs/1904.10923}{{\tt arXiv:1904.10923}}.

\bibitem{Itzykson1980}
C.~Itzykson and J.~Zuber, {\em Quantum Field Theory}.
\newblock Dover Books on Physics. Dover Publications, 2012.

\bibitem{Strathdee1994}
J.~Strathdee, ``{QED Lecture Notes}.'' ICTP, 1995.
  \url{http://streaming.ictp.it/preprints/P/95/315.pdf}.

\end{thebibliography}\endgroup
\end{document}